\documentclass[reprint,aps,pra,twocolumn,english,showpacs,superscriptaddress,amssymb,amsfonts]{revtex4-1}

\usepackage{graphicx}
\usepackage{epsfig}
\usepackage{color}
\usepackage{amsmath}
\usepackage{amsfonts}
\usepackage{mathrsfs}
\usepackage{txfonts}
\usepackage{color}

\usepackage[breaklinks]{hyperref} 
\hypersetup{colorlinks=true, linkcolor=blue, citecolor=blue, filecolor=blue, urlcolor=blue} 

\newcommand{\pone}{{\normalsize \textcircled{\scriptsize{1}}}}
\newcommand{\ptwo}{{\normalsize \textcircled{\scriptsize{2}}}}
\newcommand{\pthree}{{\normalsize \textcircled{\scriptsize{3}}}}
\newcommand{\pfour}{{\normalsize \textcircled{\scriptsize{4}}}}

\begin{document}
	\title{Floquet Higher-Order Topological Insulators with Anomalous Dynamical Polarization}
	\author{Biao Huang}
	\email{phys.huang.biao@gmail.com} 
	\affiliation{Department of Physics and Astronomy, University of Pittsburgh, Pittsburgh PA 15260, USA}
	\author{W. Vincent Liu}
	\email{wvliu@pitt.edu}
	\affiliation{Department of Physics and Astronomy, University of Pittsburgh, Pittsburgh PA 15260, USA}
	\affiliation{
		Wilczek Quantum Center, School of Physics and Astronomy and T. D. Lee Institute,  Shanghai Jiao Tong University, Shanghai 200240, China}
	\affiliation{
		Shenzhen Institute for Quantum Science and Engineering and Department of Physics, Southern University of Science and Technology, Shenzhen 518055, China
		}
	\date{\today}
	
	\begin{abstract}
		Higher order topological insulators (HOTI) have emerged as a new class of phases, whose robust in-gap ``corner'' modes arise from the bulk higher-order multipoles beyond the dipoles in conventional topological insulators. Here, we incorporate Floquet driving into HOTI's, and report for the first time a dynamical polarization theory with anomalous non-equilibrium multipoles. Further, a proposal to detect not only corner states but also their dynamical origin in cold atoms is demonstrated, with the latter one never achieved before. Experimental determination of anomalous Floquet corner modes are also proposed.
	\end{abstract}
	\maketitle
	
	\noindent{\em\color{blue} Introduction} --- Highly nonequilibrium systems challenged our understanding of many-body physics recently. The two cornerstones --- spontaneous symmetry breaking and topology --- both witnessed major conceptual revolution as time is introduced to play unexpected essential roles. Novel examples include the Floquet time crystals~\cite{Khemani2016,Else2016a,Yao2017a,Zhang2017,Choi2017,Rovny2018,Sacha2015,Ho2017,Huang2018,Sacha2017,Yao2018}, with broken time-translation symmetry enforced by spectral pairing~\cite{Keyserlingk2016} rather than the usual energetics. Also, periodic-driven systems support ``anomalous'' edge states, whose topological origin transcends any description by effective Hamiltonians~\cite{Rudner2013,Roy2016,Yao2017,Mukherjee2017,Peng2016,Maczewsky2016,Keyserlingk2016a,Else2016,Potirniche2016}.
	
	
	Meanwhile, rapid progress in cold atom experiments are made regarding dynamical controls~\cite{Eckardt2017,Cooper2019}. High tunability of lattice quench not only reveals fundamental quantities like momentum resolved Berry curvature~\cite{Hauke2014,Flaeschner2016} and Wilson lines~\cite{Li2016} inaccessible before, but also introduces new topological objects such as Bloch-wave dynamical vortices characterized by spacetime linking number~\cite{Flaeschner2018,Sun2018,Wang2017,Zhang2018a}. More recently, the advent of quantum gas microscopes further enables  dressing and probing with single-site accuracy~\cite{Endres2016,Mazurenko2017,Tai2017,Brown2018}.
	
	Inspired by the development, we explore a new class of non-equilibrium phase, dubbed the Floquet higher-order topological insulators (FHOTI), and expose its unique signatures accessible via latest technologies. Here, ``higher-order'' refers to the Bloch-wave multipole polarization compared with dipole ones in conventional topological insulators, resulting in unusual corner/hinge states with reduced dimensionality. Soon after the pioneering works~\cite{Benalcazar2017,Benalcazar2017a}, extensive experimental breakthroughs~\cite{Noh2018,Serra-Garcia2017,Peterson2017,Imhof2017,WangZJ2018,Schindler2018a} and theoretical generalizations~\cite{Kunst2017,Schindler2018,Song2017,Langbehn2017,Trifunovic2018,Calugaru2018,Queiroz2018,Ezawa2018,Lin2017,Ezawa2017,Ezawa2018a,Shapourian2017,Wang2018a,Khalaf2018,Zhu2018,Yan2018,Wang2018,Liu2018a,Dwivedi2018,You2018,Rasmussen2018} have emerged~\footnote{See also earlier related works~\cite{Zhang2013a,Seradjeh2008a}}. Yet, researches are stuck to static cases as the concept of ``dynamical multipoles" still awaits definition. Also, it is non-trivial to prove experimentally not only the existence but also the dynamical origin of corner/hinge states, with the latter one never achieved before. 
	
	This Letter addresses three critical problems to initiate researches on FHOTI. {\bf (1)} A highly solvable model with rich and analytically obtainable phase diagram is constructed. It serves as direct experimental blueprints in multiple platforms. {\bf (2)} The theory of anomalous dynamical  polarization is obtained for the first time, featuring both conceptual and phenomenological differences from static ones. {\bf (3)} Combining microscope and band-tomography techniques, we design the experimental scheme to conclusively prove both the existence of corner states and triviality of Floquet operators. Theoretically, these results pave the way to incorporating dynamical multipoles to, for instance, quantum spin liquids~\cite{Po2017}, Floquet symmetry protected topological phases~\cite{Keyserlingk2016a,Else2016,Potirniche2016}, and holographic Floquet matters in topological semimetals and superconductors~\cite{Hashimoto2017,Kinoshita2018,Ishii2018}. Experimentally, 
	our detection scheme in cold atom may help realize the long-sought goal of  distinguishing dynamical boundary states from static ones. Recently, there has developed deep connections between Floquet phases and other central issues,  including localization~\cite{Keyserlingk2016a,Titum2016,Potter2016,Fidkowski2017,Shapiro2018,Reiss2018} and quantum computations with dynamical Majorana modes~\cite{Liu2013,Thakurathi2017,Bauer2018}. Introducing multipole features into Floquet physics would surely add a new dimension towards phenomena of intrinsically non-equilibrium nature.

	\begin{figure}
		[h]
		\parbox{2.5cm}{\includegraphics[width=2.5cm]{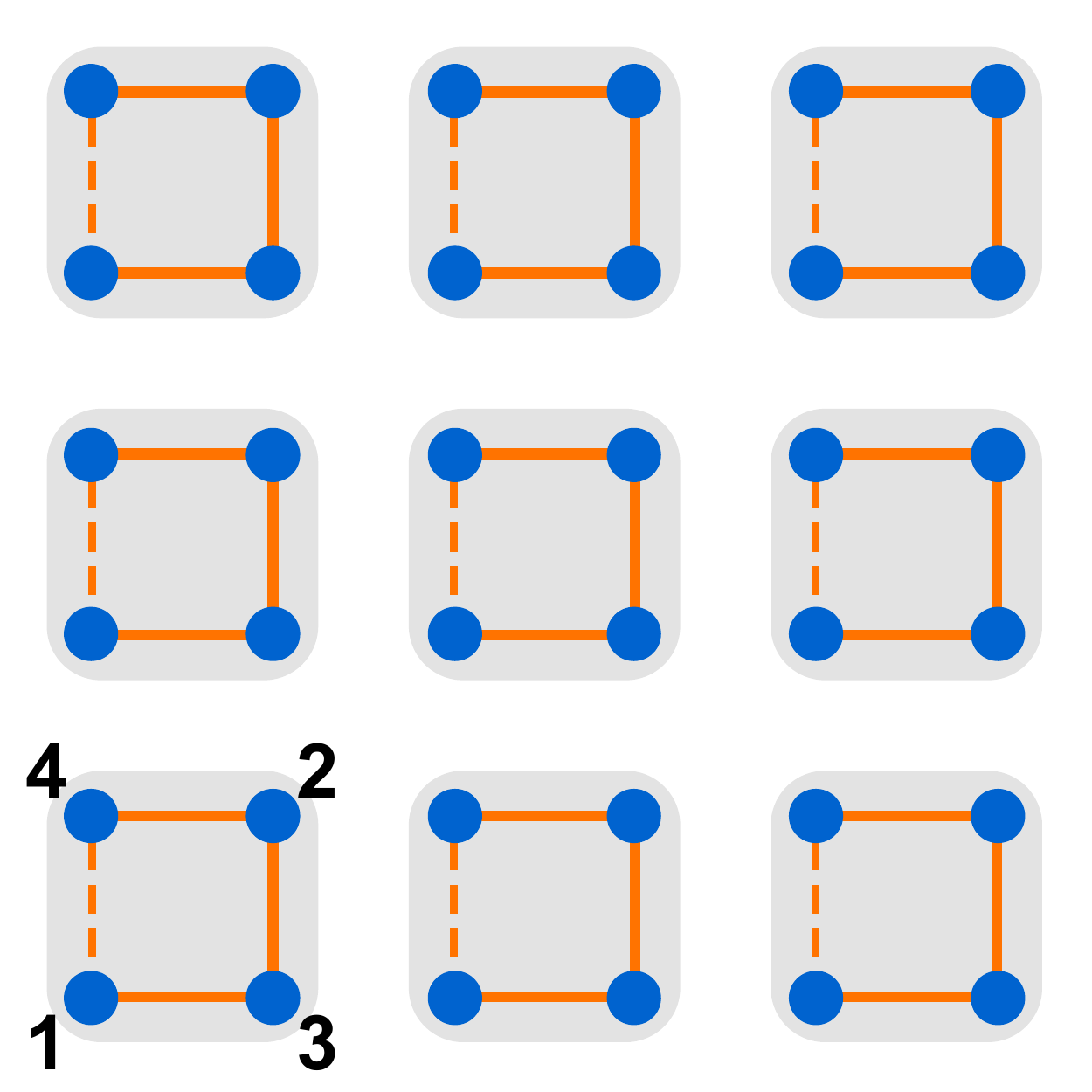}\\
			(a) $ h_{1} $}
		\parbox{2.5cm}{\includegraphics[width=2.5cm]{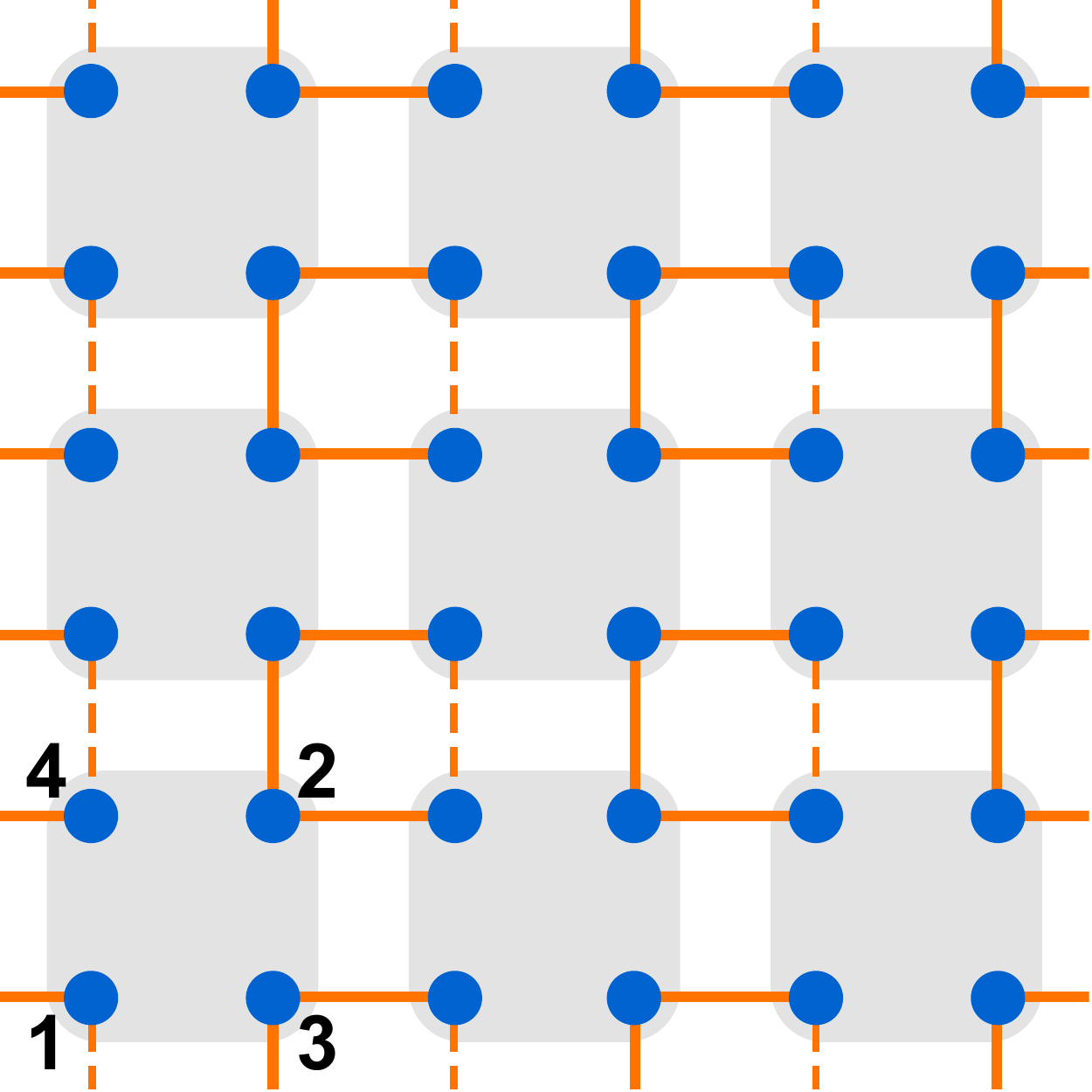}\\
			(b) $ h_{2\boldsymbol{k}} $}
		\parbox{3.4cm}{
		\includegraphics[width=3.2cm]{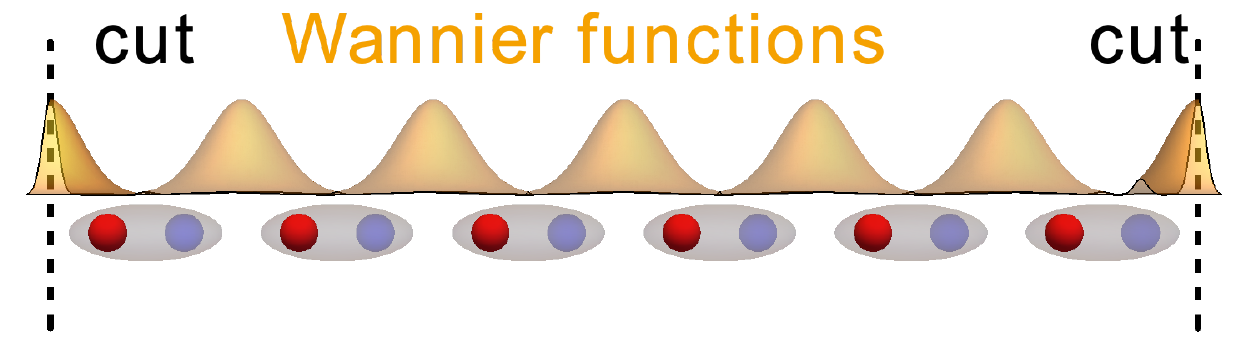}\\
		(c) Static  for $ [U(\boldsymbol{k},T)]_\varepsilon^{t/T} $\\
		\quad \\
		\includegraphics[width=3.2cm]{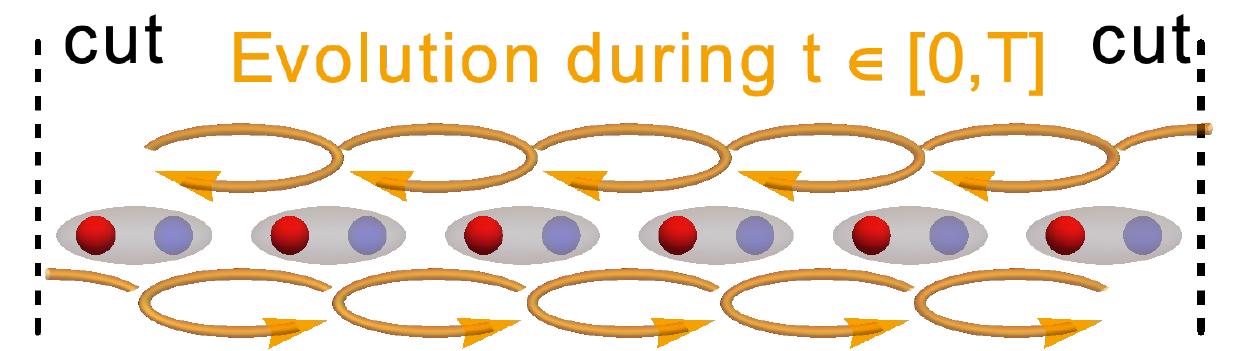}\\
		(d) Dynamical for $ U_\varepsilon(\boldsymbol{k},t) $
		}
		\caption{\label{fig:model} Connected tetramers during periodic drive at (a) step 1, 3 and (b) step 2. Solid/dashed lines represent positive/negative signs for the hopping with homogeneous magnitudes. (c) Static versus (d) dynamical polarizations. }
	\end{figure}
	\noindent {\em \color{blue} Models and the limitation of static polarization} --- Consider the binary Floquet drive 
	$ H(\boldsymbol{k},t+T) = H(\boldsymbol{k},t) $
	in Fig.~\ref{fig:model}, where
	\begin{align}\label{eq:hamiltonian}
	H(\boldsymbol{k},t) = \left\{
		\begin{array}{ll}
		\gamma' h_1, & t\in [0, T/4];\\
		\lambda' h_{2\boldsymbol{k}}, & t\in(T/4,  3T/4];\\
		\gamma' h_1, & t\in(3T/4,T]
		\end{array}
		\right. .
	\end{align}
	Here, time origin is shifted to make it transparent $ H(\boldsymbol{k},t)=H(\boldsymbol{k},-t) $, 
	and $ \gamma', \lambda' $ are hopping constants.
	The system is controlled by dimensionless numbers $ (\gamma,\lambda)\equiv (\gamma' T/2\hbar, \lambda' T/2\hbar) $, and all energy (or time) carry units $ 2\hbar/T $ (or $ T/2 $). The matrices $ h_{1} = \tau_1\sigma_0 - \tau_2\sigma_2 $, and $ h_{2\boldsymbol{k}} = \cos k_x \tau_1\sigma_0 - \sin k_x \tau_2\sigma_3 - \cos k_y\tau_2\sigma_2 - \sin k_y \tau_2\sigma_1 $, where $ \boldsymbol{\tau}, \boldsymbol{\sigma} $ are Pauli matrices spanning the basis for 4 sublattices $ 
	(\hat{c}_{\boldsymbol{k}1}, \hat{c}_{\boldsymbol{k}2}, \hat{c}_{\boldsymbol{k}3}, \hat{c}_{\boldsymbol{k}4})^T $~\footnote{The direct product of Pauli matrices $ \boldsymbol{\tau}, \boldsymbol{\sigma} $ can be understood as, i.e. $ \tau_3\sigma_1 = \begin{pmatrix}
		\sigma_1 & 0 \\ 0 & -\sigma_1
	\end{pmatrix}$.}.
	Then, the evolution operator
	$ U(\boldsymbol{k},t) = P_\tau e^{-(i/\hbar) \int_{0}^t H(\boldsymbol{k},\tau) d\tau } $,
	with $ P_\tau $ time ordering. Such a tetramer model enjoys high solvability and rich phase diagrams, as we shall see.

	``Normal" and ``anomalous" Floquet physics can be separated by decomposing ~\cite{Rudner2013,Roy2016,Yao2017} 
	\begin{align}\label{eq:twous}
	U(\boldsymbol{k},t) = U_\varepsilon(\boldsymbol{k},t) [U(\boldsymbol{k},T)]_\varepsilon^{t/T}.
	\end{align}
	Here,  the ``normal" part $ [U(\boldsymbol{k},T)]_\varepsilon^{t/T} $ takes the $ t/T $ exponential with branch cut $ \varepsilon $, $ (e^{iE_{\alpha,\boldsymbol{k}}})_\varepsilon^{t/T} $, on eigenvalues of the Floquet operator $ [U(\boldsymbol{k},T) ]_{mn} = \sum_\alpha [u_{\alpha,\boldsymbol{k}} ]_m e^{ iE_{\alpha{\boldsymbol{k}}}} [ u_{\alpha,\boldsymbol{k}} ]^*_n $. $ m,n$ (or $ \alpha $) $ =1,\dots,4 $ are sublattice (or band) indices.
	With accumulative band physics factored out to $ [U(\boldsymbol{k},T)]_\varepsilon^{t/T} $, the periodic part $ U_\varepsilon(\boldsymbol{k},t+T)=U_\varepsilon(\boldsymbol{k},t) $ describes ``anomalous" Floquet topology from purely dynamical processes. Choosing $ \varepsilon $ in certain bulk gap for $ U(\boldsymbol{k},T) $ means investigating possible in-gap open-boundary states 
	resulting from $ U_\varepsilon(\boldsymbol{k},t) $.
	
	Higher order polarization theory lies at the heart and gives rise to the name ``higher-order" topological insulators. Thus, it is worth reviewing its conventional static version~\cite{Benalcazar2017} to understand the limitations. Static polarization focuses on position operators projected to occupied {\em bands} $ \alpha $, $ \hat{y}_{\text{occ}}= \hat{P}_{\text{occ}} \hat{y} \hat{P}_{\text{occ}} $, where $ \hat{P}_{\text{occ}} = \sum_{\alpha=1}^{N_{\text{occ}}} \hat{\gamma}_{\alpha\boldsymbol{k}}^\dagger |0\rangle \langle 0 | \hat{\gamma}_{\alpha\boldsymbol{k}} $ are projectors with $ \hat{\gamma}_{\alpha\boldsymbol{k}}^\dagger = \sum_m [u_{\alpha\boldsymbol{k}}]_m \hat{c}_{\boldsymbol{k}m}^\dagger $ band creation operators. $ \hat{y}=\sum_{im} \hat{c}_{im}^\dagger |0\rangle e^{-i\Delta_y y_i} \langle 0 | \hat{c}_{im} $ is Resta's~\cite{Resta1998} position operator satisfying periodic boundary condition, with $ i=(x_i,y_i) $ denoting unit cells on $ L_x\times L_y $ lattices, and $ \Delta_y=2\pi/L_y $.
	Essential physics is that collective (Wannier) wave functions in the occupied bands may be polarized away from unit cells by $ \nu_j $ and center in between, as illustrated in Fig.~\ref{fig:model}(c). This is captured by the eigenvalues of $ \hat{y}_{\text{occ}} = \sum_{j,y_i} |w_j(y_i) \rangle e^{-i\Delta_y(y_i+\nu_j)} \langle w_j(y_i)|  $, compared with $ e^{-i\Delta_y y_i} $ for $ \hat{y} $. The static polarization is then $ p_y(x_i)= (2\pi L_y)^{-1} \sum_{jy_ik_ym} | \langle 0| \hat{c}_{k_y,mx_i} | w_j(y_i)\rangle |^2 \nu_j $~(see Supplemental Material (SM)~\cite{supp} for computation details).
	
	Clearly, the very definition of static polarization necessarily involves (quasi-energy) bands applicable only to the Floquet operator $ U(\boldsymbol{k},T) $ (or the ``normal" part $ [U(\boldsymbol{k},T)]_\varepsilon^{t/T} $). 
	In contrast, the ``anomalous" $ U_\varepsilon(\boldsymbol{k},t) $ reduces to identity operators at $ t=T $ and 
	could be gapless
	at any instant $ t\in(0,T) $~\footnote{In fact, $ U_\varepsilon(\boldsymbol{k},t) $ being gapless at certain $ (\boldsymbol{k},t) $ leads to the well-known concept of ``dynamical singularity''~\cite{Nathan2015}}. A completely new concept of dynamical higher-order polarization
	for $ U_\varepsilon(\boldsymbol{k},t) $,
	missing in current literature, needs to be introduced later.

	\begin{figure}
		[h]
		\boxed{
			\parbox{2.4cm}{
				\includegraphics[width=2.5cm]{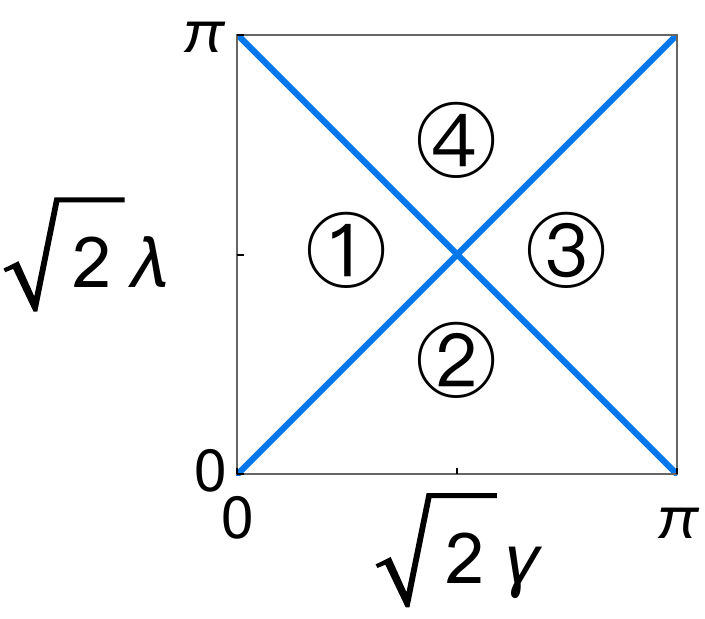}\\
				(a) Phase diagram
		}}
		\boxed{
			\parbox{2.5cm}{
				\includegraphics[width=2.6cm]{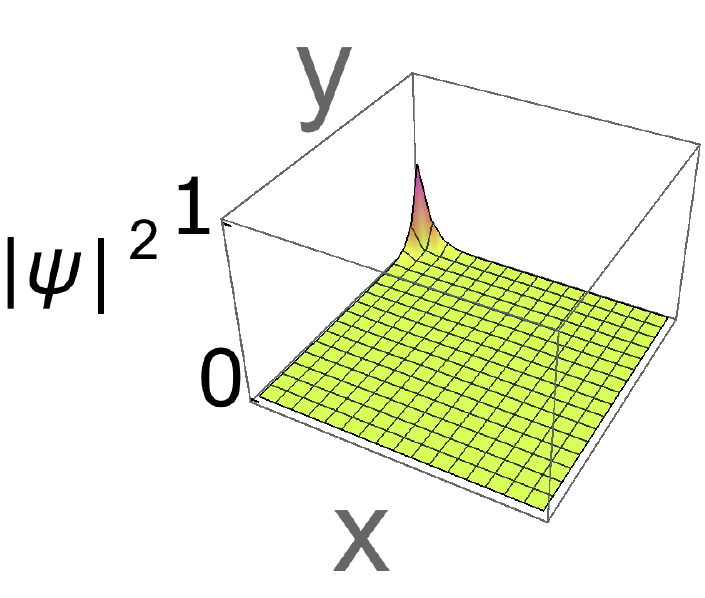}
				(b) Corner state
		}}
		\boxed{
		\parbox{2.8cm}{
			\setlength{\unitlength}{1cm}
			\begin{picture}(2.7,2.2)
			\put(-0.05,-0.35){ \includegraphics[width=3cm,trim=1.5cm 0 0 1.2cm]{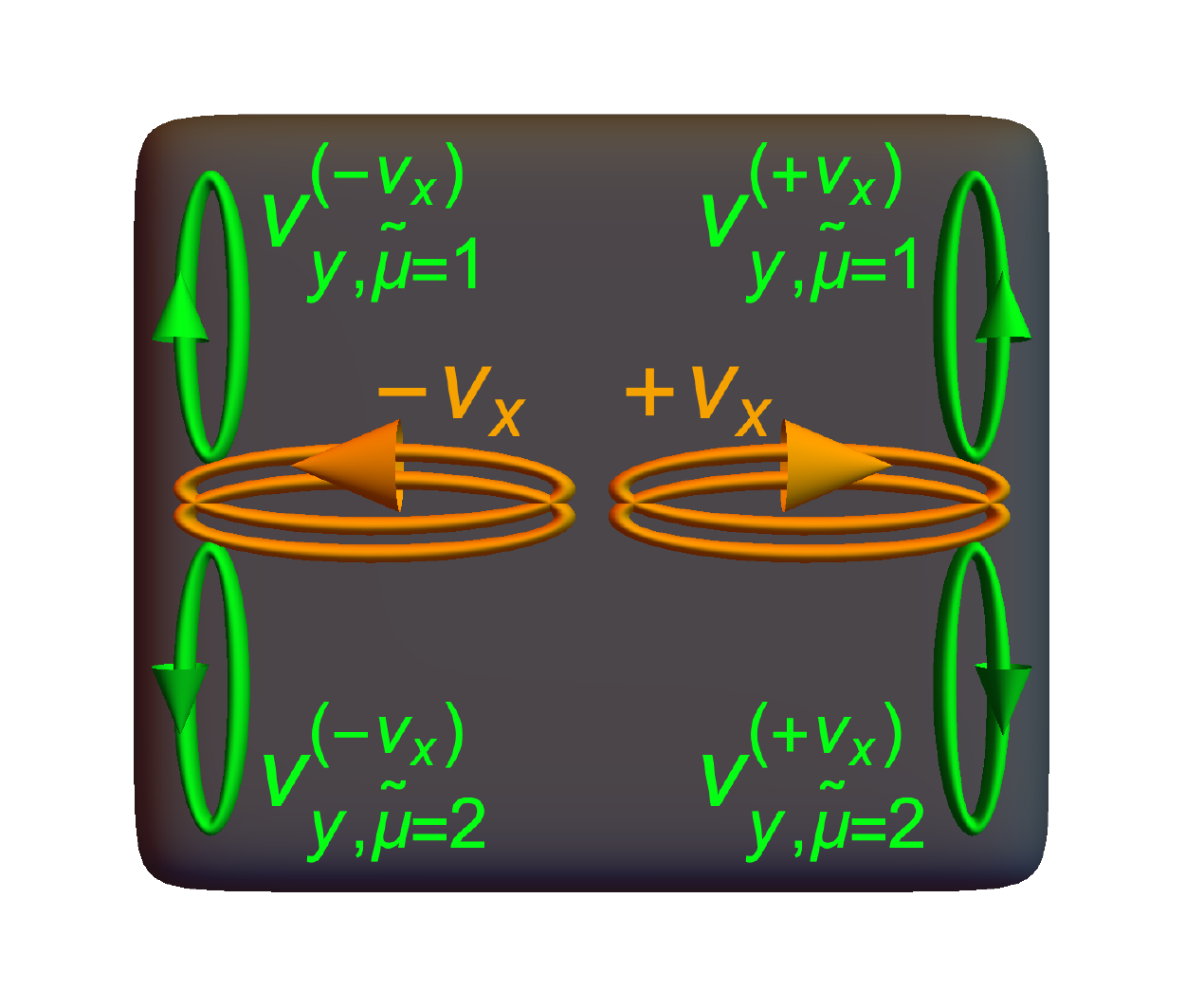}  }
			\end{picture}\\
			(c) Motion quadruplet
		}
		}
		\\
		\boxed{
			\includegraphics[width=2.5cm]{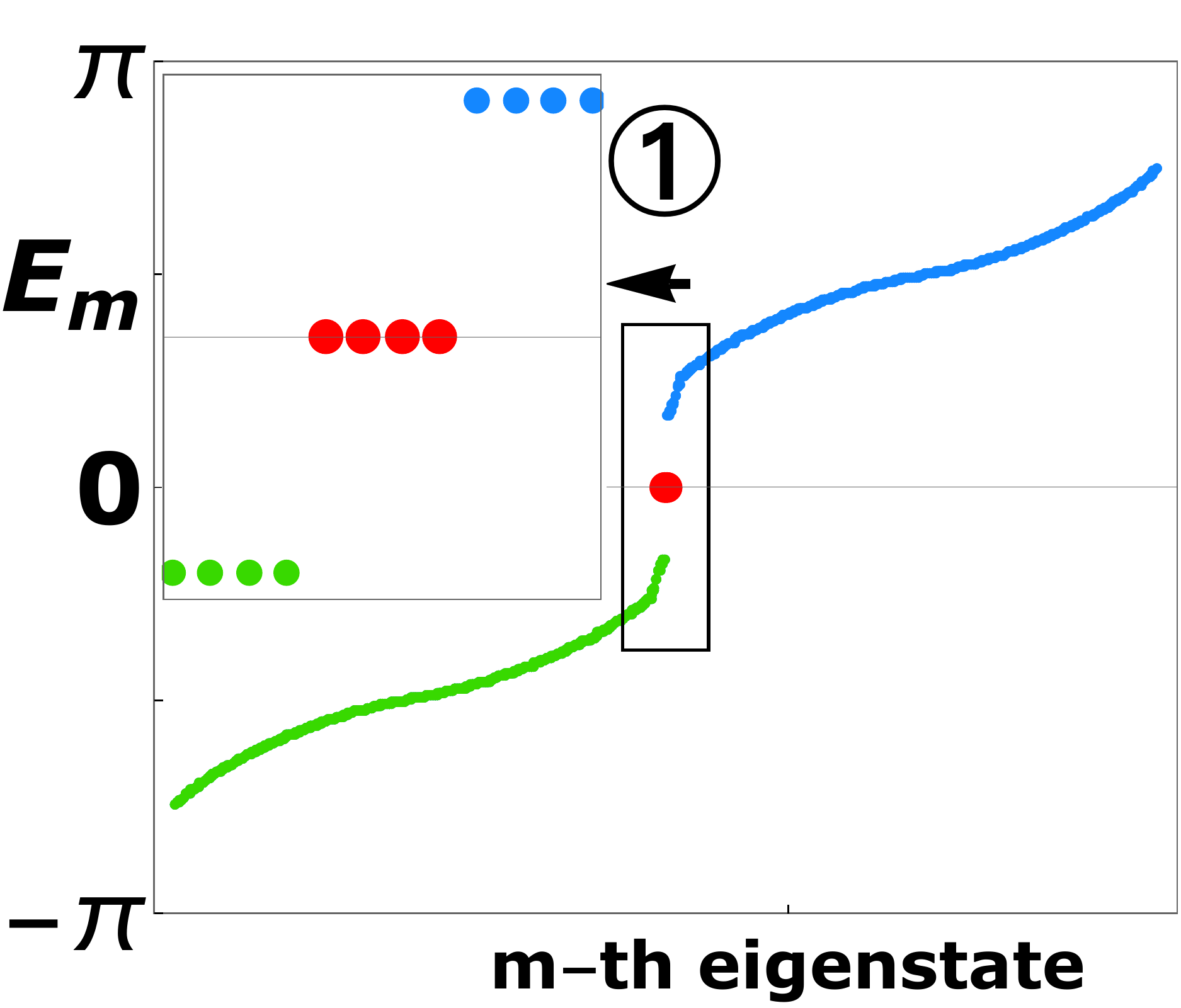}\,\,
			\includegraphics[width=1.4cm]{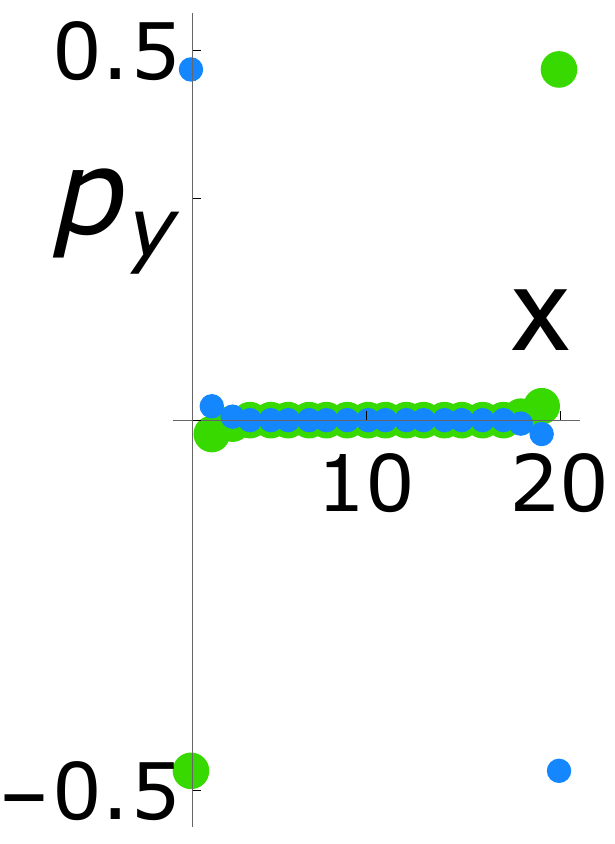}
		}
		\boxed{
			\includegraphics[width=2.5cm]{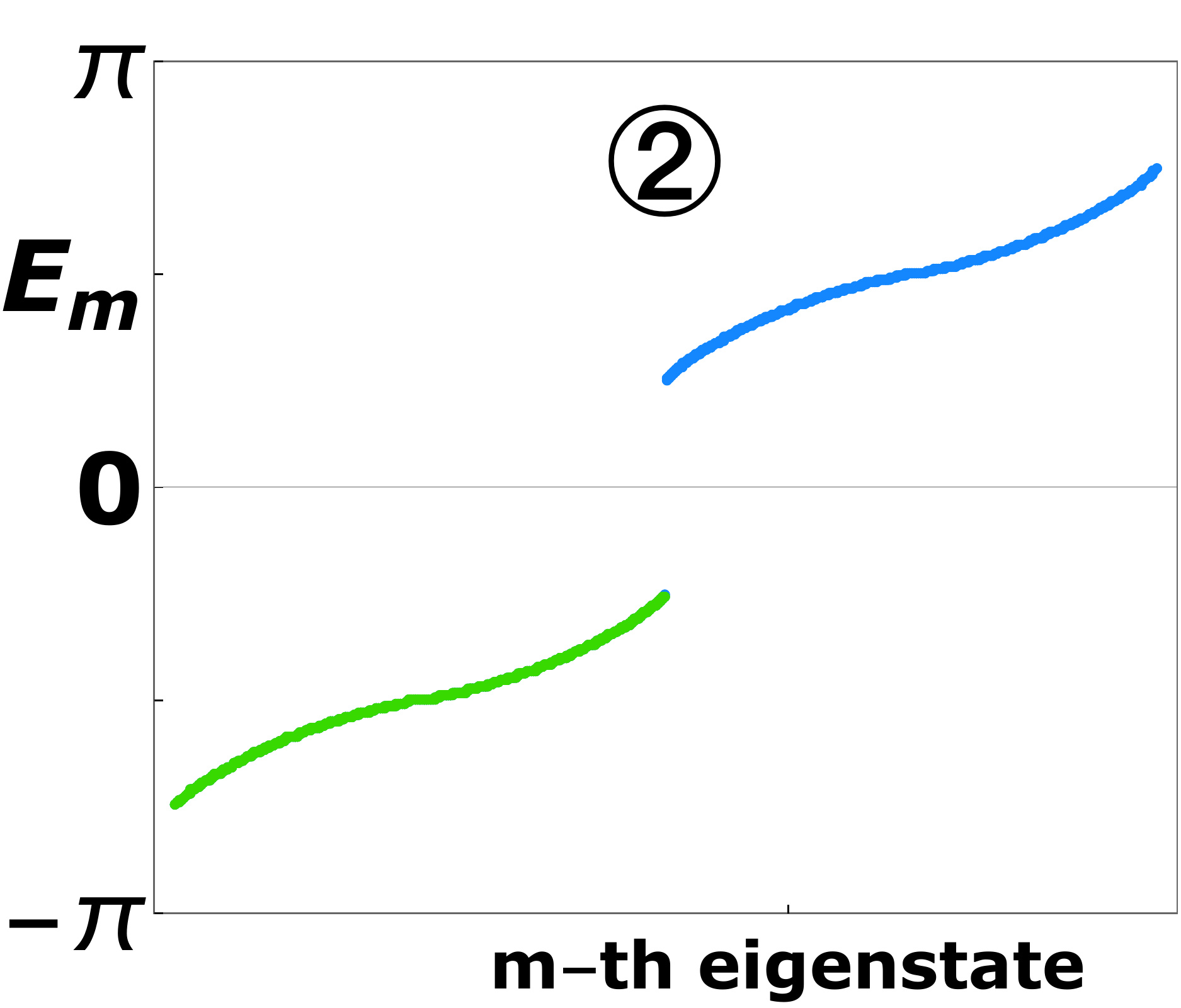}\,\,
			\includegraphics[width=1.4cm]{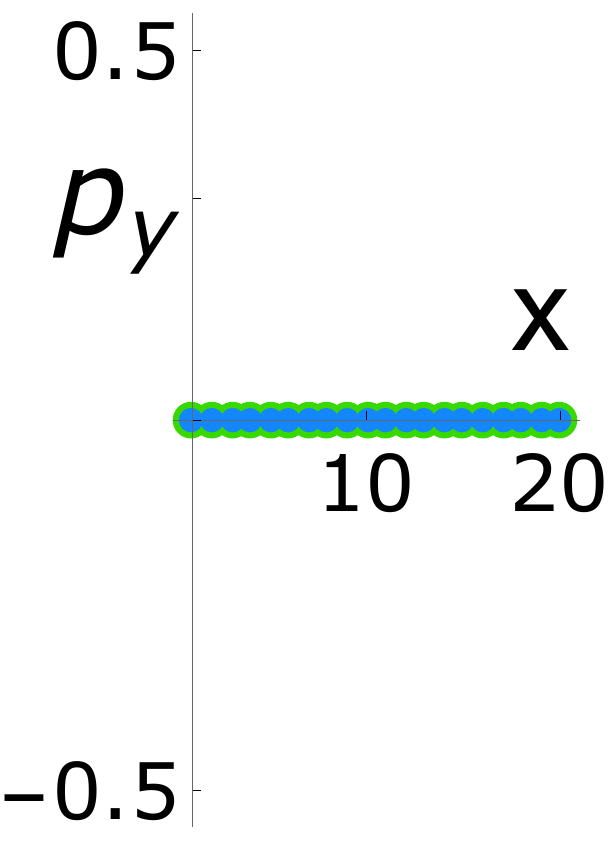}
		}\\
		\boxed{
			\includegraphics[width=2.5cm]{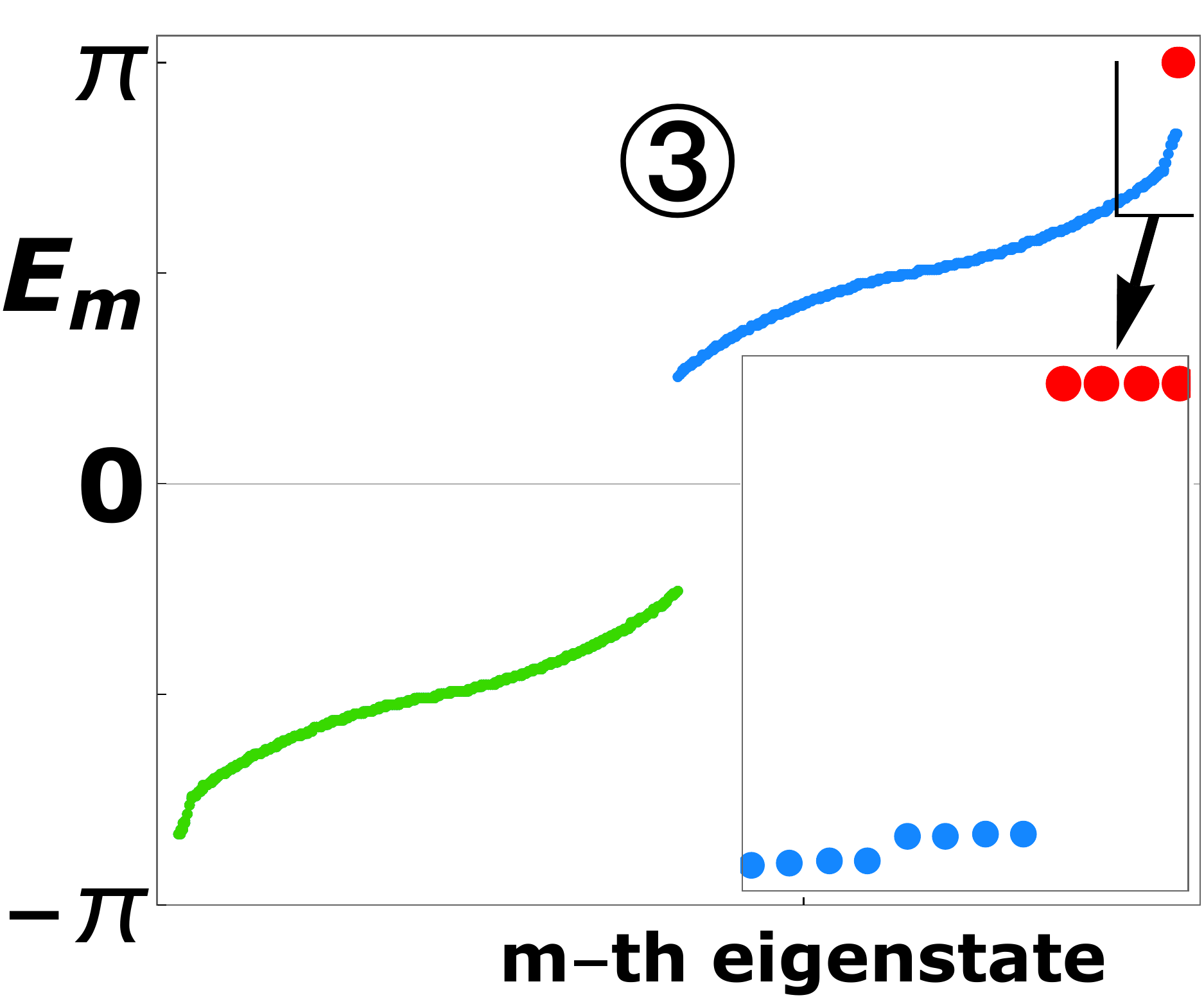}\,\,
			\includegraphics[width=1.4cm]{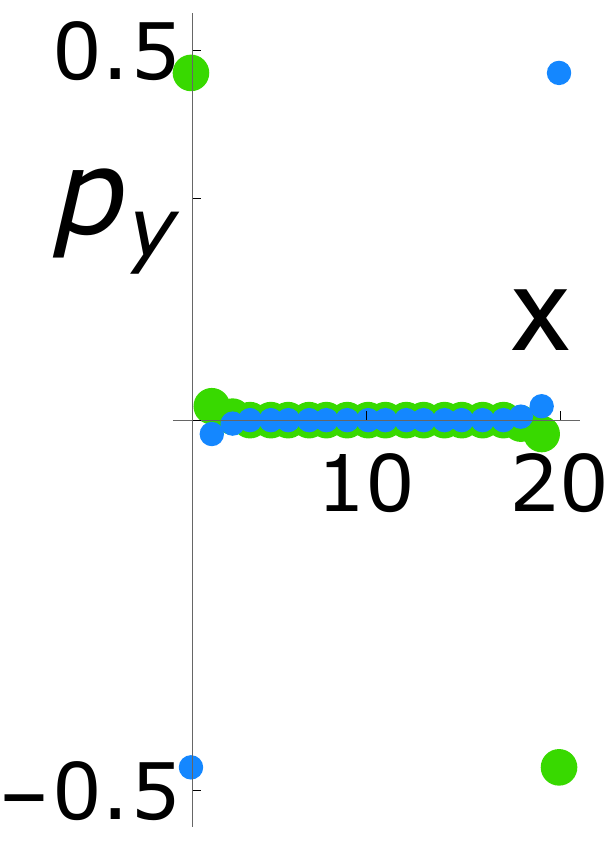}
		}
		\boxed{
			\includegraphics[width=2.5cm]{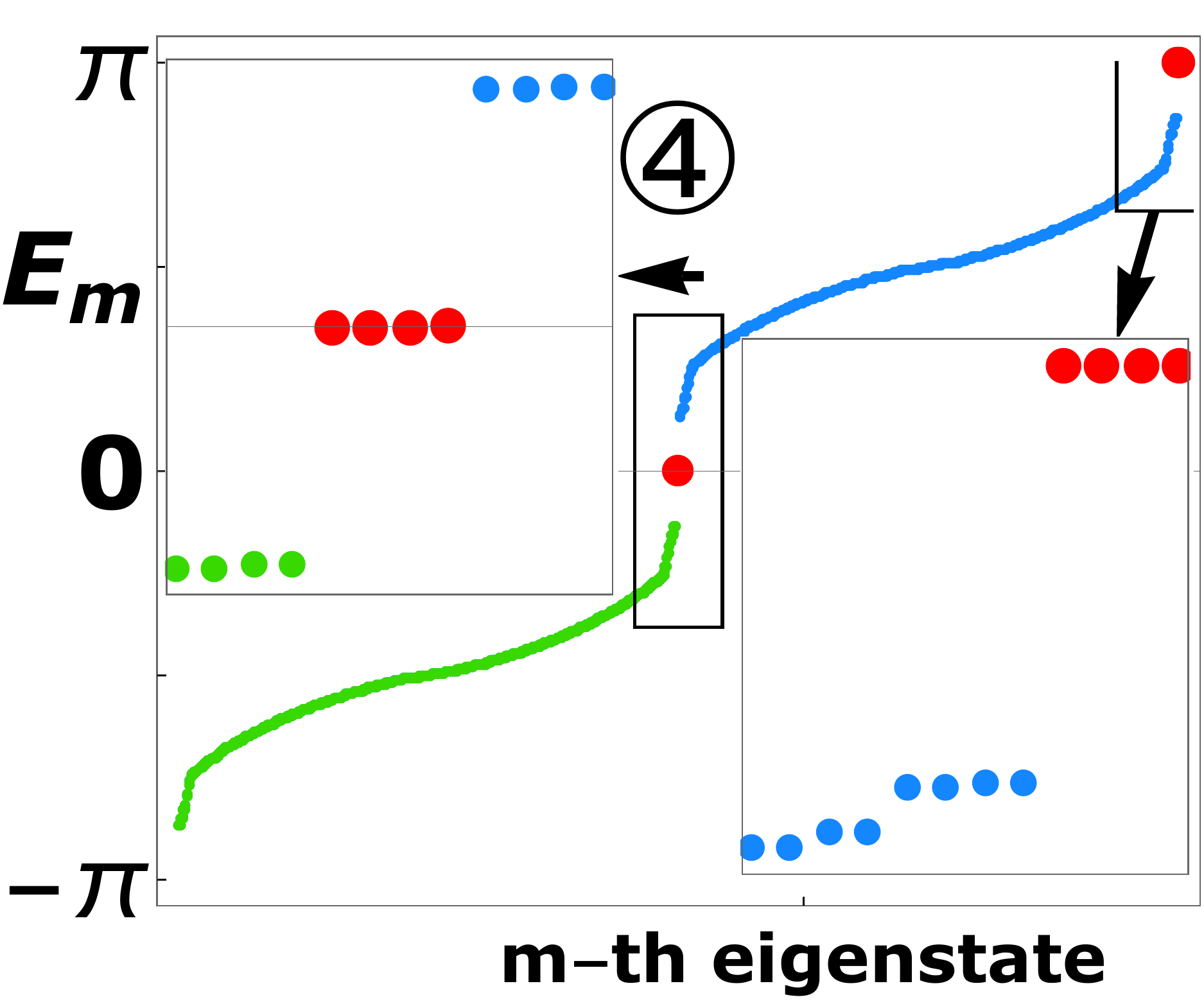}\,\,
			\includegraphics[width=1.4cm]{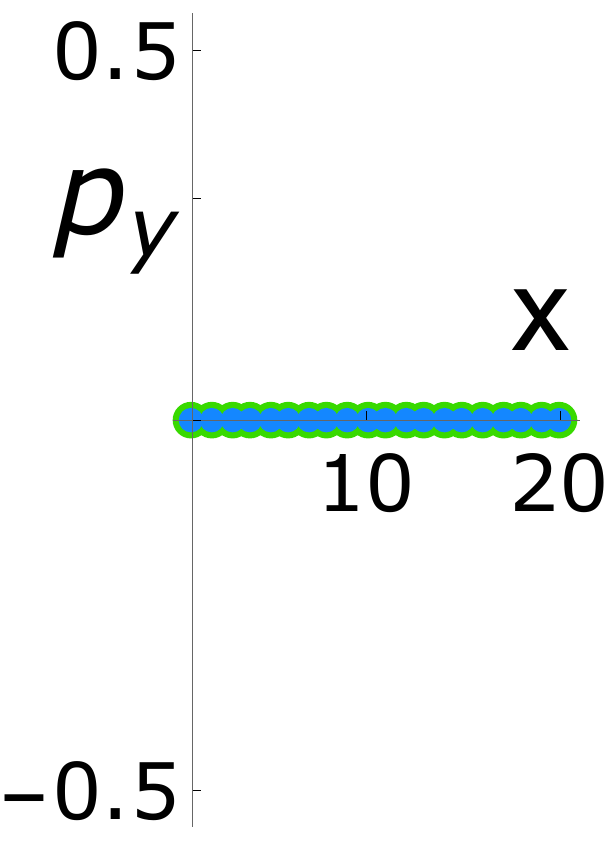}
		}\\
		(d) Spectrum and the static polarization $ p_y(x_i) $ for upper (blue) and lower (green) bands, with in-gap corner states denoted by red.
		\caption{\label{fig:spectrum}(a) Phase diagram parameterized by dimensionless $ (\gamma,\lambda)\equiv (\gamma' T/2\hbar, \lambda' T/2\hbar) $. (b) One representative in-gap corner state amplitude in phase \pfour, with others in different corners.
		(c) Dynamical quadrupoles beyond the description of static $ p_y(x_i) $.
		(d) Open-boundary spectrum (left, $L_x\times L_y=20\times20$) and static polarization in a strip  periodic along $ y $ while open at $ x=1, L_x $ (right, $ 20\times100 $) in each panel. Parameters are $ \sqrt2(\gamma,\lambda) = (\frac{\pi}{4},\frac{\pi}{2}), 
		(\frac{\pi}{2}, \frac{\pi}{4}), (\frac{3\pi}{4}, \frac{\pi}{2}), (\frac{\pi}{2}, \frac{3\pi}{4}) $ for phases \pone$ \sim $\pfour.
		}
	\end{figure}

	As a concrete example, we check the static polarization for
	Eq.~(\ref{eq:hamiltonian}). $ U(\boldsymbol{k},T) $ here possesses two quasi-energy bands $ \pm E_{\boldsymbol{k}} $~(see SM~\cite{supp}), each being doubly degenerate. The concise expression $ E_{\boldsymbol{k}} = \arccos \left[ \cos(\sqrt2\gamma) \cos(\sqrt2\lambda) -\frac{\cos k_x + \cos k_y}{2} \sin(\sqrt2\gamma) \sin(\sqrt2 \lambda)\right] $ analytically determines the phase boundaries at $ E_{\boldsymbol{k}}=0 $:
	\begin{align}\label{eq:phaseboundary}
	\sqrt2\lambda = \pm \sqrt2\gamma + n\pi, \qquad n\in \mathbb{Z}.
	\end{align} 
	Among the phases labeled in Fig.~\ref{fig:spectrum}, \pone\,-- \pthree\, are normal ones with corner states fully characterized by static polarization near $ x_i=1, L_x $. In particular, \pone\, and \ptwo\, are connected to high frequency limits $ \gamma,\lambda \rightarrow 0 $ where they reduce to the two static phases found in Ref.~\cite{Benalcazar2017}. Contrarily, the anomalous Floquet phase \pfour\, exhibits corner states in both gaps $ 0 $ and $ \pi $ well beyond the description of static polarization. 

	\noindent {\em\color{blue} Anomalous dynamical polarization} --- 
	The absence of band physics for anomalous evolution $ U_\varepsilon(\boldsymbol{k},t) $ in Eq.~(\ref{eq:twous}) indicates its intrinsic dynamical nature, and  calls for a new quantity to capture the {\em evolution} of polarization within $ t\in[0,T] $. One naive guess would be  $ \hat{x}(t) = \hat{U}_\varepsilon^\dagger (t) \hat{x} \hat{U}_\varepsilon(t) $, where $ \hat{U}_\varepsilon(t) \equiv \sum_{\boldsymbol{k}mn} \hat{c}_{\boldsymbol{k}m}^\dagger |0\rangle [U_\varepsilon(\boldsymbol{k},t)]_{mn} \langle 0| \hat{c}_{\boldsymbol{k}n} $ and $ \hat{x}=\sum_{im} \hat{c}^\dagger_{im} |0\rangle e^{-i\Delta_x x_i} \langle 0| \hat{c}_{im} $  ($ \Delta_x=2\pi/L_x, x_i=1,\dots,L_x $). However,
	$ \hat{x}(t) $ turns out to possess exactly the same eigenvalues 
	as $ \hat{x} $ (see SM~\cite{supp}). Indeed, as $ \hat{x} $ includes all sites, evolutions $ \hat{U}_\varepsilon(t) $ to any {\em single} moment merely perform a gauge transformation relabeling site indices. It highlights the important lesson that dynamical polarization is to compare {\em relative} motion of particles at two instants, motivating the definition of dynamical mean polarization
	\begin{align}\label{eq:defxmean}
	\hat{x}_{\text{mean}}(t) = (\hat{x}(t) + \hat{x}(0))/2.
	\end{align}
	
	To gain intuition of the newly defined quantity, we consider an idealized example. Suppose an evolution takes any site to its neighboring unit cell $ x_{i+1}=x_i+1 $, $ \hat{U}_\varepsilon(t_0) = \sum_{im} \hat{c}_{i+1,m}^\dagger |0\rangle \langle 0|\hat{c}_{im} $. Then $ \hat{x}(t_0) = \sum_{im} \hat{c}_{im}^\dagger |0\rangle e^{-i\Delta_x(x_i+1)} \langle 0| \hat{c}_{im} $, which is identical to $ \hat{x}(0) $ upon relabeling $ \hat{c}_{im}\rightarrow \hat{c}_{i+1,m} $. In contrast, $ \hat{x}_{\text{mean}} (t_0) = \sum_{im} \hat{c}_{im}^\dagger |0\rangle e^{-i\Delta_x(x_i+1/2)} \cos(\pi/L_x) \langle 0|\hat{c}_{im} $. The eigenvalues of $ \hat{x}_{\text{mean}}^{L_x}(t_0) $ (related to dynamical Wilson loop defined later), after taking the thermodynamic limit $ \lim_{L_x\rightarrow\infty} \cos^{L_x}(\pi/L_x) = 1 $, reads $ e^{-2\pi i \nu } $ with $ \nu=1/2 $, giving the eigenvalues of $ \hat{x}_{\text{mean}}(t_0) $ as $ e^{-i\Delta_x(x_i+\nu)} $. Importantly, $ \nu $ characterizes the {\em movement} of particles from $ 0 $ to $ t_0 $ by $ 2\nu $, rather than a static center for Wannier functions. A general proof for the unitarity of $ \hat{x}_{\text{mean}}(t) $ is given in SM~\cite{supp}. 
	
	The above example illustrates the new bulk-boundary correspondence for dynamical polarization in Fig.~\ref{fig:model} (d). Since $ \hat{U}_\varepsilon(T)=\hat{U}_\varepsilon(0) $ are identity operators, particles must undergo a round trip during a cycle. But there could be non-trivial fixed point $ t\in(0,T) $ where particles move by {\em integer}  $ 2\nu $ of unit cells, enforced by symmetry and topology in $ \hat{U}_\varepsilon(t) $. Then, the boundary serves as obstructions for {\em motion}, leaving out an immobile edge/corner state. This picture differs conceptually from conventional band polarization in Fig.~\ref{fig:model}(c) where boundaries ``cuts" the {\em static} Wannier functions centering between unit cells, leaving out a decohered edge/corner state.


	Rigorous constructions of dynamical polarization confirm exactly the above intuition. (See SM~\cite{supp} for algebras). First, $\hat{x}_{\text{mean}}^{L_x}(t) = \sum_{\boldsymbol{k}mn} \hat{c}^\dagger_{\boldsymbol{k}m} |0\rangle [W_{x,\boldsymbol{k}}(t)]_{mn} \langle 0| \hat{c}_{\boldsymbol{k}n} $, where the dynamical Wilson loop at $ L_x\rightarrow\infty $ reads
	\begin{align}\label{eq:wilson1}
	W_{x,\boldsymbol{k}}(t) = P_{k'_x} e^{-\frac{1}{2}\int_{\boldsymbol{k}}^{\boldsymbol{k}+2\pi \boldsymbol{e}_x} dk'_x U_\varepsilon^\dagger (\boldsymbol{k}',t) \partial_{k'_x} U_{\varepsilon} (\boldsymbol{k}',t)},
	\end{align}
	$ P_{k'_x} $ being path-ordering.
	Diagonalizing $ [W_{x,\boldsymbol{k}}(t)]_{mn} = \sum_{\mu} [\nu_{x,\mu} (\boldsymbol{k},t) ]_m e^{-2\pi i \nu_{x,\mu}(k_y,t)} [ \nu_{x,\mu} (\boldsymbol{k},t)]_n^* $, one obtains the {\em dynamical branches} $ \mu $ for motions along $ x $: $ \hat{x}_{\text{mean}}(t) |b_{x,\mu}(x_i,k_y,t)\rangle = e^{-i\Delta_x(x_i+\nu_{x,\mu} ({k_y,t}))} |b_{x,\mu} (x_i,k_y,t)\rangle $. 
	Crucially, the branch eigenstates $ |b_{x,\mu}(x_i,k_y,t)\rangle 
	$ should not be confused with static ``edge Wannier bands" non-existent for $ U_{\varepsilon}(\boldsymbol{k},t) $. Instead, it represents the interference pattern of polarization at two different moments, characterizing collective motions of particles from  $ 0 $ to $ t $. This is also underscored by the number of dynamical branches $ \mu $'s which {\em equals} sublattice numbers (i.e. 4 in our case), as different sublattices (or their linear combinations) can support independent Bloch wave motions. It differs from static edge Wannier bands that halves sublattice number due to projection onto occupied bands~\cite{Benalcazar2017}, and such a difference plays vital roles in all following analysis.
	
	Marching towards higher orders, we group the branches $ \mu $'s into sets $ \nu_x $'s, where $ \nu_{x,\mu_1}(k_y,t)\ne \nu_{x,\mu_2}(k_y,t) $ for $ \mu_1,\mu_2 $ in different $ \nu_x $'s. Each set $ \nu_x $ then defines a separable motion along $ x $. The second order dynamical polarization describes perpendicular motions within each $ \nu_x $, $ \hat{y}^{(\nu_x)}_{\text{mean}}(t) \equiv \hat{P}_{\nu_x}(t) \hat{y}_{\text{mean}}(t) \hat{P}_{\nu_x}(t) $, where branch set projectors $ \hat{P}_{\nu_x}(t) = \sum_{x_i,k_y,\mu \in \nu_x} |b_{x,\mu}(x_i, k_y,t)\rangle \langle b_{x,\mu}(x_i, k_y,t)|  $, and $ \hat{y}_{\text{mean}}(t) $ takes the form as Eq.~(\ref{eq:defxmean}) with $ x\rightarrow y $. 
	Now, $ \left( \hat{y}^{(\nu_x)}_{\text{mean}}(t) \right)^{L_y} $ involves nested dynamical Wilson loops (at $ L_y\rightarrow\infty $)~\cite{supp},
	\begin{align}\label{eq:wilson2}
	W^{(\nu_x)}_{y,\boldsymbol{k}} (t) &= P_{k'_y} e^{-\int_{\boldsymbol{k}}^{\boldsymbol{k}+2\pi\boldsymbol{e}_y} dk'_y 
		A^{(\nu_x)}_{y,\boldsymbol{k}'}(t)
	},
	\end{align}
	where the non-Abelian Berry curvature $ 
[A^{(\nu_x)}_{y,\boldsymbol{k}}(t)]_{\mu_1\mu_2} = \sum_{mn}\frac{1}{2} [ \nu_{x,\mu_1}(\boldsymbol{k},t) ]^*_m [U^\dagger_{\varepsilon}(\boldsymbol{k},t) \partial_{k_y} U_\varepsilon(\boldsymbol{k},t)]_{mn} [\nu_{x,\mu_2}(\boldsymbol{k},t)]_n  + \sum_m [\nu_{x,\mu_1}(\boldsymbol{k},t)]_m^* \partial_{k_y} [ \nu_{x,\mu_2}(\boldsymbol{k},t)]_m  $. Eigenvalues of $ W^{(\nu_x)}_{y,\boldsymbol{k}}(t)$, $ e^{-2\pi i \nu^{(\nu_x)}_{y,\tilde{\mu}}(k_x,t) } $, define the dynamical quadrupole  branches $ \tilde{\mu} $'s,
	as $ \nu^{(\nu_x)}_{y,\tilde{\mu}}(k_x,t) $ describes the co-movement along $ y $ within the first-order branch set $ \nu_x $ moving along $ x $. See SM~\cite{supp} for generalization to arbitrary orders in arbitrary dimensions.
	\begin{figure}
		[t]
		\parbox{1.65cm}{
			\includegraphics[width=1.65cm]{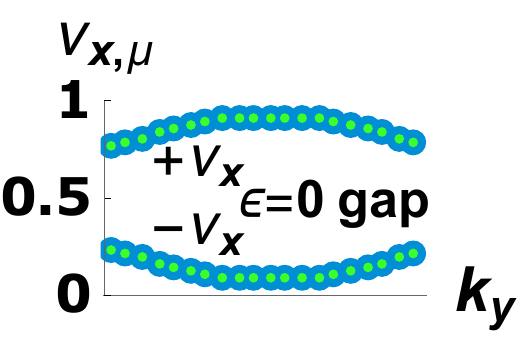}
			\includegraphics[width=1.65cm]{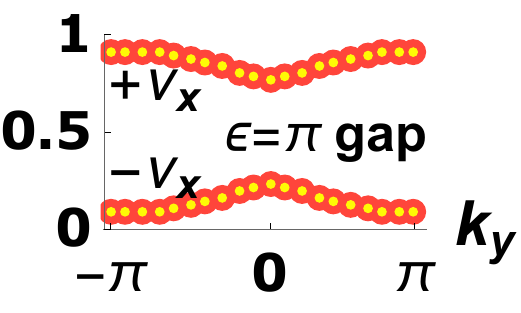}	\\
			\quad (a)
		}
		\parbox{1.65cm}{
			\includegraphics[width=1.9cm]{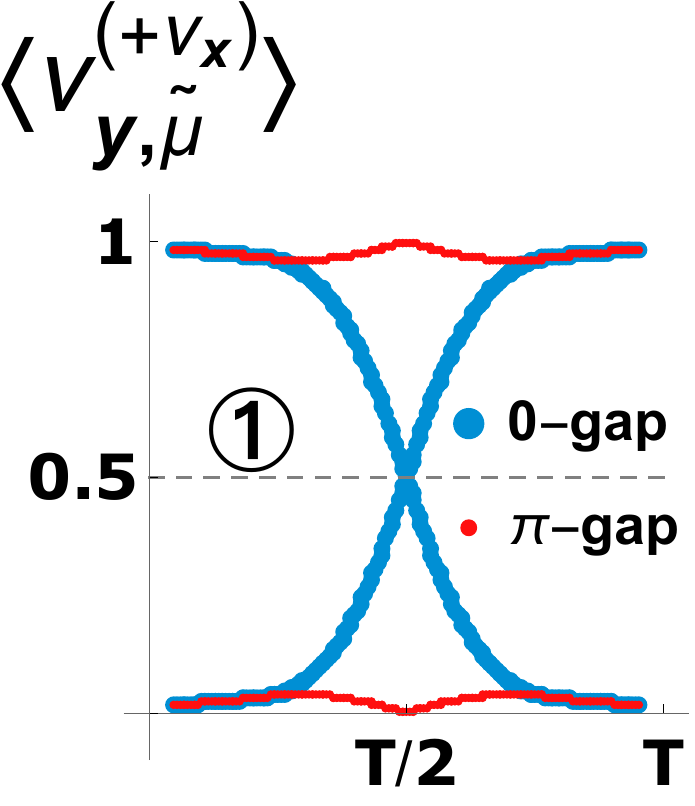}\\
			\quad(b)
		}
		\parbox{1.55cm}{
			\includegraphics[width=1.8cm]{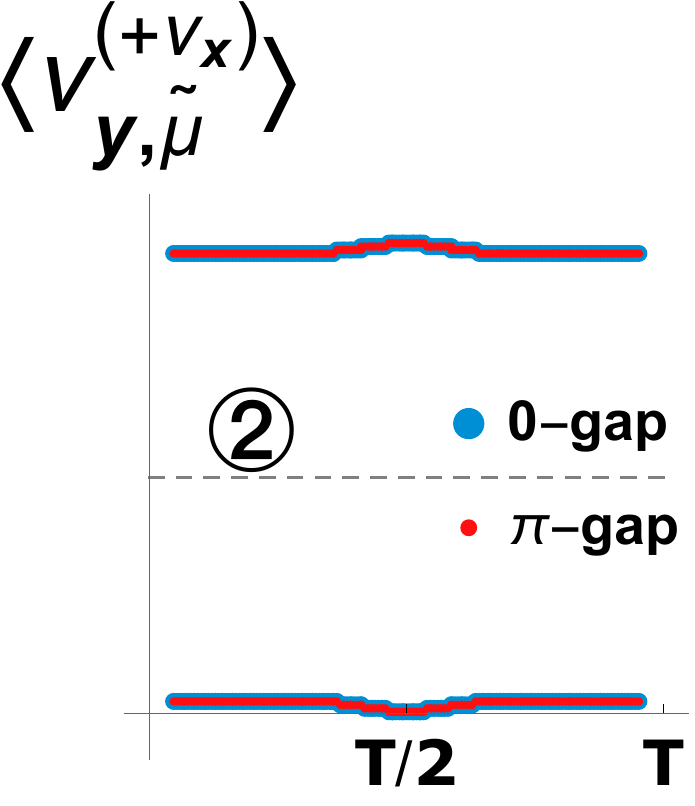}\\
			\quad (c)
		}
		\parbox{1.55cm}{
			\includegraphics[width=1.8cm]{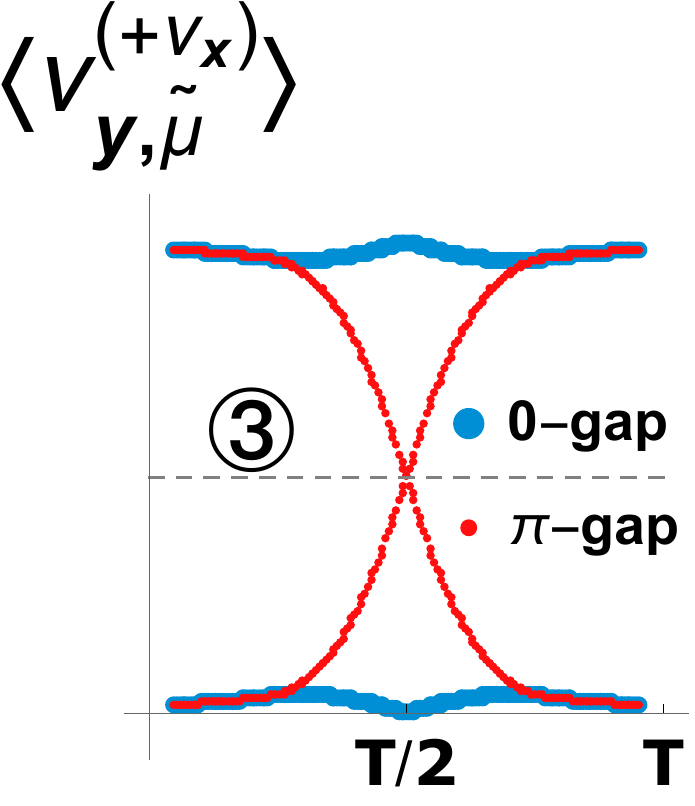}\\
			\quad (d)
		}
		\parbox{1.85cm}{
			\includegraphics[width=1.9cm]{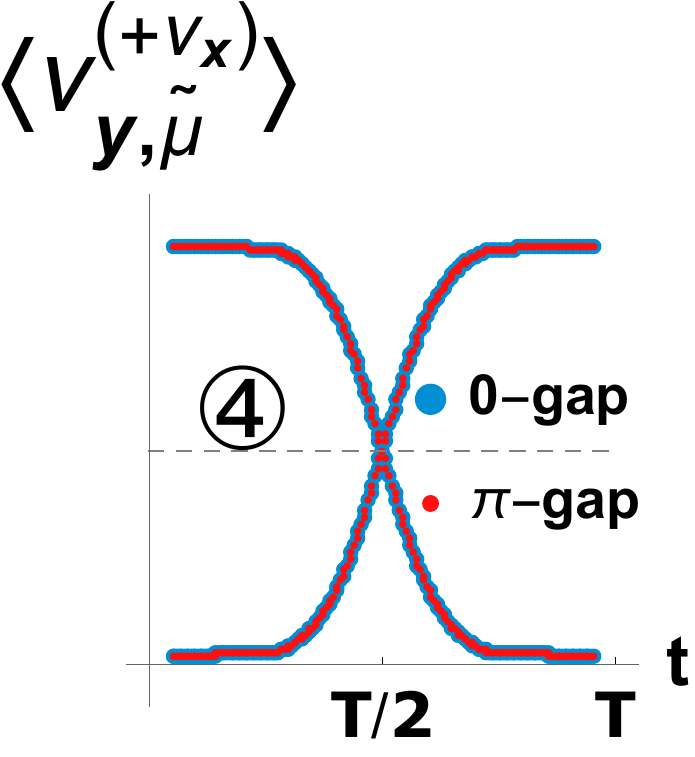}\\
			(e)
		}
		\caption{\label{fig:dynqxy} 
			 (a) Exemplary first-order dynamical branches $ \nu_{x,\mu}(k_y,t) $ in phase \pfour\, at $ t=T/2 $. Four branches for {\em each} $ \varepsilon $-gap separate into two doubly degenerate sets $ \pm \nu_{x} $. Other phases/instants $ t $ exhibit $ \nu_{x,\mu}(k_y,t) $ of similar structures. (b)--(e) Quadrupolar motions $ \langle \nu^{(\nu_x)}_{y,\tilde{\mu}} \rangle  (t) $ in a cycle, with the same parameters as in Fig.~\ref{fig:spectrum} for each phase respectively. Generically the 1/2 crossing needs not to occur at $ t=T/2 $.
		}
	\end{figure}

	Instantaneous $ \nu_{x,\mu}(k_y,t) $ and $ \nu_{y,\tilde{\mu}}^{(\nu_x)}(k_x,t) $ are constrained by mirror symmetries $ \hat{M}_{x/y}: x/y \rightarrow -x/y $~\cite{supp}. $ \hat{M}_x $ forces the 4 first-order branches to appear in pairs $ \pm \nu_{x,\mu}(k_y,t) $, while $ M_y $ formats their shapes $ \nu_{x,\mu}(k_y,t) = \nu_{x,\mu}(-k_y,t) $.  Such constraints allow for the separable sets $ \mu=1,2\in + \nu_x $, $ \mu=3,4\in-\nu_x $ for our model as exemplified in Fig.~\ref{fig:dynqxy} (a). Vitally, each set contains 2 degenerate branches, which means the nested dynamical Wilson loops are SU(2) matrices rather than U(1) numbers in static situation~\cite{Benalcazar2017}. $ \hat{M}_x $ then dictates $ \nu_{y,\tilde{\mu}}^{(+\nu_x)} (k_x,t) = \nu_{y,\tilde{\mu}}^{(-\nu_x)}(k_x,t) $ (so checking $ +\nu_x $ is enough), while $ \hat{M}_y $ further ensures $ \nu_{y,\tilde{\mu}=1}^{(+\nu_x)} (k_x,t) = - \nu_{y,\tilde{\mu}=2}^{(+\nu_x)} (k_x,t) $ (so they can only meet at $ 0, 1/2 $ mod 1). Recall that $ \nu_{x,\mu}(k_y,t) $ (or $ \nu_{y,\tilde{\mu}}^{(\nu_x)}(k_x,t)  $)   implies motions along $ x $ (or $ y $); those constraints translates to the intuition of identical motion quadruplet along $ \pm \boldsymbol{e}_x \pm \boldsymbol{e}_y $ respectively as shown in Fig.~\ref{fig:spectrum}(c).
	
	Notably, {\em instantaneous} quadrupole branches $ \nu_{y,\tilde{\mu}=1,2}^{(+\nu_x)}(k_y,t) $'s are {\em not quantized} unlike static quadrupoles~\cite{Benalcazar2017}.
	True dynamical topology hides in the {\em evolution} of  
	averaged quadrupolar motions $ \langle \nu_{y,\tilde{\mu}}^{(+\nu_x)} \rangle (t) =\int_{-\pi}^\pi \frac{dk_x}{2\pi} \nu^{(+\nu_x)}_{y,\tilde{\mu}} (k_x,t) $ as shown in Fig.~\ref{fig:dynqxy}.  A non-trivial evolution goes from $ \langle \nu_{y,\tilde{\mu}}^{(+\nu_x)}\rangle (t\rightarrow 0) = 0 $ to $ \langle \nu_{y,\tilde{\mu}}^{(+\nu_x)} \rangle (t\rightarrow T) = 1 $, or vice versa, such as for the $ \varepsilon=0 $ gap of phase \pone. Then, the two branches meeting at $ 1/2 $ (mod 1) for $ t\in(0,T) $ exactly means particles move from $ y_i $ to $ y_i\pm 1 $ (Fig.~\ref{fig:model}(d) at ``half-way"). Due to indistinguishability of particles, the net polarization in open boundary sample is from $ y_i=1 $ (or $ L_y $) all the way to $ y_i=L_y $ (or $ 1 $) with motions obstructed there, resulting in corner states at $ y_i=1, L_y $ ($ x_i=L_x $  for the $ +\nu_x $ branch set). Note the two branches, or corner states at $ y_i=1,L_y $, are related by $ \hat{M}_y $~(see SM~\cite{supp} for mathematical meaning).
	Spatial separation and mirror symmetry then protect the corner states or the $ 1/2 $ crossing for $ \langle \nu_{y,\tilde{\mu}=1,2}^{(+\nu_x)} \rangle (t) $ against hybridization, leading to quantized dynamical quadrupole 
	\begin{align}
	\tilde{P}^{(+\nu_x)}_{xy,\tilde{\mu}} = \int_0^T  dt \left( \partial_t  \langle \nu^{(+\nu_x)}_{y,\tilde{\mu}} \rangle  (t)  \right)  ( \text{mod 1} ) = 0, 1 \in \mathbb{Z}_2,
	\end{align}
	with 1 being topologically non-trivial. 
    Should the winding number $ \tilde{P}^{(+\nu_x)}_{xy,\tilde{\mu}} =2  $,
	the two $ \langle \nu_{y,\tilde{\mu}=1,2}^{(+\nu_x)} \rangle (t_0) $ would cross at 0 mod 1 at certain $ t_0\ne 0, T $, which is not a stably quantized instant and $ \langle \nu_{y,\tilde{\mu}}^{(+\nu_x)} \rangle (t_0)  $ can be perturbed away from $ 0, 1 $. Then $ \tilde{P}^{(+\nu_x)}_{xy,\tilde{\mu}} = 0 $ again, reducing winding numbers $ \tilde{P}^{(+\nu_x)}_{xy,\tilde{\mu}} $ from $ \mathbb{Z} $ to $ \mathbb{Z}_2 $. Physically, it means two particles polarized to the same corner can surely hybridize and gap out each other.

	$ \langle \nu_{y,\tilde{\mu}}^{(+\nu_x)}\rangle (t) $ for phases in Fig.~\ref{fig:spectrum} are shown in Fig.~\ref{fig:dynqxy} (b)--(e), which undoubtedly captures both normal and anomalous corner states. The difference between dynamical quadrupoles in two gaps $ \varepsilon=0, \pi $ gives the static quadrupole of the band sandwiched by the gaps, explaining the vanishing of static quadrupoles in phase \pfour. 
	
	Two remarks are in order. First, only mirror symmetries are required. In SM~\cite{supp}, extended models verify the validity of dynamical polarization when {\em all} accidental symmetries in Eq.~(\ref{eq:hamiltonian}) (time-reversal $ \Theta $, particle-hole $ \Gamma $, chiral $ S $ and four-fold rotations) are violated. In contrast, breaking mirror symmetries immediately destroys the corner states together with $ \langle \nu_{y,\tilde{\mu}}^{(+\nu_x)}\rangle (t) =1/2 $.  Second, with only one mirror symmetry i.e. $ \hat{M}_y $, indices for our main model will change to $ \mathbb{Z} $ because the sets $ \pm\nu_x $ then could move in the same direction along $ x $, each contributing a $ \mathbb{Z}_2 $ type of corner state adding up to $ \mathbb{Z} $. In that case~\cite{Langbehn2017}, classification for $ d+1 $ dimensional second-order insulators is completely the same as $ d $ dimensional first-order ones determined by $ \Theta, \Gamma, S $, unlike ours.

	\noindent{\em\color{blue} Scheme for detection} --- Static Hamiltonians exactly as Eq.~(\ref{eq:hamiltonian}) have {\em already} been engineered in photonic/phononic/electric circuit experiments~\cite{Peterson2017,Serra-Garcia2017,Imhof2017}, which are also platforms for earlier realizations of first-order anomalous Floquet insulators driven in the form of bond-connectivity~\cite{Mukherjee2017,Peng2016,Maczewsky2016}. Furthermore, experiments realizing Abelian/non-Abelian synthetic gauge fields by shaking or Raman scheme have been abundant for cold atoms~\cite{Cooper2019,Duca2015,Aidelsburger2015,Wu2016}, with the $ \pi $-flux model one of the earliest prototype~\cite{Aidelsburger2013}. For completeness, we also discussed a specific lattice quench scheme in SM~\cite{supp}. Thus, engineering our model in diverse ranges of platforms appears immediate and we focus chiefly on detections.
	
	First, with the digital mirror devices achieving chemical potential  engineering with single-site accuracy~\cite{Tai2017}, one could detect localized corner states via initially injecting particles around a potential barrier $ V(x,y) = V_0\Theta(-x)\Theta(-y) $ and observing the particle density evolution, as shown in Fig.~\ref{fig:expt} (a)--(e). ``Softer" corners with more smeared out potentials (a possible outcome of lattice shaking) is considered in SM~\cite{supp}, which do not affect the qualitative features as expected.

	 
	 A more challenging task is to prove the triviality of Floquet operators $ U(\boldsymbol{k},T) $ and confirm the dynamical origin for corner states.  
	 Our model possesses the appealing feature that parameters $ \sqrt2\lambda $  and $ \pi+\sqrt2\lambda $ give exactly the same $ U(\boldsymbol{k},T) $~\cite{supp}. That means one can populate certain Floquet band of phase \pfour\, by preparing the system in static limit $ \lambda\rightarrow0 $, and then quench to $ \lambda\rightarrow \pi/\sqrt2 $ while maintaining band populations. (Reversed processes can be used before band-mapping to determine band populations). Further, in both limits the system exhibits flat-bands $ \pm E_{\boldsymbol{k}} = \pm \sqrt2\gamma $, ideal for populating certain Floquet band homogeneously by high-temperature gases~\cite{Flaeschner2016,Li2016,Aidelsburger2015}. Then,  one could perform similar tomography experiment previously carried out in two-band static cases~\cite{Hauke2014,Flaeschner2016}.

	 Specifically, $ U(\boldsymbol{k},T) $ for our model are parameterized by three angles $ \chi_{\boldsymbol{k}}, \theta_{\boldsymbol{k}}\in[0,\pi], \varphi_{\boldsymbol{k}}\in(-\pi,\pi] $ in $ S^3 $. After equilibration, the lattice depth is ramped up and the system evolves under static chemical potentials $ \mu_m $ for sublattices $ m=1\sim 4 $; three profiles of $ \{\mu_i\} $ could fully determine $ (\chi_{\boldsymbol{k}},\theta_{\boldsymbol{k}}, \varphi_{\boldsymbol{k}}) $~\footnote{There are infinitely many choices to back up the angles. We demonstrate one such choice in SM~\cite{supp}}. Finally, momentum density $ n_{\boldsymbol{k}} $ is measured after time-of-flight through real-space density $ n(\boldsymbol{r}=\hbar\boldsymbol{k}t/m) $. It is clear from Fig.~\ref{fig:expt}(f) that the weak momentum dependence of all angles in phase \pfour\, signals the vanishing static quadrupoles. 

	\begin{figure}
		[h]
		\parbox{1.9cm}{
			\includegraphics[width=1.9cm, angle=90]{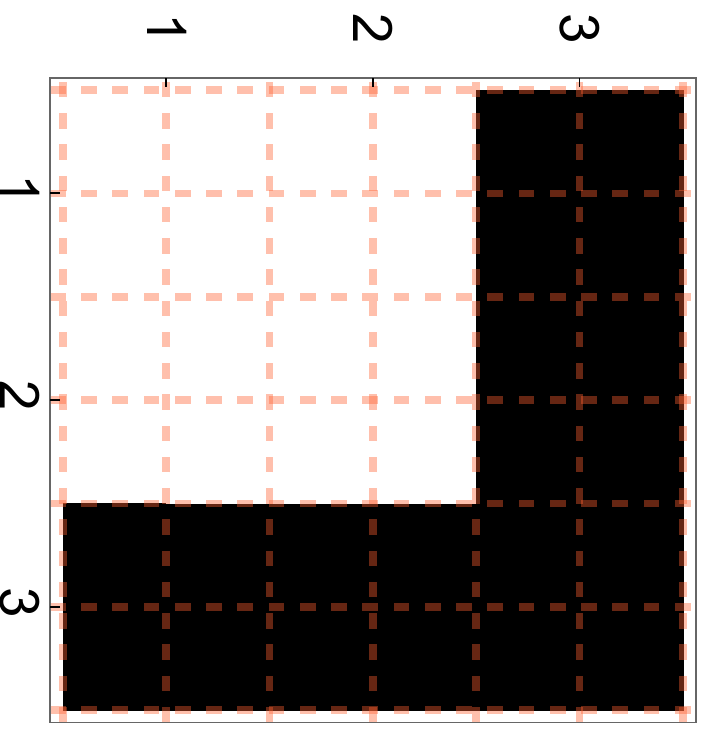}\\
		(a) Initial}
		\parbox{1.9cm}{
			\includegraphics[width=1.9cm, angle=90]{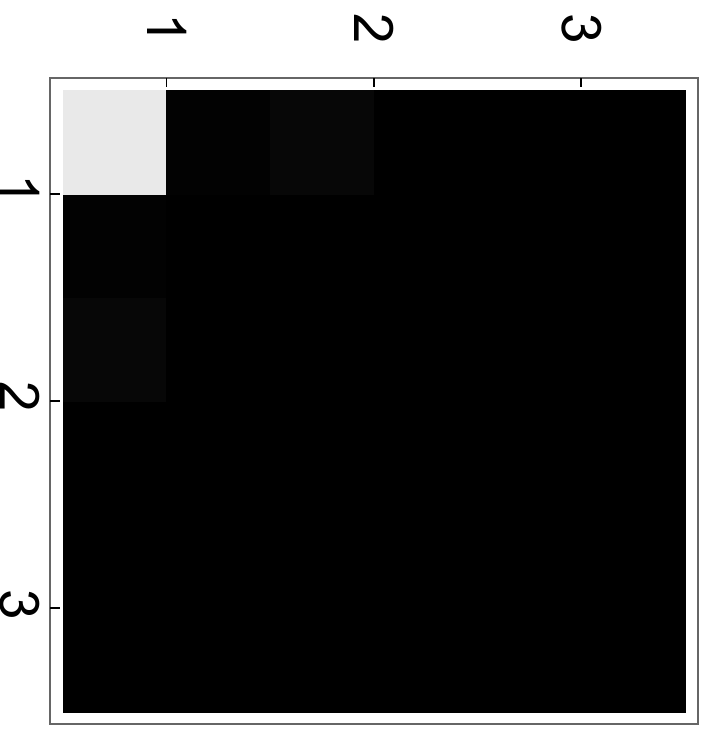}\\
		(b) Phase \pone\pthree}
		\parbox{1.9cm}{
			\includegraphics[width=1.9cm, angle=90]{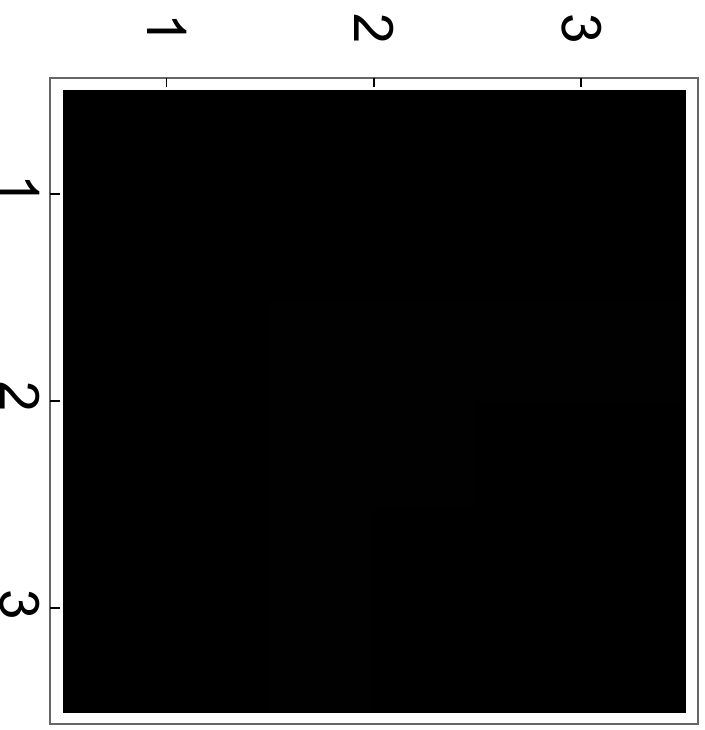}\\
		(c) Phase \ptwo}
		\parbox{1.9cm}{
			\includegraphics[width=1.9cm, angle=90]{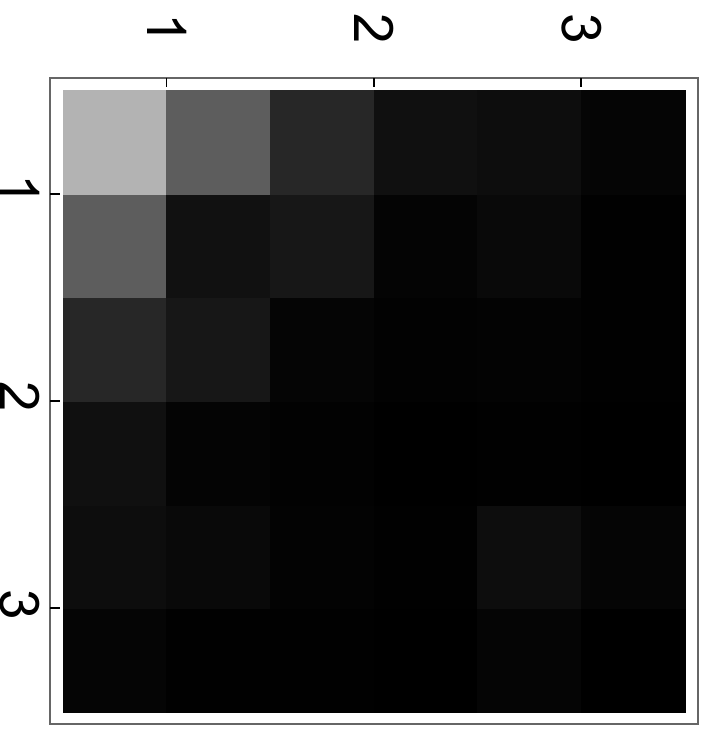}\\
		(d) Phase \pfour}
		\parbox{0.4cm}{
			\includegraphics[width=0.45cm]{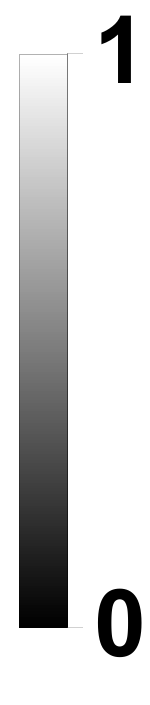}\\\quad }
		\\
		\parbox{3.5cm}{\includegraphics[width=3.45cm]{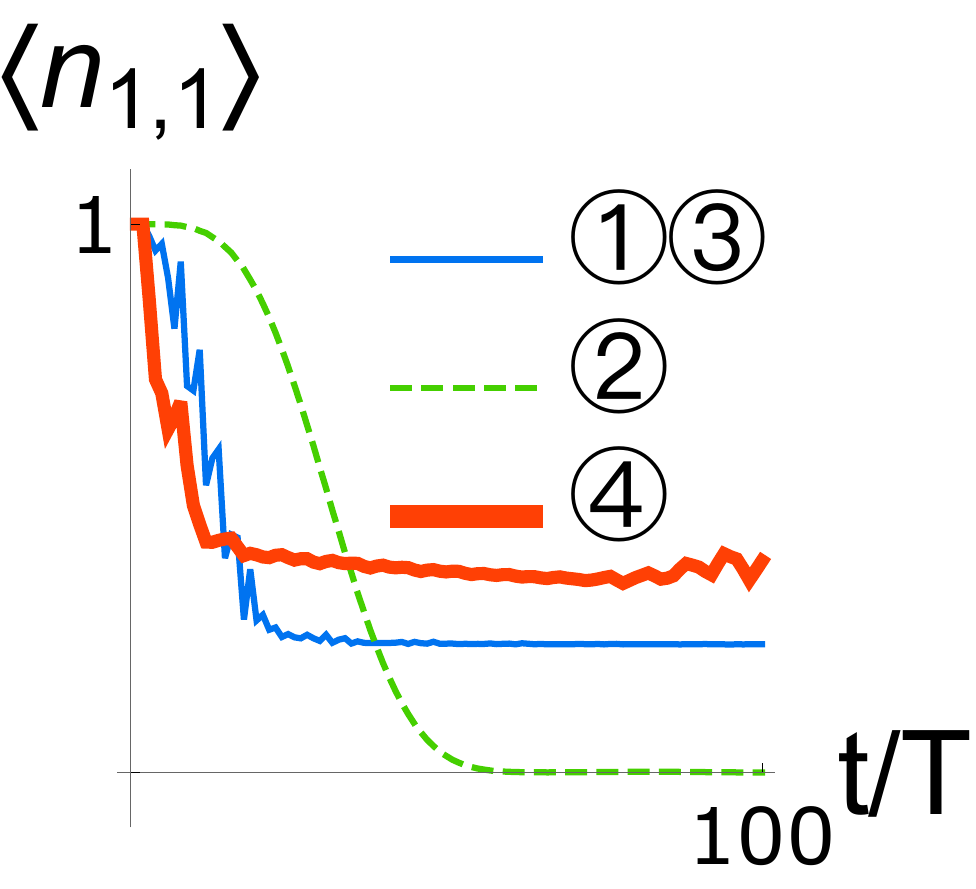}\\ (e) Density dynamics}
		\parbox{4.2cm}{
			\includegraphics[width=4cm]{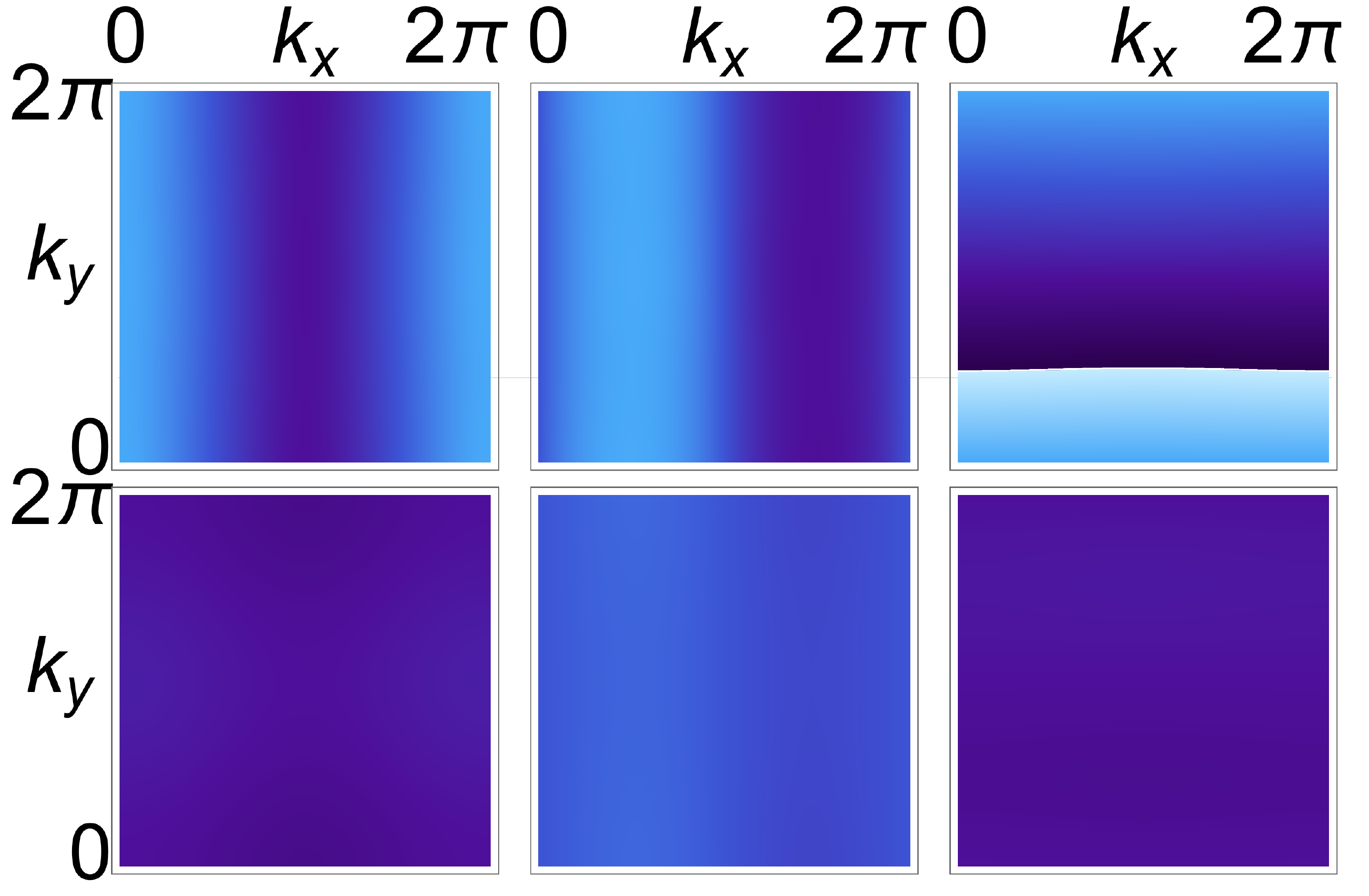}\\
			\parbox{2cm}{$ \scriptsize\chi_{\boldsymbol{k}}$ \qquad\quad $\scriptsize\theta_{\boldsymbol{k}} $\\ \includegraphics[width=1.5cm]{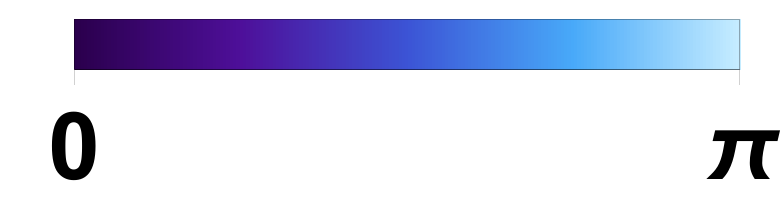}}
			\parbox{1cm}{$\qquad \scriptsize\varphi_{\boldsymbol{k}} $
			\includegraphics[width=1.5cm]{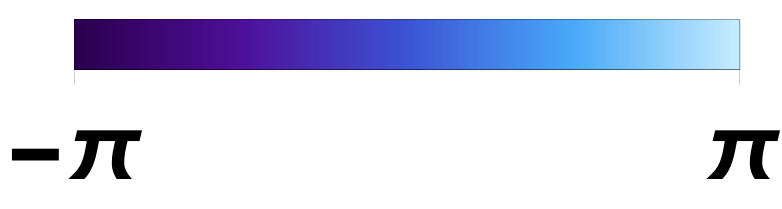}}
			\\
			(f) TOF signatures
			} 
		\parbox{0.05cm}{\pone \\  \quad\\ \quad\\
			 \pfour\\\quad \\\quad \\}
		\caption{\label{fig:expt}
		Simulation of detection signatures. (a)-(d) Density evolution from (a) $ t=0 $ to (b)--(d) $ t=100T $ in $ 20 \times 20 $ unit cells (1600 sites). Only the corner $ 3\times3 $ cells are shown. Each pixel denotes one lattice site, and axis ticks denote unit cells. (e) Average particle density in the unit cell $ (x,y)=(1,1) $. (f) The TOF signature for $ \chi_{\boldsymbol{k}} $ (left), $ \theta_{\boldsymbol{k}} $ (middle) and $ \varphi_{\boldsymbol{k}} $ (right) in phase \pone\, (upper row) and phase \pfour\, (lower row). Parameters $ (\sqrt2/\pi)(\gamma,\lambda) = (0.5,0.1), (0.5,0.05), (0.9,0.5), (0.5,0.95) $ for \pone$ \sim $\pfour\, respectively.
		}
	\end{figure}

	\noindent{\em\color{blue} Conclusion} --- A dynamical polarization theory is formulated here with a whole new spectrum of concepts and techniques. The central quantity, dynamical mean polarization $ \hat{x}_{\text{mean}}(t) $  in Eq.~(\ref{eq:defxmean}),
	depicts the {\em temporal} interference pattern for Bloch-wave evolutions.
	It serves as a natural and qualitative generalization of 
	static polarization 
	(corresponding to $ \hat{y}_{\text{occ}} $)
	characterizing {\em spatial} interference patterns for Bloch waves of occupied bands in different sublattices. Validity of dynamical polarization theory in Eqs.~(\ref{eq:wilson1}) and (\ref{eq:wilson2}) transcends specific dimensions and/or symmetries, offering a generic way to define and study dynamical multipoles in driven or quenched systems. The first proposal to unambiguously detect the anomalous character of Floquet phases beyond band descriptions is also provided. 
	

	\noindent{\em \color{blue} Update} --- During submission, we noted two preprints~\cite{Bomantara2018,Rodriguez-Vega2018} constructing model/symmetry-dependent topological invariants, and another on classification with a single mirror symmetry~\cite{Peng2018}. Also, there appeared a complementary proposal~\cite{Unal2018} to measure first-order Floquet topology. 

	\noindent{\em\color{blue} Acknowledgment} --- This work is supported by AFOSR Grant No.  FA9550-16-1-0006, ARO Grant No. W911NF-11-1-0230, MURI-ARO Grant No. W911NF-17-1-0323, and NSF of China Overseas Scholar Collaborative Program Grant No. 11429402 sponsored by Peking University.


	%
	
	\pagebreak
	
	\widetext
	\clearpage
	
	\setcounter{equation}{0}
	
	\begin{center}
		{\bf\Large Supplemental Material}
	\end{center}

	\renewcommand{\thesection}{S-\arabic{section}}
	\renewcommand{\theequation}{S\arabic{equation}}
	\setcounter{equation}{0}  
	\renewcommand{\thefigure}{S\arabic{figure}}
	\setcounter{figure}{0}  

	\title{Supplemental Materials for ``Higher-Order Floquet Topological Insulators with Anomalous Corner States''}
	\author{Biao Huang}
	\email{phys.huang.biao@gmail.com} 
	\affiliation{Department of Physics and Astronomy, University of Pittsburgh, Pittsburgh PA 15260, USA}
	\author{W. Vincent Liu}
	\email{wvliu@pitt.edu}
	\affiliation{Department of Physics and Astronomy, University of Pittsburgh, Pittsburgh PA 15260, USA}
	\affiliation{
		Wilczek Quantum Center, School of Physics and Astronomy and T. D. Lee Institute,  Shanghai Jiao Tong University, Shanghai 200240, China}
	\date{\today}
	\maketitle

	\tableofcontents

	\section{Theory of anomalous dynamical polarization}
	In this section, we give a detailed account of the anomalous Floquet dynamical polarization theory. Such a polarization theory is different from static ones right from the definition at the very beginning.
	
	\subsection{Preliminary: Procedure to obtain periodized evolution operator $ U_\varepsilon(t) $}
	
	To be self-contained, we briefly review the decomposition of evolution operator into the ``normal" and ``anomalous" parts~\cite{Rudner2013,Roy2016,Yao2017}. In a periodically driven system $ H(\boldsymbol{k},t+T) = H(\boldsymbol{k},t) $, the evolution operator $ U(\boldsymbol{k},t) $ can be decomposed into a periodic part $ U_\varepsilon(\boldsymbol{k},t+T)=U_\varepsilon(\boldsymbol{k},t) $ (``anomalous") and an accumulative static part $ [U(\boldsymbol{k},T)]_\varepsilon^{t/T} $ (``normal"),
	\begin{align}\label{eqsupp:ut}
	U(\boldsymbol{k},t) \equiv P_{t'} e^{-i\int_0^t dt' H(\boldsymbol{k},t')} = U_\varepsilon(\boldsymbol{k},t) [U(\boldsymbol{k},T)]_\varepsilon^{t/T}.
	\end{align}
	To perform such a decomposition, one first obtain the effective static Hamiltonian by diagonalizing $ U(\boldsymbol{k},T) $
	\begin{align}
	U(\boldsymbol{k},T) |u_{\alpha,\boldsymbol{k}}\rangle = e^{i E_{\alpha,\boldsymbol{k}}} |u_{\alpha,\boldsymbol{k}} \rangle , \qquad
	\Rightarrow \qquad
	H_\varepsilon(\boldsymbol{k}) = \sum_\alpha | u_{\alpha,\boldsymbol{k}} \rangle \left( -i \ln_{\varepsilon} e^{iE_{\alpha,\boldsymbol{k}}} \right) \langle u_{\alpha,\boldsymbol{k}} |
	\end{align}
	Here, $ \varepsilon $ is the position of branch cut. Second, the static Hamiltonian evolution means
	\begin{align}\label{eqsupp:floquetU}
	[U(\boldsymbol{k},T)]_\varepsilon^{t/T} = e^{i H_{\varepsilon}(\boldsymbol{k}) t/T} = \sum_{\alpha} |u_{\alpha,\boldsymbol{k}}\rangle \left( e^{iE_{\alpha,\boldsymbol{k}}} \right)_\varepsilon^{t/T} \langle u_{\alpha,\boldsymbol{k}} |,
	\end{align}
	where $ \,_\varepsilon $ is a reminder for branch cuts when taking $ t/T $ roots. It becomes the identity $ \mathbb{I} $ at $ t=0 $ and gives the Floquet operator at period ends $ t=T $. Its spectrum resembles that of $ U(\boldsymbol{k},T) $ but with an artificial linear expansion $ \left( e^{iE_{\alpha,\boldsymbol{k}}} \right)_\varepsilon^{t/T}  $. Finally, the periodic part is obtained by a re-statement of Eq.~(\ref{eqsupp:ut}):
	\begin{align}\label{eqsupp:uep}
	U_\varepsilon(\boldsymbol{k},t) \equiv U(\boldsymbol{k},t) [U(\boldsymbol{k},T)]^{-t/T}_\varepsilon,
	\end{align}
	where $ [U(\boldsymbol{k},T)]_\varepsilon^{-t/T} $ is the inverse of Eq.~(\ref{eqsupp:floquetU}).
	From this form, one could easily obtain the periodicity $ U_\varepsilon(\boldsymbol{k},t+T) = U_\varepsilon(\boldsymbol{k},t) $, and in particular, $ U_\varepsilon(\boldsymbol{k},0) = U_{\varepsilon}(\boldsymbol{k},T) = \mathbb{I} $, the identity matrix.

	Let us practice calculations with a simple example, $ U(T) = i\tau_x $, which has eigenvalues $ \pm i $ and eigenvectors $ |\pm\rangle = (1/\sqrt2) (1,\pm1)^T $, $ \,^T$ is transpose. Then the corresponding $ H_{\varepsilon=0} = |+\rangle \frac{\pi}{2} \langle +| + |-\rangle \frac{3\pi}{2} \langle -| = \frac{\pi}{2} \left(\begin{array}{cc}
	2 & -1 \\ -1 & 2
	\end{array} \right) $, and 
	$ H_{\varepsilon=\pi} = |+\rangle \frac{\pi}{2} \langle +| + |-\rangle (-\frac{\pi}{2}) \langle - | = \frac{\pi}{2} \begin{pmatrix}
	0 & 1 \\ 1 & 0
	\end{pmatrix} $, which give rise to $ [U(T)]_{\varepsilon=0}^{t/T} = e^{i\pi t/T} e^{-i (\pi/2) (t/T) \tau_x} = e^{i\pi t/T} \begin{pmatrix}
	\cos\frac{\pi t}{2T} & -i\sin\frac{\pi t}{2T} \\
	-i\sin\frac{\pi t}{2T} & \cos\frac{\pi t}{2T}
	\end{pmatrix} $, while $ [U(T)]_{\varepsilon=\pi}^{t/T} = e^{i(\pi/2) (t/T)\tau_x} = \begin{pmatrix}
	\cos\frac{\pi t}{2T} & i \sin\frac{\pi t}{2T} \\
	i\sin\frac{\pi t}{2T} & \cos\frac{\pi t}{2T}
	\end{pmatrix} $. Note that only at the period end $ t=T $ any $ \varepsilon $ give the same result, $ \left( [U(T)]_\varepsilon^{t/T}\right)|_{t=T} =  U(T)=i\tau_x $, but they are generally different during the cycle $ t\in(0,T) $. Finally, the periodized evolution operator $ U_\varepsilon(t) $ is obtained by acting on the full evolution operator $ U(t) $ the inverse of $ [U(T)]^{t/T}_\varepsilon $ obtained above, for different branch cuts $ \varepsilon $ as in Eq.~(\ref{eqsupp:uep}).
	
	It is worth emphasizing that to define the branch cut, one {\bf only needs the ``static" $ U(\boldsymbol{k},T) $ to be gapped at $ \varepsilon $. The periodized part $ U_\varepsilon(\boldsymbol{k},t) $ does {\em not} need to be gapped at any moment.} Thus, there is no concept of ``bands" for $ U_\varepsilon(\boldsymbol{k},t) $. In fact, $ U_\varepsilon(\boldsymbol{k},t) $ being gapless at certain $ (\boldsymbol{k},t) $ underlies the renowned concept of ``dynamical singularity" serving as alternative characterization of anomalous Floquet insulators~\cite{Nathan2015}.

	\subsection{Proving $ (\hat{x}(t))^{L_x}=\hat{\mathbb{I}}  $ and $ \hat{x}_{\text{mean}}(t) \hat{x}_{\text{mean}}^\dagger (t) = \hat{\mathbb{I}} $. }
	
	Let us fix the notations below, where we write operators in the second quantized form and adopt the Fourier convention
	\begin{align}
	& \hat{c}_{im} = \frac{1}{\sqrt{N_{\text{cell}}}} \sum_{\boldsymbol{k}} e^{-i\boldsymbol{k}\cdot\boldsymbol{r}_i} \hat{c}_{\boldsymbol{k}m}, \qquad
	\qquad
	\hat{c}_{\boldsymbol{k}m} = \frac{1}{\sqrt{N_{\text{cell}}}} \sum_{\boldsymbol{k}} e^{i\boldsymbol{k}\cdot\boldsymbol{r}_i} \hat{c}_{im}, 
	\qquad\qquad\qquad N_{\text{cell}} = L_xL_y\dots,
	\\
	& \boldsymbol{r}_i = x_i\boldsymbol{e}_x + y_i\boldsymbol{e}_y + \dots, \quad
	x_i=1,\dots, L_x, \,\, y_i=1,\dots,L_y,\dots;\quad 
	\boldsymbol{\Delta} = \frac{2\pi}{L_x}\boldsymbol{e}_x + \frac{2\pi}{L_y} \boldsymbol{e}_y + \dots \equiv \Delta_x\boldsymbol{e}_x +  \Delta_y\boldsymbol{e}_y + \dots,
	\\
	\label{supp:eqxpol}
	& \hat{x} = \sum_{im} \hat{c}_{im}^\dagger |0\rangle e^{ -i \Delta_x x_i} \langle 0| \hat{c}_{im} 
	= \sum_{\boldsymbol{k},m} \hat{c}^\dagger_{\boldsymbol{k}+\Delta_x \boldsymbol{e}_x,m} |0\rangle \langle 0| \hat{c}_{\boldsymbol{k}m},
	\\
	& \hat{U}_\varepsilon(t) = \sum_{\boldsymbol{k}mn} \hat{c}_{\boldsymbol{k}m}^\dagger |0\rangle [U_\varepsilon(\boldsymbol{k},t)]_{mn}  \langle 0| \hat{c}_{\boldsymbol{k}n},\\ \label{supp:eqxtuu}
	& \hat{x}(t) = \hat{U}_\varepsilon^\dagger (t) \hat{x} \hat{U}_\varepsilon(t) 
	= 
	\sum_{\boldsymbol{k}mn} \hat{c}^\dagger_{\boldsymbol{k}+\Delta_x\boldsymbol{e}_x m} |0\rangle [U_{\varepsilon}^\dagger (\boldsymbol{k}+\Delta_x\boldsymbol{e}_x,t) U_{\varepsilon}(\boldsymbol{k}, t)]_{mn} \langle 0| \hat{c}_{\boldsymbol{k}n},\\ \label{supp:eqxmeandef}
	& \hat{x}_{\text{mean}}(t) = (\hat{x}(t) + \hat{x}(0))/2.
	\end{align}
	Polarization along other directions, i.e. $ \hat{y} $, can be defined similar to Eq.~(\ref{supp:eqxpol}) by replacing $ (x_i,\Delta_x) \rightarrow (y_i,\Delta_y) $ etc.. It is worth emphasizing that $ m, n $ here denote the sublattice indices, {\em not} the band indices as for static polarization. Throughout our discussions we do not invoke the band indices anywhere, as there is no concept of bands for $ U_\varepsilon(\boldsymbol{k},t) $. Also, in this section of supplemental material, we would reserve the ``bra" and ``ket" notations for second-quantized operators (emphasized by $ \hat{\,} $ ), i.e. $ \hat{c}_{\boldsymbol{k}m}^\dagger |0\rangle $. The eigenstates for matrices (without $ \hat{\,} $ ), instead, will be denoted explicitly by their components.
	
	We next prove two relations mentioned in the main text that help justify using $ \hat{x}_{\text{mean}}(t) $:
	\\
	\noindent {\bf (1) $ (\hat{x}(t))^{L_x} = \hat{\mathbb{I}}$, the identity operator ($ \hat{\mathbb{I}} = \sum_{im} \hat{c}_{im}^\dagger |0\rangle \delta_{im,jn} \langle  0| \hat{c}_{jn} = \sum_{\boldsymbol{k}m} \hat{c}_{\boldsymbol{k}m}^\dagger |0\rangle \langle 0| \hat{c}_{\boldsymbol{k}m} $) at any time $ t $.\\
		(2) $ \hat{x}_{\text{mean}}(t) $ is a unitary operator  in the thermodynamic limit, $ \hat{x}^\dagger_{\text{mean}}(t) \hat{x}_{\text{mean}} (t) = \hat{\mathbb{I}} $ at any time $ t $.
	}
	
	{\bf Proof (1):} 
	The form of $ \hat{x}(t) $ in Eq.~(\ref{supp:eqxtuu}) directly gives
	\begin{align}\nonumber
	(\hat{x}(t))^{L_x} &= \sum_{kmn} \hat{c}^\dagger_{\boldsymbol{k}m} |0\rangle \left[ 
	\left(U^\dagger_{\varepsilon}(\boldsymbol{k}+L_x\Delta_x \boldsymbol{e}_x,t) U_\varepsilon(\boldsymbol{k}+(L_x-1)\Delta_k \boldsymbol{e}_x,t)\right)
	\right.
	\dots
	\\
	& \qquad\qquad \left.
	\cdot \left(U^\dagger_{\varepsilon}(\boldsymbol{k}+2\Delta_x \boldsymbol{e}_x,t) U_\varepsilon(\boldsymbol{k}+\Delta_x \boldsymbol{e}_x ,t)\right)
	\left(U^\dagger_{\varepsilon}(\boldsymbol{k}+\Delta_x\boldsymbol{e}_x,t) U_\varepsilon(\boldsymbol{k},t)\right) 
	\right]_{mn} \langle 0| \hat{c}_{\boldsymbol{k}n}\\
	&= \sum_{\boldsymbol{k}m} \hat{c}_{\boldsymbol{k}m}^\dagger |0\rangle \langle 0| \hat{c}_{\boldsymbol{k}m},
	\end{align}
	where we have used the unitarity $ [U_\varepsilon(\boldsymbol{k},t) U_{\varepsilon}^\dagger (\boldsymbol{k},t)]_{mn} = \delta_{mn} $ and that $ L_x\Delta_x=2\pi $, which is the period of Brillouin zone $ U_\varepsilon(\boldsymbol{k}+2\pi\boldsymbol{e}_x,t) = U_\varepsilon(\boldsymbol{k},t) $. Thus, the eigenvalues of $ \hat{x}(t) $ is again $ 1^{1/L_x} = (e^{-2\pi i x_i})^{1/L_x}=e^{-i\Delta_k x_i} $. That means at any {\em single} moment, the evolution of polarization for all lattice is not a well-defined concept, and one has to compare multiple moments, i.e. $ \hat{x}_{\text{mean}}(t) $.
	
	{\bf Proof (2):} 
	From Eqs.~(\ref{supp:eqxpol}), (\ref{supp:eqxtuu}) and (\ref{supp:eqxmeandef}),
	\begin{align}
	\nonumber
	\hat{x}_{\text{mean}}^\dagger (t) \hat{x}_{\text{mean}}(t)
	&= 
	\frac{1}{4} (2\hat{\mathbb{I}} + \hat{x}^\dagger  \hat{U}_{\varepsilon}^\dagger(t) \hat{x} \hat{U}_\varepsilon(t) + \hat{U}_\varepsilon^\dagger(t) \hat{x}^\dagger \hat{U}_\varepsilon(t) \hat{x}) =\sum_{\boldsymbol{k}m} \hat{c}_{\boldsymbol{k}m} |0\rangle 
	[B_{\boldsymbol{k}}]_{mn} \langle 0| \hat{c}_{\boldsymbol{k}n},\\ \nonumber
	B_{\boldsymbol{k}} &= \frac{1}{4} (2+ U_\varepsilon^\dagger (\boldsymbol{k}+\Delta_x\boldsymbol{e}_x,t) U_\varepsilon(\boldsymbol{k},t) + U_\varepsilon^\dagger (\boldsymbol{k}-\Delta_x\boldsymbol{e}_x,t) U_\varepsilon(\boldsymbol{k},t) )\\\nonumber
	\nonumber
	&= \frac{1}{4} (2+ (1+(\partial_{k_x}U^\dagger_\varepsilon(\boldsymbol{k},t)) U_\varepsilon(\boldsymbol{k},t) \Delta_x + O(\Delta_x^2) ) + (1-(\partial_{k_x}U^\dagger_\varepsilon(\boldsymbol{k},t)) U_\varepsilon(\boldsymbol{k},t) \Delta_x + O(\Delta_x^2)) 
	\\
	&= 1 + O(\Delta_x^2) \xrightarrow{L_x\rightarrow\infty } 1 \\ \nonumber
	\nonumber
	\left( \hat{x}_{\text{mean}}^\dagger (t) \right)^{L_x} \left( \hat{x}_{\text{mean}}(t) \right)^{L_x}  &= 1 + L_x O(\Delta_x^2) \xrightarrow{L_x\rightarrow\infty} 1, \qquad\qquad (\Delta_x=2\pi/L_x).
	\end{align}
	In the last equation, we used repeatedly $ \hat{x}_{\text{mean}}^\dagger (t) O(\Delta_x^2) \hat{x}_{\text{mean}}(t) = O(\Delta^2_x) $, where $ O(\Delta_x^2) $ means any terms of quadratic order $ \Delta_x^2 $ or higher.
	Thus, in the thermodynamic limit, both $ \hat{x}_{\text{mean}} $ and $ \hat{x}_{\text{mean}}^{L_x} $ (the latter is related to the dynamical Wilson loop) converge to unitary operators.

	\subsection{Higher-order dynamical multipoles: general constructions \label{secsupp:hoti}}
	
	In this subsection, we present the algebraic details for the rigorous construction of dynamical polarization. To be general, we consider {\bf arbitrary dimensions}, for polarizations to {\bf arbitrary orders}. Specific applications to 2D will be discussed in the next subsection regarding effects of mirror symmetries.

	First, from Eqs.~(\ref{supp:eqxpol}), (\ref{supp:eqxtuu}) and (\ref{supp:eqxmeandef}), we have the dynamical mean polarization
	\begin{align}\label{supp:eqQxt}
	\hat{x}_{\text{mean}}(t) = \sum_{\boldsymbol{k}mn} \hat{c}_{\boldsymbol{k}+\Delta_x \boldsymbol{e}_x, m}^\dagger |0\rangle [Q_{x,\boldsymbol{k}}(t)]_{mn} \langle 0| \hat{c}_{\boldsymbol{k}m}, \qquad
	\boxed{Q_{x,\boldsymbol{k}}(t) = \frac{\mathbb{I}+U_\varepsilon^\dagger (\boldsymbol{k}+\Delta_x \boldsymbol{e}_x, t) U_\varepsilon^\dagger (\boldsymbol{k}, t)}{2}}
	\end{align}
	The eigen-problem of $ \hat{x}_{\text{mean}} $ can be solved by considering its $ L_x $-th power
	\begin{align}\label{supp:eqWxt}
	& \hat{x}_{\text{mean}}^{L_x}(t) = \sum_{\boldsymbol{k}mn} \hat{c}^\dagger_{\boldsymbol{k}m} |0\rangle [W_{x,\boldsymbol{k}}(t)]_{mn} \langle 0| \hat{c}_{\boldsymbol{k}n},
	\qquad
	\boxed{ 
		W_{x,\boldsymbol{k}}(t)=  Q_{x,\boldsymbol{k}+(L_x-1) \Delta_x\boldsymbol{e}_x}(t) \dots Q_{x,\boldsymbol{k}+\Delta_x\boldsymbol{e}_x}(t) Q_{x,\boldsymbol{k}}(t),
	}
	\end{align}
	which is diagonal in $ \boldsymbol{k} $. Further obtaining the eigenstates/eigenvalues of the unitary, time-dependent Wilson loop $ W_{x,\boldsymbol{k}}(t) $,
	\begin{align}\label{supp:eqWeig}
	[W_{x,\boldsymbol{k}}(t) ]_{mn} = \sum_{\mu} [\nu_{x,\mu}(\boldsymbol{k},t) ]_{m} e^{-2\pi i \nu_{x,\mu} (k_{l\ne x},t)} [ \nu_{x,\mu} (\boldsymbol{k},t)]_n^*, && 
	\delta_{\mu_1\mu_2} = \sum_{m} [\nu_{x,\mu_1}(\boldsymbol{k})]_m^* [\nu_{x,\mu_2}(\boldsymbol{k})]_m.
	\end{align}
	we can solve the eigen-problem of $ \hat{x}_{\text{mean}}(t) $. The eigenvalues of $ \hat{x}_{\text{mean}}(t) $ are the $ L_x $ roots of $ e^{-2\pi i \nu_{x,\mu}(\boldsymbol{k},t)} $ for each branch $ \mu $, labeled by $ x_i=1,\dots,L_x $,
	\begin{align}\label{supp:eqxb}
	\hat{x}_{\text{mean}}(t) |b_{x,\mu}(x_i, k_{l\ne x},t)\rangle = e^{-i\Delta_x (x_i + \nu_{x,\mu}(k_{l\ne x},t))}|b_{x,\mu}(x_i, k_{l\ne x},t) \rangle .
	\end{align}
	Compared with eigenvalues $ e^{-i\Delta_x x_i} $ for $ \hat{x} $, we see the relative ``shift" $ \nu_{x,\mu}(k_{l\ne x},t) $ denoting the motion of $ 2\nu_{x,\mu}(k_{l\ne x},t) $ along $ x $ from $ t=0 $ to $ t $, as explained in the main text. Eigenvectors $ |b_{x,\mu}(x_i, k_{l\ne x}) \rangle $ can be explicitly written as
	\begin{align}\label{supp:eqb1st}
	|b_{x,\mu}(x_i, k_{l\ne x},t) \rangle = \frac{1}{\sqrt L_x} \sum_{k_x,m} \hat{c}_{\boldsymbol{k}m}^\dagger |0\rangle  e^{i k_x  x_i } [\nu_{x,\mu}(\boldsymbol{k},t)]_m
	\end{align}
	To verify, we first note that 
	\begin{align}
	\hat{x}_{\text{mean}}^{L_x+1}(t)=\hat{x}_{\text{mean}}^{L_x}(t) \hat{x}_{\text{mean}}(t)=\hat{x}_{\text{mean}} (t) \hat{x}_{\text{mean}}^{L_x}(t) \qquad
	\Rightarrow Q_{x,\boldsymbol{k}}(t) W_{x,\boldsymbol{k}}(t) = W_{x,\boldsymbol{k}+\Delta_x\boldsymbol{e}_x} (t) Q_{x,\boldsymbol{k}}(t).
	\end{align}
	Combined with Eqs.~(\ref{supp:eqWxt}) and (\ref{supp:eqWeig}), we see the general form for $ Q_{\boldsymbol{k}}(t) $ reads
	\begin{align}
	[Q_{x,\boldsymbol{k}}]_{mn} = e^{i\Delta_x x'_i} \sum_{\mu} [\nu_{x,\mu}(\boldsymbol{k}+ \Delta_x \boldsymbol{e}_x,t)]_{m} e^{-i\Delta_x \nu_{x,\mu}(k_{l\ne x},t) } [\nu_{x,\mu}(\boldsymbol{k},t)]_{n}^*,\qquad\qquad
	x_i'\in \mathbb{Z}.
	\end{align}
	Here, the global phase $ e^{i\Delta_k x'_i} $ can always be absorbed into $ [\nu_{x,\mu}(\boldsymbol{k},t)]_m $ by the gauge transformation $ [\nu_{x,\mu}(\boldsymbol{k},t)]_m \rightarrow e^{-ik_x x_i'} [\nu_{x,\mu}(\boldsymbol{k},t)]_m $, so we set $ x_i'=0 $. 
	Then,
	\begin{align}\nonumber
	\hat{x}_{\text{mean}}(t) |b_{x,\mu}(x_i,k_{l\ne x},t)\rangle &=
	\frac{1}{\sqrt L_x} \sum_{k_xmn\mu}  \hat{c}^\dagger_{\boldsymbol{k}+\Delta_x \boldsymbol{e}_x,m} |0\rangle [Q_{x,\boldsymbol{k}}]_{mn}[\nu_{x,\mu}(\boldsymbol{k},t)]_n e^{ik_x x_i} \\
	&=
	\frac{1}{\sqrt L_x} \sum_{k_xm\mu}  \hat{c}^\dagger_{\boldsymbol{k}+\Delta_x \boldsymbol{e}_x,m}|0\rangle e^{-i\Delta_x\nu_{x,\mu}(k_{l\ne x},t)} [\nu_{x,\mu}(\boldsymbol{k}+\Delta_x \boldsymbol{e}_x, t)]_m e^{ik_x x_i} = e^{-i\Delta_x(x_i+\nu_\mu)} |b_{x,\mu}(x_i,k_{l\ne x},t)\rangle,
	\end{align}
	where in the last step we have replaced the dummy index $ k_x \rightarrow k_x -\Delta_x $.

	The physical picture behind the above formulations should be noted, especially its difference from the static polarization~\cite{Benalcazar2017}:
	\begin{enumerate}
		\item $ |b_{x,\mu}(x_i,k_{l\ne x},t)\rangle $ cannot be understood as  static ``edge Wannier band" eigenfunctions, as even the notion of bands for $ \hat{U}_\varepsilon(t) $ does not exist, let alone edge ones. Indeed, $ \hat{x}_{\text{mean}}(t) = (\hat{x}(t)+\hat{x}(0))/2 $ describes the {\em interference} for polarizations patterns at two different moments $ 0 $ and $ t $, so its eigenstates are not the edge ``bands" at any specific moment. Instead, the interference patterns $ |b_{x,\mu}(x_i,k_{l\ne x},t)\rangle $ characterize the evolution of polarization from $ 0\rightarrow t $, and therefore we dub $ \mu $ in $ |b_{x,\mu}(x_i,k_{l\ne x},t)\rangle $'s as a {\bf dynamical branches} signaling different relative motions across unit cells from $ 0\rightarrow t $. 
		\item As no ``band projection" is invoked inhere, the number of dynamical branches $ \mu $ equals the number of sublattices $ m $, i.e. 4 sublattices would result in 4 dynamical branches. This is to be contrasted with the static situation that because of projection to filled bands at half filling, 4 sublattices (4 bands) would results in only 2 edge Wannier bands. Such a difference further affects the following discussion on second-order polarization, where the dynamical nested Wilson loop~Eq.~(\ref{supp:eqW2yt}) is an SU(2) matrix for 4 sublattices in the main text, in contrast to static nested Wilson loop which is a U(1) number~\cite{Benalcazar2017}.
		
	\end{enumerate}
	
	The first-order dynamical branches $ \mu $'s can be grouped into several separable sets $ \nu_x $'s. Two branches $ \mu_1, \mu_2 $ in the same set $ \nu_x $ would have $ \nu_{x,\mu_1}(k_{l\ne y},t) = \nu_{x,\mu_2}(k_{l\ne x},t)  $ intersecting at certain $ (k_{l\ne x},t) $, or being degenerate. $ \mu_1, \mu_2 $ in different sets $ \nu_x $'s are completely separable $  \nu_{x,\mu_1}(k_{l\ne y},t) \ne \nu_{x,\mu_2}(k_{l\ne x},t) $ at any $ k_{l\ne x} $ throughout $ t\in (0,T) $. Intuitively, different branch sets $ \nu_x $ denote separable magnitude of motion along $ x $ at certain $ t $. Thus, one could defined the branch set projectors 
	\begin{align}\label{supp:eqP1}
	P_{\nu_x} = \sum_{x_i, k_{l\ne x};\mu\in \nu_x} |b_{x,\mu}(x_i,k_{l\ne x},t)\rangle \langle b_{x,\mu}(x_i,k_{l\ne x},t)| = \sum_{\mu\in\nu_x} \sum_{\boldsymbol{k}mn} \hat{c}_{\boldsymbol{k}m}^\dagger |0\rangle [\nu_{x,\mu}(\boldsymbol{k},t)]_m [\nu_{x,\mu}(\boldsymbol{k},t)]_n^* \langle 0| \hat{c}_{\boldsymbol{k}n}.
	\end{align}
	Introduce the dynamical branch annihilation/creation operators
	\begin{align}
	\hat{b}_{\boldsymbol{k}\mu}(t) = \sum_{m} \hat{c}_{\boldsymbol{k},m} [\nu_{x,\mu}(\boldsymbol{k},t)]^*_m, &&
	\hat{b}^\dagger_{\boldsymbol{k}\mu} (t)  = \sum_{m} \hat{c}_{\boldsymbol{k},m}^\dagger [\nu_{x,\mu}(\boldsymbol{k},t)]_m,
	&&
	\{ \hat{b}_{\boldsymbol{k},\mu_1} (t) , \hat{b}_{\boldsymbol{k}',\mu_2}^\dagger(t) \} = \delta_{\boldsymbol{k},\boldsymbol{k}'} \delta_{\mu_1\mu_2},
	\end{align}
	the projectors can also be written as
	\begin{align}
	\hat{P}_{\nu_x} = \sum_{\mu\in \nu_x} \hat{b}_{\boldsymbol{k}\mu} |0\rangle \langle 0 |\hat{b}_{\boldsymbol{k}\mu}.
	\end{align}
	The second order dynamical polarization corresponds to motions perpendicular to $ x $ within each $ \nu_x $ set of branches respectively. Here, we take $ y $ as the second direction orthogonal to $ x $. The dynamical mean polarization along $ y $ can be similarly defined,
	\begin{align}\label{supp:eqQy}
	\hat{y}_{\text{mean}} = \frac{\hat{y}(t) + \hat{y}(0)}{2} = \sum_{\boldsymbol{k}mn} \hat{c}_{\boldsymbol{k}+\Delta_y \boldsymbol{e}_y, m}^\dagger |0\rangle [Q_{y,\boldsymbol{k}}]_{mn} \langle 0| \hat{c}_{\boldsymbol{k}m}, \qquad 
	Q_{y,\boldsymbol{k}} = \frac{\mathbb{I}+U_\varepsilon^\dagger (\boldsymbol{k}+\Delta_y \boldsymbol{e}_y, t) U_\varepsilon(\boldsymbol{k}, t) }{2}.
	\end{align}
	Projecting to the branch set $ \nu_x $, we have the second order dynamical polarization, described by $ \hat{y}^{(\nu_x)}_{\text{mean}}(t) $, as
	\begin{align}\label{supp:ynux}
	\hat{y}^{(\nu_x)}_{\text{mean}}(t) &= \hat{P}_{\nu_x}(t) \hat{y}_{\text{mean}}(t) \hat{P}_{\nu_x}(t)
	=\sum_{\boldsymbol{k},\mu_1\mu_2\in\nu_{x,\mu_i}} \hat{b}_{\boldsymbol{k}+\Delta_y \boldsymbol{e}_y,\mu_1}(t) |0\rangle  
	[Q^{(\nu_x)}_{y,\boldsymbol{k}}(t)]_{\mu_1\mu_2}
	\langle 0| \hat{b}_{\boldsymbol{k}\mu_2}(t),\\  
	[Q^{(\nu_x)}_{y,\boldsymbol{k}}(t)]_{\mu_1\mu_2} &= \sum_{m_1m_2} 
	[\nu_{x,\mu_1}(\boldsymbol{k}+\Delta_y\boldsymbol{e}_y,t)]_{m_1}^* [Q_{y,\boldsymbol{k}}(t)]_{m_1m_2} [\nu_{x,\mu_2}(\boldsymbol{k},t)]_{m_2} 
	\end{align}
	Combined with Eq.~(\ref{supp:eqQy}),
	\begin{align}\label{supp:eqQ2yt}
	\boxed{
		[Q_{y,\boldsymbol{k}}^{(\nu_x)}(t)]_{\mu_1\mu_2} = \sum_{m_1m_2} 
		[\nu_{x,\mu_1}(\boldsymbol{k}+\Delta_y\boldsymbol{e}_y,t )]_{m_1}^* \left[\frac{\mathbb{I}+U_\varepsilon^\dagger (\boldsymbol{k}+\Delta_y \boldsymbol{e}_y, t) U_\varepsilon(\boldsymbol{k}, t) }{2} \right]_{m_1m_2} [\nu_{x,\mu_2}(\boldsymbol{k},t )]_{m_2} 
		.}
	\end{align}
	To solve the quadrupole eigenstate problem, we similarly consider
	\begin{align}\label{supp:eqW2yt}
	\left(\hat{y}^{(\nu_x)}_{\text{mean}}(t)\right)^{L_y} = \sum_{\boldsymbol{k},\mu_{1}\mu_2\in \nu_x } 
	\hat{b}_{\boldsymbol{k}\mu_{1}}^\dagger(t) |0\rangle [W^{(\nu_x)}_{y,\boldsymbol{k}}(t)]_{\mu_{1}\mu_2} \langle 0| \hat{b}_{\boldsymbol{k}\mu_2} (t), &&
	\boxed{
		W^{(\nu_x)}_{y,\boldsymbol{k}}(t)= 
		Q^{(\nu_x)}_{y,\boldsymbol{k}+(L_y-1) \Delta_y \boldsymbol{e}_y} (t) \,
		\dots \, Q^{(\nu_x)}_{y,\boldsymbol{k}+\Delta_y \boldsymbol{e}_y} (t) \, Q^{(\nu_x)}_{y,\boldsymbol{k}}(t)
	}
	\end{align}
	Then, we diagonalize the dynamical nested Wilson loop (the second-order (quadrupole) branches are denoted with $ \tilde{\mu} $, while the first-order (dipole) ones with $ \mu $)
	\begin{align}\label{supp:eqW2ytdiag}
	\boxed{
		[W^{(\nu_x)}_{y,\boldsymbol{k}}(t)]_{\mu_1\mu_2} = \sum_{\tilde{\mu}} [\nu^{(\nu_x)}_{y,\tilde{\mu}}(\boldsymbol{k},t)]_{\mu_1} 
		e^{2\pi i \nu^{(\nu_x)}_{y,\tilde{\mu}}(k_{l\ne y},t) } [\nu^{(\nu_x)}_{y,\tilde{\mu}}(\boldsymbol{k},t)]^*_{\mu_2}  .
	}
	\end{align}
	Note $ \mu_1,\mu_2 \in \nu_x $ for $ [W_{y,\boldsymbol{k}}^{(\nu_x)}(t)]_{\mu_1\mu_2} $, so the dimension of {\em nested} $ W_{y,\boldsymbol{k}}^{(\nu_x)}(t) $ is smaller than the number of sublattice, unlike the first order one $ W_{x,\boldsymbol{k}}(t) $. Finally, the dynamical quadrupole eigen-problem can be solved similar to Eqs.~(\ref{supp:eqxb}) and (\ref{supp:eqb1st}),
	\begin{align}
	& \hat{y}^{(\nu_x)}_{\text{mean}}(t) |b^{(\nu_x)}_{y,\tilde{\mu}}(y_i, k_{l\ne y},t)\rangle = e^{i\Delta_y(y_i+\tilde{p}^{(\nu_x)}_{y,\tilde{\mu}}(k_{l\ne y},t) )} |b^{(\nu_x)}_{y,\tilde{\mu}} (y_i,k_{l\ne y},t)\rangle,\\ \label{supp:eqb2yt}
	& |b^{(\nu_x)}_{y,\tilde{\mu}}(y_i,k_{l\ne y},t)\rangle = \frac{1}{\sqrt{L_y}} \sum_{k_y,\mu\in\nu_x} \hat{b}_{\boldsymbol{k}\mu}^\dagger |0\rangle e^{-ik_y y_i} [\nu^{(\nu_x)}_{y,\tilde{\mu}}(\boldsymbol{k},t)]_\mu. 
	\end{align}
	Limited to 2D, the averaged quadrupolar motion then reads 
	\begin{align}\label{supp:averagednu}
	\boxed{
		\langle \nu^{(\nu_x)}_{y,\tilde{\mu}}\rangle (t) = \frac{1}{L_x} \sum_{k_x} \nu^{(\nu_x)}_{y,\tilde{\mu}} (k_{x},t) }
	\end{align}

	From the above discussions, we have achieved a closed form to describe dynamical polarization to arbitrary orders. For instance, the third order polarization would constitute of the $\hat{z}_{\text{mean}}(t)$ projected to certain set of quadrupole branches $ P_{\nu_{xy}}  = \sum_{y_i, k_{l\ne y}, \tilde{\mu}\in \nu_{xy}} |b^{(\nu_x)}_{y,\tilde{\mu}}(y_i,k_{l\ne y},t)\rangle  \langle b^{(\nu_x)}_{y,\tilde{\mu}}(y_i,k_{l\ne y},t)| $ in Eq.~(\ref{supp:eqb2yt}). Then, one can carry out the similar procedure starting from Eq.~(\ref{supp:ynux}) to construct the third order dynamical Wilson loop, and end up with $ W_{z,\boldsymbol{k}}^{(\nu_{xy})} $ similar to Eq.~(\ref{supp:eqW2yt}). Its diagonalization then yields the solution to octupole eigen-problem similar to Eq.~(\ref{supp:eqb2yt}), from which one can construct the octupole branch set projectors to compute the fourth order polarization, so on and so forth.

	Finally, we consider thermodynamic limit formalism, $ L_x, L_y\dots \rightarrow \infty $. Using $ f(x+dx) \approx f(x) + (\partial f(x)/\partial x) dx $, they become
	\begin{align}
	& 
	Q_{x,\boldsymbol{k}}(t) \approx
	\mathbb{I} - \frac{1}{2} U_\varepsilon^\dagger (\boldsymbol{k},t) \partial_{k_y} U_\varepsilon(\boldsymbol{k},t) dk_x \xrightarrow{L_x\rightarrow\infty}  
	\exp( -A_{\boldsymbol{k}}(t) dk), \qquad
	A_{\boldsymbol{k}}(t) = - \frac{1}{2} U_\varepsilon^\dagger (\boldsymbol{k},t) \partial_{k_y} U_\varepsilon(\boldsymbol{k},t)
	\\
	&W_{x,\boldsymbol{k}}(t) \xrightarrow{L_x\rightarrow\infty}  
	\exp(-A_{\boldsymbol{k}+ 2\pi \boldsymbol{e}_x }(t)dk) \dots \exp(-A_{\boldsymbol{k}+dk\boldsymbol{e}_x} dk ) \exp(-A_{\boldsymbol{k}}(t) dk)
	\equiv
	P_{k'_x} e^{-\int_{\boldsymbol{k}}^{\boldsymbol{k}+2\pi \boldsymbol{e}_x} A_{\boldsymbol{k}}(t) dk'_x} 
	\\	\nonumber
	& [Q_{y,\boldsymbol{k}}^{(\nu_x)}(t)]_{\mu_1\mu_2} 
	\xrightarrow{L_y\rightarrow\infty}  \delta_{\mu_1\mu_0} - [A^{(\nu_x)}_{y,\boldsymbol{k}}]_{\mu_1\mu_2} dk_y 
	\approx
	[\exp(-A^{(\nu_x)}_{y,\boldsymbol{k}}dk)]_{\mu_1\mu_2},
	\\	
	& [A^{(\nu_x)}_{y,\boldsymbol{k}}(t)]_{\mu_1\mu_2} = \sum_{mn}\frac{1}{2} [ \nu_{x,\mu_1}(\boldsymbol{k},t) ]^*_m [U^\dagger_{\varepsilon}(\boldsymbol{k},t) \partial_{k'_y} U_\varepsilon(\boldsymbol{k},t)]_{mn}  [\nu_{x,\mu_2}(\boldsymbol{k},t)]_n  + \sum_m [\nu_{x,\mu_1}(\boldsymbol{k},t)]_m^* \partial_{k'_y} [ \nu_{x,\mu_2}(\boldsymbol{k},t)]_m 
	\\ \label{eqsupp:wxyk}
	&
	W^{(\nu_x)}_{y,\boldsymbol{k}} \xrightarrow{L_y\rightarrow\infty}
	\exp\left(-A^{(\nu_x)}_{y,\boldsymbol{k}+2\pi \boldsymbol{e}_y}(t) dk_y \right)
	\dots
	\exp\left(-A^{(\nu_x)}_{y,\boldsymbol{k}+dk_y \boldsymbol{e}_y}(t) dk_y\right) \exp\left(-A^{(\nu_x)}_{y,\boldsymbol{k}}(t) dk_y \right)
	\equiv
	P_{k'_y} e^{-\int_{\boldsymbol{k}}^{\boldsymbol{k}+2\pi \boldsymbol{e}_y} dk'_y A^{(\nu_x)}_{y,\boldsymbol{k}'}(t) 
	} 
	\\
	&
	\langle \nu_{y,\tilde{\mu}}^{(\nu_x)} \rangle (t) \xrightarrow{L\rightarrow\infty}  \int_{-\pi}^\pi \frac{d k_x}{2\pi} \nu_{y,\tilde{\mu}}^{(\nu_x)} (k_x,t)
	\end{align}
	These thermodynamic limit results are the ones presented in the main text, while their finite-size expressions are Eqs.~(\ref{supp:eqQxt}), (\ref{supp:eqWxt}), (\ref{supp:eqQ2yt}), (\ref{supp:eqW2yt}) and (\ref{supp:averagednu}) respectively.

	\subsection{Constraints enforced by mirror symmetries in two dimensions}
	From this section on, we limit our discussion to 2D for the second-order dynamical polarization, and discuss the constraints imposed by the two mirror symmetries $ \hat{M}_x, \hat{M}_y $. Their actions on a lattice, denoted by unit-cell indices $ x_i, y_i\in \mathbb{Z} $ and sublattice indices $ m $, are 
	\begin{align}
	\hat{M}_x: (x_i,y_i,m) \rightarrow (-x_i, y_i, M_x(m)), &&
	\hat{M}_y: (x_i, y_i, m)\rightarrow (x_i, -y_i,M_y(m)).
	\end{align}
	Formally, these operators read
	\begin{align}\label{supp:tempmx}
	\hat{M}_x &= \sum_{x_i,y_i,m} \hat{c}^\dagger_{-x_i,y_i,M_x(m)} |0\rangle \langle 0| \hat{c}_{x_i,y_i,m} e^{i\phi^x_m} 
	= \sum_{\boldsymbol{k},m}  \hat{c}^\dagger_{-k_x,k_y,M_x(m)} |0\rangle \langle 0|  \hat{c}_{k_x,k_y,m} e^{i\phi^x_m} \equiv
	\sum_{\boldsymbol{k},mn} \hat{c}^\dagger_{-k_x,k_y,m} |0\rangle  [M_x]_{mn} \langle 0| \hat{c}_{k_x,k_y,n},\\
	\label{supp:tempxy}
	\hat{M}_y &= \sum_{x_i,y_i,m} \hat{c}^\dagger_{x_i,-y_i,M_y(m)}|0\rangle \langle 0|  \hat{c}_{x_i,y_i,m} e^{i\phi^y_m} 
	= \sum_{\boldsymbol{k},m}  \hat{c}^\dagger_{k_x,-k_y,M_y(m)}|0\rangle \langle 0|  \hat{c}_{k_x,k_y,m} e^{i\phi^y_m} \equiv
	\sum_{\boldsymbol{k},mn} \hat{c}^\dagger_{k_x,-k_y,m} |0\rangle  [M_y]_{mn} \langle 0|\hat{c}_{k_x,k_y,n},
	\end{align}
	both being unitary $ \hat{M}_{x,y}\hat{M}_{x,y}^\dagger = 1 $. Note the matrices $ M_x,M_y $ only denote the sublattice transformations and do not carry momentum $ \boldsymbol{k} $ dependence. Here, the phase factors $ e^{i\phi^{x,y}_m} $ are chosen such that
	\begin{align}\label{eqsupp:mut}
	\hat{M}_{x,y} \hat{U}_\varepsilon(t) \hat{M}_{x,y}^\dagger = \hat{U}_\varepsilon(t).
	\end{align}
	Now, Eq.~(\ref{eqsupp:mut}) gives
	\begin{align}
	\sum_{\boldsymbol{k}mn} \hat{c}^\dagger_{\boldsymbol{k}m} |0\rangle [U_\varepsilon(\boldsymbol{k},t)]_{mn} \langle 0 | \hat{c}_{\boldsymbol{k}n} &= \sum_{\boldsymbol{k}mn}  \hat{c}^\dagger_{-k_x,k_y,m} |0\rangle  [ M_x U_\varepsilon (\boldsymbol{k},t) M_x^\dagger]_{mn} \langle 0| \hat{c}_{-k_x,k_y,n},\\
	&= \sum_{\boldsymbol{k}mn}  \hat{c}^\dagger_{k_x,-k_y,m} |0\rangle  [ M_y U_\varepsilon (\boldsymbol{k},t) M_y^\dagger]_{mn} \langle 0| \hat{c}_{k_x,-k_y,n},
	\end{align}
	implying the relation for matrices
	\begin{align}\label{eqsupp:matmum}
	M_x U_\varepsilon(\boldsymbol{k},t) M_x^\dagger = U_\varepsilon(-k_x,k_y,t), &&
	M_y U_\varepsilon(\boldsymbol{k},t) M_y^\dagger = U_\varepsilon(k_x,-k_y,t).
	\end{align}

	
	We first check the constraints on the first-order dynamical branches $ \nu_{x,\mu}(k_y,t) $. From Eq.~(\ref{eqsupp:matmum}),
	\begin{align}
	&
	M_x Q_{x,\boldsymbol{k}}(t) M_x^\dagger = \frac{1}{2} \left(\mathbb{I} + U^\dagger_\varepsilon(-k_x-\Delta_x,k_y,t) U_\varepsilon(-k_x,k_y,t) \right) = Q_{-x,-k_x,k_y}(t),\\ \label{supp:eqmxwmx}
	\Rightarrow\quad &
	M_x W_{x,\boldsymbol{k}}(t) M_x^\dagger = (M_xQ_{x,\boldsymbol{k}+(L_x-1)\Delta_x \boldsymbol{e}_x} (t) M_x^\dagger ) \dots 
	(M_xQ_{x,\boldsymbol{k}}(t)M_x^\dagger ) = W_{-x, -k_x,k_y}(t).
	\end{align}
	In the thermodynamic limit, recall the path-ordering definition (the matrix $ A_{\boldsymbol{k}-2\pi\boldsymbol{e}_x}=A_{\boldsymbol{k}} $)
	\begin{align}
	P_{+k'_x} e^{\int_{\boldsymbol{k}}^{\boldsymbol{k}+2\pi \boldsymbol{e}_x} dk'_x A_{(k'_x,k_y)} } &\equiv e^{A_{\boldsymbol{k}+2\pi \boldsymbol{e}_x}dk_x}\dots e^{A_{\boldsymbol{k}+dk_x\, \boldsymbol{e}_x}dk_x}  e^{A_{\boldsymbol{k}} dk_x},\\\nonumber
	P_{-k'_x} e^{\int_{\boldsymbol{k}}^{\boldsymbol{k}-2\pi \boldsymbol{e}_x} dk' A_{(k'_x,k_y)} } &\equiv  e^{A_{\boldsymbol{k}-2\pi \boldsymbol{e}_x} d(-k)} e^{A_{\boldsymbol{k}-(2\pi-dk)\boldsymbol{e}_x}d(-k_x)} \dots e^{A_{\boldsymbol{k}-dk_x \boldsymbol{e}_x}d(-k_x)} e^{A_{\boldsymbol{k}}d(-k_x)}\\
	&=
	e^{-A_{\boldsymbol{k}} dk_x} e^{-A_{\boldsymbol{k}+dk \boldsymbol{e}_x}dk_x} \dots e^{-A_{\boldsymbol{k}+2\pi \boldsymbol{e}_x}dk_x}.
	\end{align}
	That means
	$ \left( P_{+k'_x} e^{\int_{\boldsymbol{k}}^{\boldsymbol{k}+2\pi \boldsymbol{e}_x} dk'_x F_{(k',k_y)}} \right)^\dagger =  P_{-k'_x} e^{-\int_{\boldsymbol{k}}^{\boldsymbol{k}-2\pi \boldsymbol{e}_x} dk'_x F_{\boldsymbol{k}}^\dagger } $, and we have	
	\begin{align}
	(W_{x,\boldsymbol{k}})^\dagger = \left(
	P_{k'_x} e^{-\frac{1}{2}\int_{\boldsymbol{k}}^{\boldsymbol{k}+2\pi \boldsymbol{e}_x} dk'_x U^\dagger_\varepsilon(\boldsymbol{k}',t) \partial_{k'_x} U_\varepsilon(\boldsymbol{k}',t)
	}
	\right)^\dagger
	=
	P_{-k'_x} e^{+\frac{1}{2} \int_{\boldsymbol{k}}^{\boldsymbol{k}-2\pi \boldsymbol{e}_x}dk' \left(U^\dagger_\varepsilon(\boldsymbol{k}',t) \partial_{k'_x} U_\varepsilon(\boldsymbol{k}',t) \right)^\dagger} = W_{-x,\boldsymbol{k}},
	\end{align}
	where $ \left(U^\dagger_\varepsilon(\boldsymbol{k}',t) \partial_{k'_x} U_\varepsilon(\boldsymbol{k}',t) \right)^\dagger = - U^\dagger_\varepsilon(\boldsymbol{k}',t) \partial_{k'_x} U_\varepsilon(\boldsymbol{k}',t) $ is used in the last step.
	Thus, Eq.~(\ref{supp:eqmxwmx}) reduces to
	\begin{align}
	M_x W_{x,\boldsymbol{k}} M_x^\dagger = W_{x,-k_x,k_y}^\dagger,
	\end{align}
	which gives
	\begin{align}\label{eqsupp:mxbranches1}
	&
	\sum_{n}[W_{x,\boldsymbol{k}}(t)]_{mn} [\nu_{x,\mu}(\boldsymbol{k},t)]_n = e^{2\pi i \nu_{x,\mu}(k_y,t)} [\nu_{x,\mu}(\boldsymbol{k},t) ]_m\\ \nonumber
	&
	\sum_{nn'} [W_{x,\boldsymbol{k}}(t)]_{mn} \left( [M_x^\dagger]_{nn'} [
	\nu_{x,\mu}(-k_x,k_y,t)
	]_ {n'} \right)
	= \sum_{n'} [M_x^\dagger W_{x,-k_x,k_y}^\dagger(t)]_{mn'} [\nu_{x,\mu}(-k_x,k_y,t)]_{n'} 
	\\\label{eqsupp:mxbranches2}
	& \qquad\qquad\qquad\qquad\qquad\qquad\qquad\qquad\,\,\,
	= e^{-2\pi i \nu_{x,\mu}(k_y,t)}  \left( \sum_{n'} [M_x^\dagger]_{mn'} [\nu_{x,\mu}(-k_x,k_y,t)]_{n'}\right).
	\end{align}
	Therefore, the first-order dynamical polarization at any instance $ t $ obey the constraints
	\begin{align}\label{supp:mxnu}
	\boxed{
		\hat{M}_x\Rightarrow \nu_{x,\mu}(k_y,t) :\qquad \text{ appear in pairs } \pm \nu_{x,\mu}(k_y,t) \text{ mod 1}.}
	\end{align}
	On the other hand, the symmetry $ \hat{M}_y $ gives $ M_y U_\varepsilon(\boldsymbol{k},t) M_y^\dagger = U_\varepsilon(k_x,-k_y,t) $, which does not change the path integral along $ x $ but only changes the base point where the integration starts. Thus,
	$ M_y W_{x,\boldsymbol{k}}(t) M_y^\dagger = W_{x,k_x,-k_y}(t)$,
	implying that
	\begin{align}\label{eqsupp:mybranches1}
	& \sum_{n} [W_{x,\boldsymbol{k}}(t)]_{mn} [\nu_{x,\mu}(\boldsymbol{k},t)]_n = e^{2\pi i \nu_{x,\mu}(k_y,t)} [\nu_{x,\mu}(\boldsymbol{k},t)]_m
	\\ \label{eqsupp:mybranches2}
	&
	\sum_{nn'} [W_{x,\boldsymbol{k}}(t)]_{mn} \left( [M_y^\dagger]_{nn'} [\nu_{x,\mu}(k_x,-k_y,t) ]_{n'} \right) = e^{2\pi i \nu_{x,\mu}(-k_y)}  \left( \sum_{n'} M_y^\dagger [\nu_{x,\mu}(k_x,-k_y,t) ]_{n'} \right) .
	\end{align}
	That means the constraint
	\begin{align}\label{supp:mynu}
	\boxed{
		\hat{M}_y\Rightarrow\nu_{x,\mu}(k_y,t): \qquad
		\text{appear in pairs } \nu_{x,\mu_1}(k_y,t) = \nu_{x,\mu_2}(-k_y,t), \qquad
		\quad
		\text{or satisfy}\quad 
		\nu_{x,\mu}(k_y,t) = \nu_{x,\mu}(-k_y,t),
	}
	\end{align}
	
	Now consider quadrupoles. Similar to $ Q_{x,\boldsymbol{k}} $ discussed above, the constraints for bare polarization along $ y $ are
	\begin{align}
	M_x Q_{y,\boldsymbol{k}} M_x^\dagger = Q_{y,-k_x,k_y},&&
	M_y Q_{y,\boldsymbol{k}} M_y^\dagger = Q_{y,k_x,-k_y}^\dagger.
	\end{align}
	Compared with the first order situation, $ Q^{(\nu_x)}_{y,\boldsymbol{k}}(t), W^{(\nu_x)}_{y,\boldsymbol{k}} (t) $ also involve the symmetry action on certain first order Wilson loop eigenfunctions $ [\nu_{x,\mu}(\boldsymbol{k})]_m $. Denote
	\begin{align}\label{supp:eqmxykt}
	[M_{x}(\boldsymbol{k},t)]_{\mu_1\mu_2} \equiv \sum_{mn}  [\nu_{x,\mu_1}(-k_x,k_y,t)]_m^* [M_{x}]_{mn} [\nu_{x,\mu_2}(\boldsymbol{k},t)]_n, 
	&&
	[M_{y}(\boldsymbol{k},t)]_{\mu_1\mu_2} \equiv \sum_{mn} [\nu_{x,\mu_1}(k_x,-k_y,t)]_m^* [M_{y}]_{mn} [\nu_{x,\mu_2}(\boldsymbol{k},t)]_n.
	\end{align}
	For $ M_x(\boldsymbol{k},t) $, Eqs.~(\ref{eqsupp:mxbranches1}) and (\ref{eqsupp:mxbranches2}) indicates that the non-vanishing elements $ [M_{x}(\boldsymbol{k},t)]_{\mu_1\mu_3}\ne 0 $ require the eigenvalues $ \nu_{x,\mu_1}(k_y,t) = - \nu_{x,\mu_3}(k_y,t) $, namely, $ \mu_1,\mu_3 $ in different sets $ \pm \nu_x $. 
	Let $ \mu_1,\mu_2 \in + \nu_x $, then
	\begin{align}\nonumber
	[Q_{y,\boldsymbol{k}}^{(+ \nu_x)}(t)]_{\mu_1\mu_2} &= \sum_{mn} [\nu_{x,\mu_1}(\boldsymbol{k}+\Delta_y\boldsymbol{e}_y,t) ]^*_{m} [M_x^\dagger  M_x Q_{y,\boldsymbol{k}}(t) M_x^\dagger M_x]_{mn} [\nu_{x,\mu_2}(\boldsymbol{k},t) ]_n
	\\
	\nonumber
	&= 
	\sum_{mn} [\nu_{x,\mu_1} (\boldsymbol{k}+\Delta_y\boldsymbol{e}_y,t)]_m^*  [M_x^\dagger  \mathbb{I} Q_{y,-k_x,k_y}(t) \mathbb{I} M_x ]_{mn} [\nu_{x,\mu_2}(\boldsymbol{k},t) ]_n 
	\\
	&=
	\sum_{\mu_3\mu_4\in - \nu_x} [M_x^\dagger(\boldsymbol{k},t)]_{\mu_1\mu_3} [Q^{(-\nu_x)}_{y,-k_x,k_y}(t)]_{\mu_3\mu_4} [M_x(\boldsymbol{k},t)]_{\mu_4\mu_2},
	\\
	\Rightarrow \quad 
	[W^{(+\nu_x)}_{y,\boldsymbol{k}}(t)]_{\mu_1\mu_2} &= \sum_{\mu_3\mu_4\in -\nu_x} [M_x^\dagger(\boldsymbol{k},t)]_{\mu_1\mu_3} [W^{(-\nu_x)}_{y,-k_x,k_y}(t)]_{\mu_3\mu_4} [M_x(\boldsymbol{k},t)]_{\mu_4\mu_2}.
	\end{align}
	Here we have used the completeness relation $ \delta_{mn} = \sum_{\mu} [\nu_{x,\mu}(\boldsymbol{k},t)]_m [\nu_{x,\mu}(\boldsymbol{k},t)]_n^* $. Then, for second-order Wilson loops 
	\begin{align}
	& \sum_{\mu_2\in +\nu_x} [W_{y,\boldsymbol{k}}^{(+\nu_x)} (t)]_{\mu_1\mu_2} [\nu_{y,\tilde{\mu}}^{(+\nu_x)}(\boldsymbol{k},t)]_{\mu_2} = e^{2\pi i \nu_{y,\tilde{\mu}}^{(+\nu_x)}(k_x,t)} [\nu_{y,\tilde{\mu}}^{(+\nu_x)}(\boldsymbol{k},t)]_{\mu_1}, && ( \mu_1\in+\nu_x ),\\
	&
	\sum_{\mu_2\in -\nu_x, \mu_3\in +\nu_x} [W^{(-\nu_x)}_{y,-k_x,k_y}(t)]_{\mu_1\mu_2} \left( [M_x(\boldsymbol{k},t)]_{\mu_2\mu_3} [\nu_{y,\tilde\mu}^{(+\nu_x)}(\boldsymbol{k},t) ]_{\mu_3} \right) =  
	e^{2\pi i \nu_{y,\tilde{\mu}}^{(+\nu_x)}(k_x,t)}
	\left( \sum_{\mu_3\in +\nu_x } [M_x(\boldsymbol{k},t)]_{\mu_1\mu_3} [\nu_{y,\mu}^{(+\nu_x)}(\boldsymbol{k})) ]_{\mu_3} \right),
	&&
	( \mu_1\in -\nu_x )
	\end{align}
	That means the two different first-order branch sets $ \pm \nu_x $ have the same quadrupole polarization along $ y $,
	\begin{align}
	\label{supp:mxquad}
	\boxed{
		\hat{M}_x\Rightarrow \nu_{y,\tilde{\mu}}^{(+\nu_x)}(k_x,t): \qquad \nu_{y,\tilde\mu}^{(+\nu_x)}(k_x,t) = \nu_{y,\tilde\mu}^{(-\nu_x)}(k_x,t) \text{ mod 1}
	}
	\end{align} 
	For $ M_y(\boldsymbol{k},t) $, Eqs.~(\ref{eqsupp:mybranches1}) and (\ref{eqsupp:mybranches2}) forces that the non-vanishing elements $ [M_y(\boldsymbol{k},t)]_{\mu_1\mu_2} $ only connect the first-order branches $ \nu_{x,\mu_1}(k_y,t) = \nu_{x,\mu_2}(-k_y,t) $ in the same branch set, i.e. $ \mu_1,\mu_2\in +\nu_x $,
	\begin{align}\nonumber
	[Q^{{(+\nu_x)}}_{y,\boldsymbol{k}}(t)]_{\mu_1\mu_2} &= 
	\sum_{mn} [\nu_{x,\mu_1}(\boldsymbol{k}+\Delta_y\boldsymbol{e}_y,t) ]_m^* [M_y^\dagger  M_y Q_{y,\boldsymbol{k}} (t) M_y^\dagger M_y]_{mn} [ \nu_{x,\mu_2}(\boldsymbol{k},t) ]_n\\ \nonumber
	&= 
	\sum_{mn} [ \nu_{x,\mu_1}(\boldsymbol{k}+\Delta_y\boldsymbol{e}_y,t) ]_m^* [M_y^\dagger \mathbb{I}  Q^\dagger_{y,k_x,-k_y} (t) \mathbb{I} M_y]_{mn} [ \nu_{x,\mu_2}(\boldsymbol{k},t) ]_n\\ \nonumber
	&=
	\sum_{\mu_3\mu_4\in +\nu_x} [M_y^\dagger(\boldsymbol{k},t)]_{\mu_1\mu_3} \left[\left(Q^{(+\nu_x)}_{y,k_x,-k_y} (t) \right)^\dagger \right]_{\mu_3\mu_4} [M_y(\boldsymbol{k},t)]_{\mu_4\mu_2},
	\\
	\Rightarrow \quad 
	[W^{(+\nu_x)}_{y,\boldsymbol{k}}(t)]_{\mu_1\mu_2} &=
	\sum_{\mu_3\mu_4\in +\nu_{x}} [M_y^\dagger (\boldsymbol{k},t)]_{\mu_1\mu_3}  \left[\left( W^{(+\nu_x)}_{y,k_x,-k_y}(t) \right)^\dagger \right]_{\mu_3\mu_4} [M_y (\boldsymbol{k},t)]_{\mu_4\mu_2}
	\end{align}
	Thus, for $ \mu_1\in+\nu_x $,
	\begin{align}\label{tempmy1}
	& \sum_{\mu_2\in+\nu_x} [W_{y,\boldsymbol{k}}^{(+\nu_x)}(t)]_{\mu_1\mu_2} [\nu_{y,\tilde{\mu}}^{(+\nu_x)}(\boldsymbol{k},t)]_{\mu_2} = e^{2\pi i \nu_{y,\tilde{\mu}}^{(+\nu_x)}(k_x,t)} [\nu_{y,\tilde{\mu}}^{(+\nu_x)}(\boldsymbol{k},t)]_{\mu_1}\\ \label{tempmy2}
	&
	\sum_{\mu_2\mu_3\in +\nu_x} [W^{(+\nu_x)}_{y,k_x,-k_y}(t)]_{\mu_1\mu_2} \left( [M_y(\boldsymbol{k},t)]_{\mu_2\mu_3} [\nu_{y,\tilde\mu}^{(+\nu_x)}(\boldsymbol{k},t)]_{\mu_3} \right) =  
	e^{-2\pi i \nu_{y,\tilde{\mu}}^{(+\nu_x)}(k_x,t)}
	\left( \sum_{\mu_3\in+\nu_x} [M_y(\boldsymbol{k},t)]_{\mu_1\mu_3} [\nu_{y,\mu}^{(+\nu_x)}(\boldsymbol{k},t)]_{\mu_3}  \right).
	\end{align}
	Thus, within each first-order branch set, i.e. $ +\nu_x $, the second-order branches
	\begin{align}
	\label{supp:myquad}
	\boxed{
		\hat{M}_y\Rightarrow \nu_{y,\tilde{\mu}}^{(+\nu_x)} (k_x,t) :
		\qquad
		\text{ appear in pairs } \pm \nu_{y,\tilde{\mu}}^{(+\nu_x)} (k_x,t) \text{  mod } 1.
	}
	\end{align}
	
	Eqs.~(\ref{supp:mxnu}), (\ref{supp:mynu}), (\ref{supp:mxquad}) and (\ref{supp:myquad}) are the general constraints on first/second order dynamical polarization imposed by mirror symmetries. Next, we apply these constraints to our model in the main text.
	
	In our model, the sublattice transformations are $ M_x: 1\leftrightarrow 3, 2\leftrightarrow 4 $, and $ M_y: 1\leftrightarrow 4, 2\leftrightarrow3  $. The phase factors in Eq.~(\ref{supp:tempmx}) and (\ref{supp:tempxy}) are $ \phi^x_{m=1,3}=0, \phi^x_{m=2,4}=\pi, \phi^y_m=0 $, giving rise to
	\begin{align}\label{eqsupp:mxmymodel}
	M_x = \tau_1 \sigma_3, && M_y = \tau_1\sigma_1, &&
	M_x^2=M_y^2=1 && M_xM_y=-M_yM_x.
	\end{align}
	Let us emphasize again that 4 sublattices render 4 first-order {\em dynamical} branches instead of 2 Wannier bands in the static case. Note that at fixed points $ t=0 $ and $ T $, the periodized evolution operator $ \hat{U}_\varepsilon(t) = \hat{\mathbb{I}} $ is trivially the identity by construction; thus, $ \hat{x}_{\text{mean}}(t) \xrightarrow{t=0,T} \hat{x} = \sum_{im} \hat{c}_{im}^\dagger |0\rangle e^{-i\Delta_x x_i} \langle 0| \hat{c}_{im} $ and $ \nu_{x,\mu} = 0 $ by definition of $ \hat{x}_{\text{mean}}(t) $ in Eq.~(\ref{supp:eqxb}). That means at $ t=0,T $, the first order polarization vanishes as $ 0 $ mod $ 1 $, which is simply the initial condition by construction.  Then, during $ t\in(0,T) $, constraints (\ref{supp:mxnu}) and (\ref{supp:mynu}) force the 4 branches to separate into two sets, denoted $ \nu_{x,\mu=1,2}(k_y,t) \in +\nu_x $ and $ \nu_{x,\mu=3,4}(k_y,t) \in -\nu_x $, where $ \nu_{x,\mu=1,2} = - \nu_{x,\mu=3,4} $ mod 1. Within each set, the branches satisfy $ \nu_{x,\mu_1}(-k_y,t) = - \nu_{x,\mu_2}(k_y,t) $, where $ \mu_1,\mu_2 $ in the same set. (In our specific model, these two branches are degenerate.) For branches in different sets, if they ever touch each other, Eq.~(\ref{supp:mxnu}) means it must occur at $ \nu_{x,\mu_1}(k_y,t) = \nu_{x,\mu_3}(k_y,t)=0 $ or $ 1/2 $ for $ \mu_1,\mu_3 $ in different sets. Then, there are 3 possible situations for the 2 sets of branches $ \nu_{x,\mu}(k_y,t) \in \pm\nu_x $:
	\begin{enumerate}
		\item They touch 0 during $ t\in(0,T) $ (in addition to $ t=0, T $). In this case, these branches are non-separable and in an anomaly-free system where all quadrupoles sum up to zero (no ``center-of-mass" motion for the whole lattice), the system has zero dynamical quadrupole. 
		\item
		They touch $ 1/2 $ during $ t\in(0,T) $. Then it immediately means the existence of dynamical dipoles as in the first-order anomalous Floquet insulators studied in previous literature.
		\item
		They do not touch either $ 0 $ or $ 1/2 $ throughout $ t\in(0,T) $ and for all $ k_y $. This is the regime where separable first-order branches can be defined for computing the second-order dynamical polarization.
	\end{enumerate}
	
	Our model in the main text corresponds to the third situation. The first order branches $ \nu_{x,\mu}(k_y,t) $ at a single instant $ t $ exhibit similar structures as the static ``gapped" Wannier bands, except for the crucial difference of 4 dynamical branches versus 2 Wannier bands due to absence/presence of band projections respectively. Such a difference is indispensable for non-trivial dynamical quadrupoles, as we show later.
	
	The second-order situation is dramatically different. Most obviously, Eqs.~(\ref{supp:mxquad}) and (\ref{supp:myquad}) do {\em not} give the quantization of $ \nu_{y,\tilde{\mu}}^{(\nu_x)}(k_x,t) $ at any $ t $, unlike in the static case. Indeed, as $ \nu_{y,\tilde{\mu}}^{(\nu_x)}(k_x,t=0,T) = 0 $ (mod 1) by construction ($ \hat{U}_\varepsilon(0 \text{ or }T)=\hat{\mathbb{I}} $), a quantization of $ \nu_{y,\tilde{\mu}}^{(\nu_x)}(k_x,t) $ would immediately imply a trivial quadrupole at all time $ \nu_{y,\tilde{\mu}}^{(\nu_x)}(k_x,t) = 0 $ by continuity. Mathematically, this is highlighted by that our dynamical nested Wilson loops are SU(2) matrices, which allows for a continuous evolution of second-order dynamical branches $ \pm \nu_{y,\tilde{\mu}}^{(\nu_x)}(k_x,t) $ in pairs by Eq.~(\ref{supp:myquad}). In contrast, static nested Wilson loops are U(1) numbers~\cite{Benalcazar2017}, which give rise to its quantization via a relation similar to Eq.~(\ref{supp:myquad}) (Sec. 6.2.2 in the supplemental materials of~\cite{Benalcazar2017}). That invites a new examination of the physical picture implied by the constrained quadrupolar motion enforced by Eq.~(\ref{supp:mxquad}) and (\ref{supp:myquad}). 
	
	First, Eq.~(\ref{supp:mxquad}) means the two first-order branch sets $ \pm\nu_x $ for our model have the same quadrupolar motions. This is natural: due to $ \hat{M}_x: x\rightarrow -x $, if there is a branch moving along $ \boldsymbol{e}_x+\boldsymbol{e}_y $ direction ($ +\nu_x $ set, with positive second order polarization along $ y $), there must be another branch moving along $ -\boldsymbol{e}_x+\boldsymbol{e}_y $ ($ -\nu_x $ set, with positive second order polarization along $ y $). Therefore, it is sufficient to check only one first-order branch set, i.e. $ +\nu_x $ as in the main text. 
	
	Second, Eq.~(\ref{supp:myquad}) for our SU(2) dynamical nested Wilson loop only enforces the pairwise $ \pm\nu_{y,\tilde{\mu}}^{(+\nu_x)} (k_x,t) $, instead of any directly quantized value. However, the additional dimension, time $ t $, allows for a meaningful definition of quantized quadrupolar motions. Note at fixed points $ t=0, T $, where $ U_\varepsilon(0/T) = \mathbb{I} $, $ \nu_{y,\tilde{\mu}}^{(+\nu_x)} = 0 $ mod 1 by construction. Thus, particles are undergoing ``round-trips" during $ t\in[0,T] $. There are two types of evolutions for $ \nu_{y,\tilde{\mu}}^{(+\nu_x)} (k_x,t) $. One is to start at $ \nu_{y,\tilde{\mu}}^{(+\nu_x)} (k_x,t=0) = 0 $ and winds back to $ \nu_{y,\tilde{\mu}}^{(+\nu_x)} (k_x,t=T)=0 $. Such a motion corresponds to a topologically trivial case as when a boundary is imposed, the motion can be smoothly distorted to zero throughout $ t\in[0,T] $. However, another type of round trip is that $ \nu_{y,\tilde{\mu}}^{(+\nu_x)} (k_x,t\rightarrow 0) \rightarrow 0 $ but end up at $ \nu_{y,\tilde{\mu}}^{(+\nu_x)} (k_x,t\rightarrow T) \rightarrow 1 $. Such motions cannot be distorted to zero as $ t=0,T $ are fixed points with quantized $ \nu_{y,\tilde{\mu}}^{(+\nu_x)} (k_x,t)=0 $ mod 1, which cannot be changed smoothly. Then by connecting $ \nu_{y,\tilde{\mu}}^{(+\nu_x)} (k_x,t\rightarrow 0)=0 $ and $ \nu_{y,\tilde{\mu}}^{(+\nu_x)} (k_x,t\rightarrow T)=1 $, the particle must go through $ 0\rightarrow 1 $ during $ t\in[0,T] $. That implies the quantized dynamical quadrupole 
	\begin{align}
	\tilde{P}^{+(\nu_x)}_{xy,\tilde{\mu}} = \int_0^T dt \left( \partial_t  \langle \nu^{(+\nu_x)}_{y,\tilde{\mu}} \rangle  (t)  \right), &&
	\langle \nu^{(\nu_x)}_{y,\tilde{\mu}} \rangle  (t) =\int_{-\pi}^\pi \frac{dk_x}{2\pi} \nu^{(+\nu_x)}_{y,\tilde{\mu}} (k_x,t) .
	\end{align}
	From Eq.~(\ref{supp:myquad}), there will be another quadrupolar motion $ -\langle \nu_{y,\tilde{\mu}}^{(+\nu_x)}\rangle (t) $ starting at $ 1 $ and end up at $ 0 $ as $ t:0\rightarrow T $; $ \pm \langle \nu_{y,\tilde{\mu}}^{(+\nu_x)} \rangle (t) $ must meet at $ 1/2 $ for some $ t $. Such a crossing is robust against perturbations when mirror symmetry is preserved. However, when mirror symmetry $ \hat{M}_y $ is present, the two motions branches $ [\nu_{y,\tilde{\mu}}^{(+\nu_x)}(\boldsymbol{k},t)]_\mu $ and $ \sum_{\mu_2} [M_y(\boldsymbol{k},t)]_{\mu\mu_2} [\nu_{y,\tilde{\mu}}^{(+\nu_x)}(\boldsymbol{k},t)]_{\mu_2} $, with eigenvalues $ \pm\nu_{y,\tilde{\mu}}^{(+\nu_x)} $ in Eqs.~(\ref{tempmy1}) and (\ref{tempmy2}) are related by $ \hat{M}_y $, corresponding to the anomalous corner states at different $ \boldsymbol{e}_x\pm\boldsymbol{e}_y $ corners similarly related by $ \hat{M}_y $. Thus, unless a perturbation is strong enough to bridge the two corner states separated by $ L_y $, such a ``crossing" at $ 1/2 $ cannot be broken. Mathematically, that means a perturbing Hamiltonian $ \hat{H}' $ must have long-range elements connecting
	\begin{align}
	\langle b_{y,\tilde{\mu}=1}^{(\nu_x)} (1,k_x,t) | \hat{H}' \hat{M}_y | b_{y,\tilde{\mu}=1}^{(\nu_x)} (1,k_x,t)\rangle = 
	\langle b_{y,\tilde{\mu}=1}^{(\nu_x)} (1,k_x,t) | \hat{H}' | b_{y,\tilde{\mu}=2}^{(\nu_x)} (L_y,k_x,t)\rangle
	\end{align}
	for the second order branch eigenstates in Eq.~(\ref{supp:eqb2yt}).
	In terms of physical process, the two branches carrying particles counter-propagate along $ \pm y $ respectively. When $ \pm \nu_{y,\tilde{\mu}}^{(+\nu_x)} (k_x,t)=1/2 $ mod 1, they both move by 1 unit cell (recall that the physical meaning of $ \nu_{y,\tilde{\mu}}^{(+\nu_x)} (k_x,t) $ is relative motion from 0 to $ t $ by $ 2\nu_{y,\tilde{\mu}}^{(+\nu_x)} (k_x,t) $). The indistinguishability of particles then means the net effect of polarization is to move a particle from $ y_i=0 $ (or $ L_y $) all the way across sample to $ y_i=L_y $ (or $ 0 $), resulting in the corner states in open boundary systems. Note that this is only possible if $ 1/2 $ is reached, when i.e. particles in $ y_i $ and $ y_i+2 $ collide into the same unit cell $ y_i+1 $, which allows for gauge transformations exchanging branch labels due to indistinguishability of particles. If $ 1/2 $ is not reached, particles in two branches remain distinguishable by spatial locations and must go back to their original location at $ t=T $, corresponding to the topologically trivial case.


	\subsection{Extensions for the main text model: only $ M_x, M_y $ matters, not $ T, C, S $}

	In previous analysis, we only invoked the two mirror reflection symmetries. But the model in Eq.~(1) of the main text contains several unnecessary accidental symmetries, such as time reversal $ \Theta=TK $, particle-hole $ \Gamma=CK $, chiral $ S=\Theta\Gamma $, and four-fold rotation symmetries. Here $ T=\tau_0\sigma_0, C=\tau_3\sigma_0, S=\tau_3\sigma_0 $ are unitary matrices, $ K $ is complex conjugation, $ Ki=-iK $. The Hamiltonian in momentum space satisfying those symmetries obeys
	\begin{align}
	TH(\boldsymbol{k},t)T^{-1} = H^*(-\boldsymbol{k},-t), &&
	CH(\boldsymbol{k},t)C^{-1} = -H^*(-\boldsymbol{k},t), && 
	SH(\boldsymbol{k},t)S^{-1} = -H(\boldsymbol{k},-t).
	\end{align}
	For extended model breaking these symmetries, we still keep the driving of the form in Eq.~(1) of main text, such that $ H(\boldsymbol{k},-t) = H(\boldsymbol{k},t) $. All possible additional terms that preserve $ M_x, M_y $ up to nearest unit cells are listed in Table~\ref{supp:tabterms}, together with whether they satisfy $ T, C, S $ symmetries respectively.

	\begin{table}[h]
		\caption{\label{supp:tabterms} Possible terms up to nearest unit cell preserving $ M_x, M_y $. Terms adopted in the main text are highlighted with bold text.}
		\begin{tabular}{l|l|ccc}
			& Satisfy $ M_x, M_y $ & $ T $? & $ C $? & $ S $?\\
			$ \tau_0\sigma_0 $ & $ \cos k_x, \cos k_y, 1 $ & $\checkmark$ & $ \times $ & $ \times $\\
			$ \tau_0\sigma_1 $ & $ \sin k_x $ & $ \times $ & $\checkmark$ & $ \times $ \\
			$ \tau_0\sigma_3 $ & $ \sin k_y $ & $ \times $ & $ \checkmark $  & $ \times $ \\
			$ \tau_1\sigma_0 $ & $ \boldsymbol{\cos k_x}, \cos k_y, \boldsymbol{1} $ & $ \checkmark $ & $ \checkmark $ & $ \checkmark $ \\
			$ \tau_1\sigma_1 $ & $ \sin k_x $ & $ \times $ & $ \times $ & $ \checkmark $ \\
			$ \tau_1\sigma_3 $ & $ \sin k_y $ & $ \times $ & $ \times $ & $ \checkmark $ \\
		\end{tabular}
		\qquad\qquad\qquad
		\begin{tabular}{l|l|ccc}
			& Satisfy $ M_x, M_y $ & $ T $? & $ C $? & $ S $?\\
			$ \tau_2\sigma_1 $ & $ \boldsymbol{\sin k_y} $ & $ \checkmark $ & $ \checkmark $ & $ \checkmark $ \\
			$ \tau_2\sigma_2 $ & $ \cos k_x, \boldsymbol{\cos k_y}, \boldsymbol{1} $  & $ \checkmark $ & $ \checkmark $ & $ \checkmark $ \\
			$ \tau_2\sigma_3 $ & $ \boldsymbol{\sin k_x} $ & $ \checkmark $ & $ \checkmark $ & $ \checkmark $\\
			$ \tau_3\sigma_1 $ & $ \sin k_y $ & $ \times $ & $ \checkmark $ & $ \times $ \\
			$ \tau_3\sigma_2 $ & $ \cos k_x, \cos k_y, 1 $ & $ \times $ & $ \checkmark $ & $ \times $ \\
			$ \tau_3\sigma_3 $ & $ \sin k_x $ & $ \times $ & $ \checkmark $ & $ \times $ 
		\end{tabular}
	\end{table}

	\begin{figure}
		[h]
		\parbox{3cm}{
			\includegraphics[width=3cm]{model_t1}		\\ (a) $ h_{1\boldsymbol{k}}' $
		}
		\parbox{3.5cm}{
			\includegraphics[width=3cm]{model_t2}	\\ (b) $ h_{2\boldsymbol{k}}  $
		}
		\parbox{3.2cm}{\quad \\
			\includegraphics[width=3.2cm]{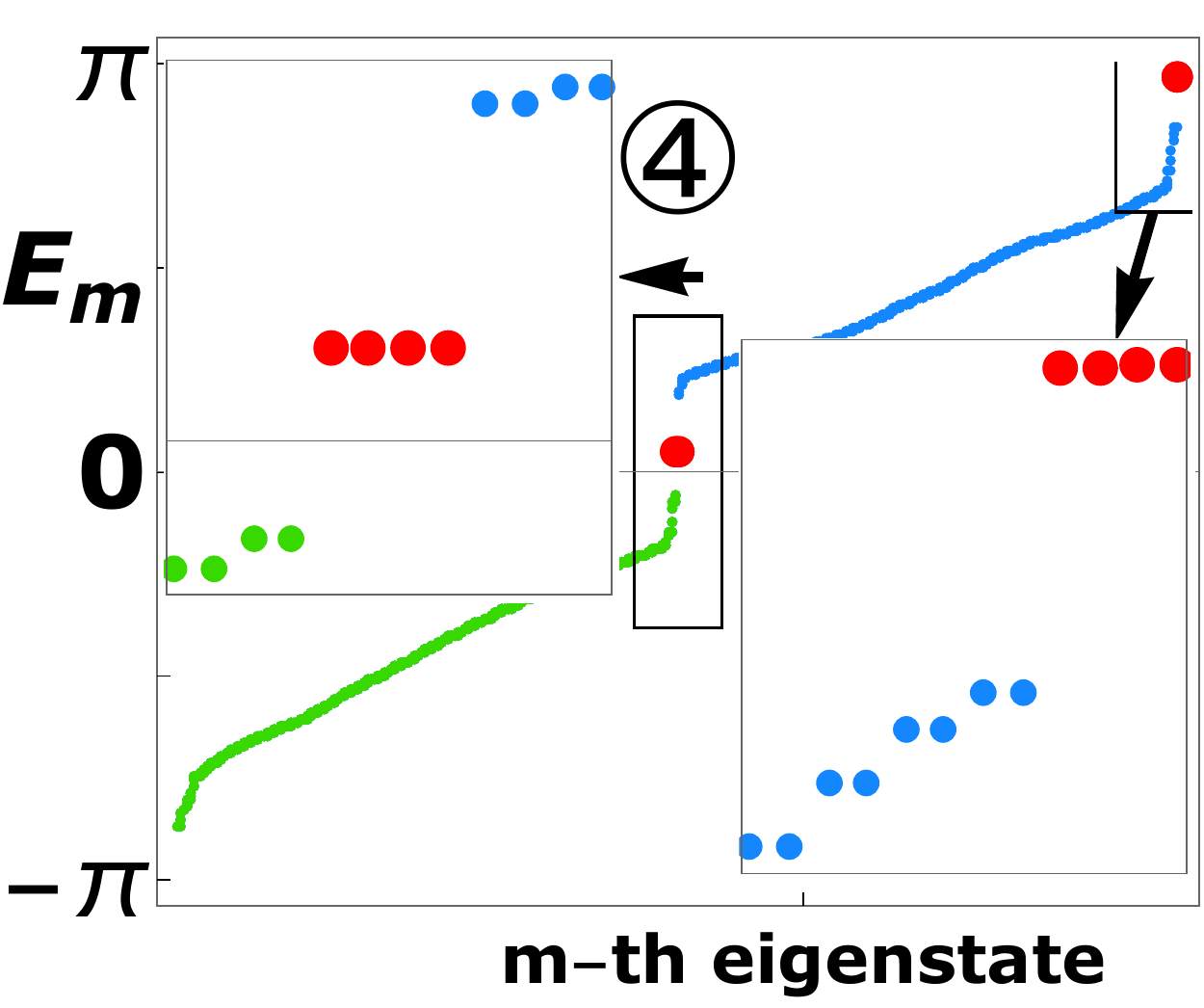} \\
			\quad \\ 
			(c)
		}
		\parbox{3.5cm}{		
			\includegraphics[width=2.5cm]{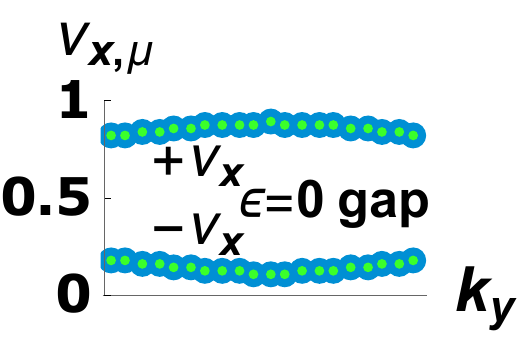}		
			\includegraphics[width=2.5cm]{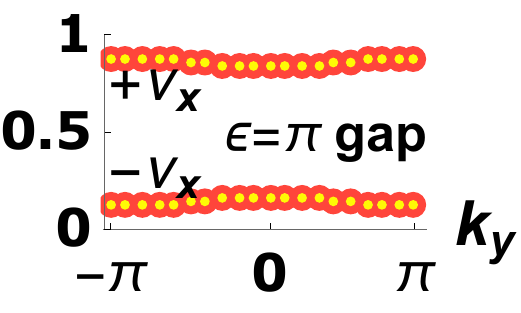}\\ (d)
		}
		\parbox{3.5cm}{
			\includegraphics[width=3cm]{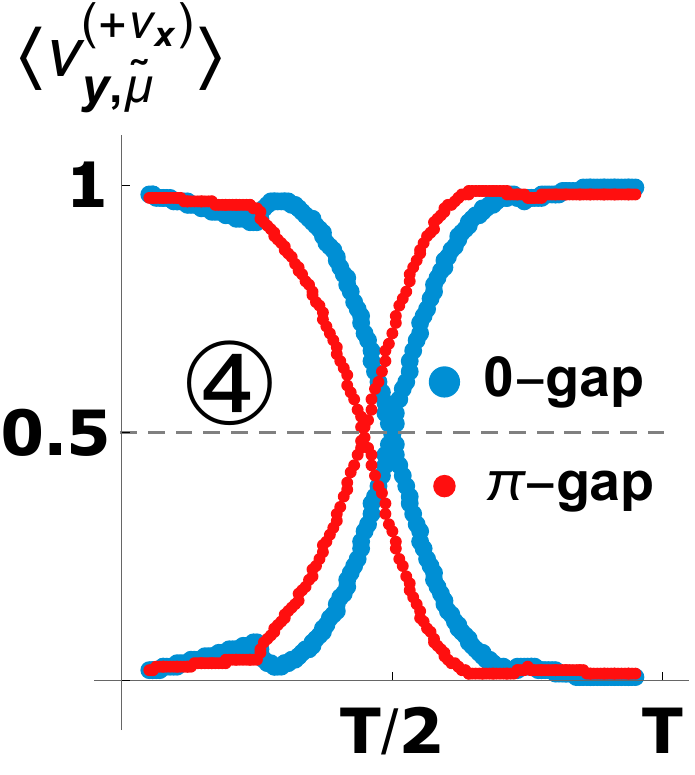} \\(e)
		}
		\caption{\label{supp:fig_addmodel} Extension of the model breaking time-reversal, chiral, and four-fold rotation symmetries. Here, only $ M_x, M_y $ are preserved while all other accidental symmetries in the main text are broken. The parameters are $ (\gamma,\lambda,\delta)=\frac{\pi}{\sqrt2} (0.5, 0.9, 0.4) \approx (1.11, 2.00, 0.89) $, where the additional dimensionless $ \delta\equiv\delta' T/2\hbar $ breaks the accidental symmetries. (a) (b) The driving Hamiltonians in real space (the term $ \tau_0\sigma_0 \cos k_y $ for homogeneous longer-range hopping among same sublattices is not drawn for clarity), where $ h_{1\boldsymbol{k}}' $ breaks all the symmetries. The direction of black arrows show the additional term with imaginary hopping $ +i $. (c) The open-boundary spectrum. (d) First order dynamical branches at $ t/T=0.5 $. (e) Quadrupolar motions. }
	\end{figure}
	Here, we show a specific example where all symmetries except for $ M_x, M_y $ are broken, so as to highlight that only mirror symmetries are needed in our analysis. This can be achieved by changing 
	\begin{align}
	\gamma' h_{1} \rightarrow  \gamma' h_{1\boldsymbol{k}}' = \gamma' h_1 + \delta' (\tau_3\sigma_1\sin k_y + \tau_3\sigma_2 \cos k_y -0.4 \tau_0\sigma_0\cos k_y )
	\end{align}
	in Eq.~(1) of the main text. Parameters here $ (\gamma',\lambda')=\frac{\pi}{\sqrt2}(0.5,0.9)\times (2\hbar/T) \approx (1.11,2.00) \times (2\hbar/T)   $, while the additional terms $ \propto \delta'=\frac{\pi}{\sqrt2} 0.4\times (2\hbar/T) \approx 0.89\times (2\hbar/T) $ break time-reversal, particle-hole, chiral, and obviously four-fold rotation symmetries $ k_y\rightarrow k_x $. From the results in Fig.~\ref{supp:fig_addmodel}, it is clear that the corner states are still characterized by dynamical polarization we introduced. Also, all the first and second order dynamical polarizations are fully characterized by the constraints Eqs.~(\ref{supp:mxnu}), (\ref{supp:mynu}), (\ref{supp:mxquad}) and (\ref{supp:myquad}). Finally, as a more generic feature than those in the main text, the ``crossing" point for the second order polarization no longer occur at $ t=T/2 $ in Fig.~\ref{supp:fig_addmodel} (e).
	
\begin{figure}
		[h]
		\parbox{8cm}{\includegraphics[width=3cm]{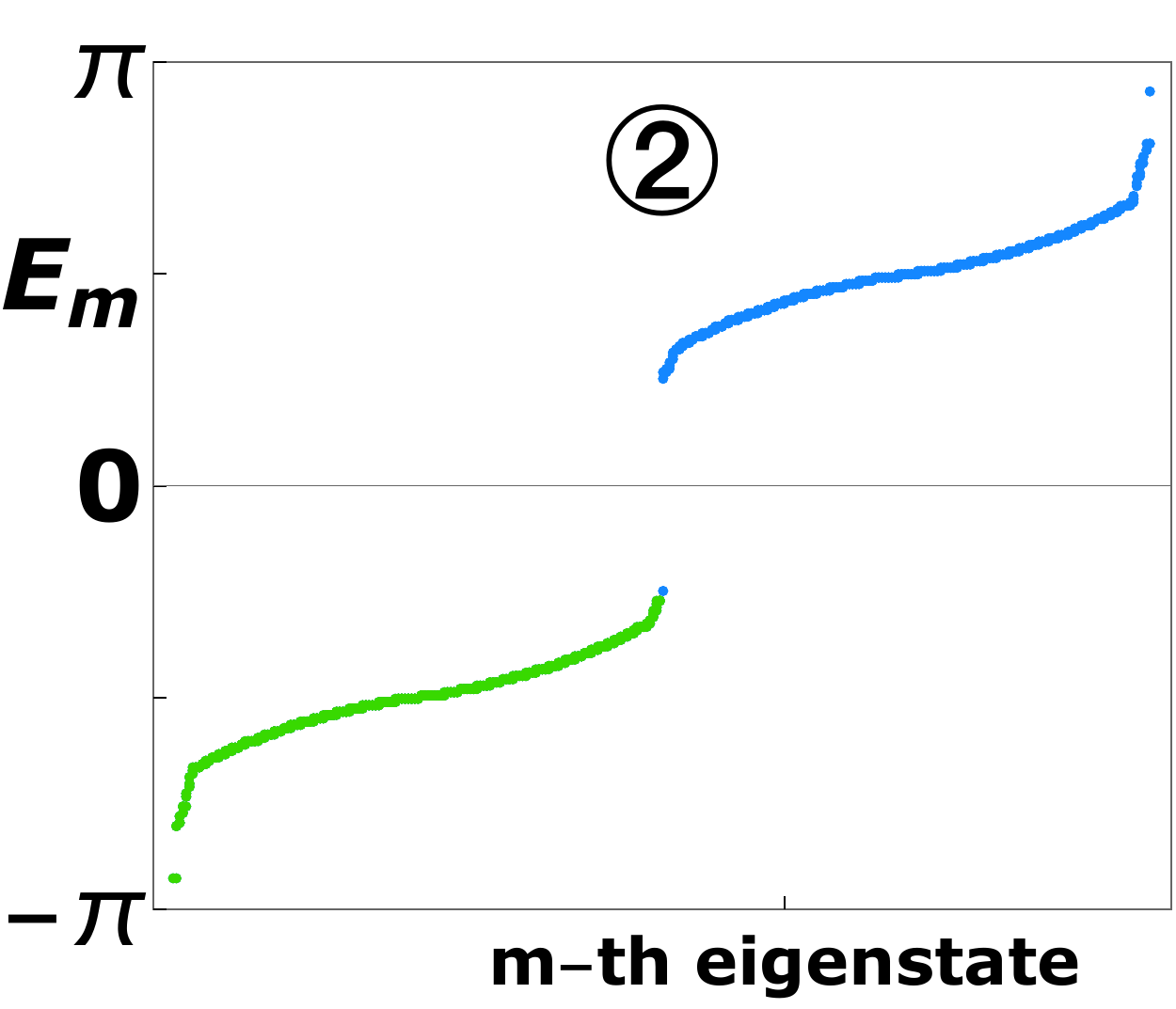}
			\includegraphics[width=3cm]{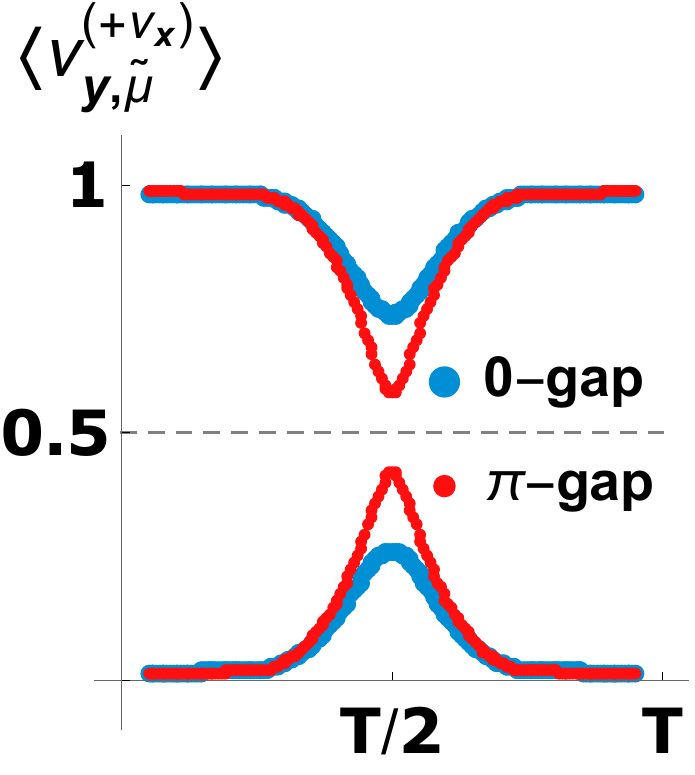} \\
			(a) Perturbing main text model
		} 
		\parbox{8cm}{\includegraphics[width=3cm]{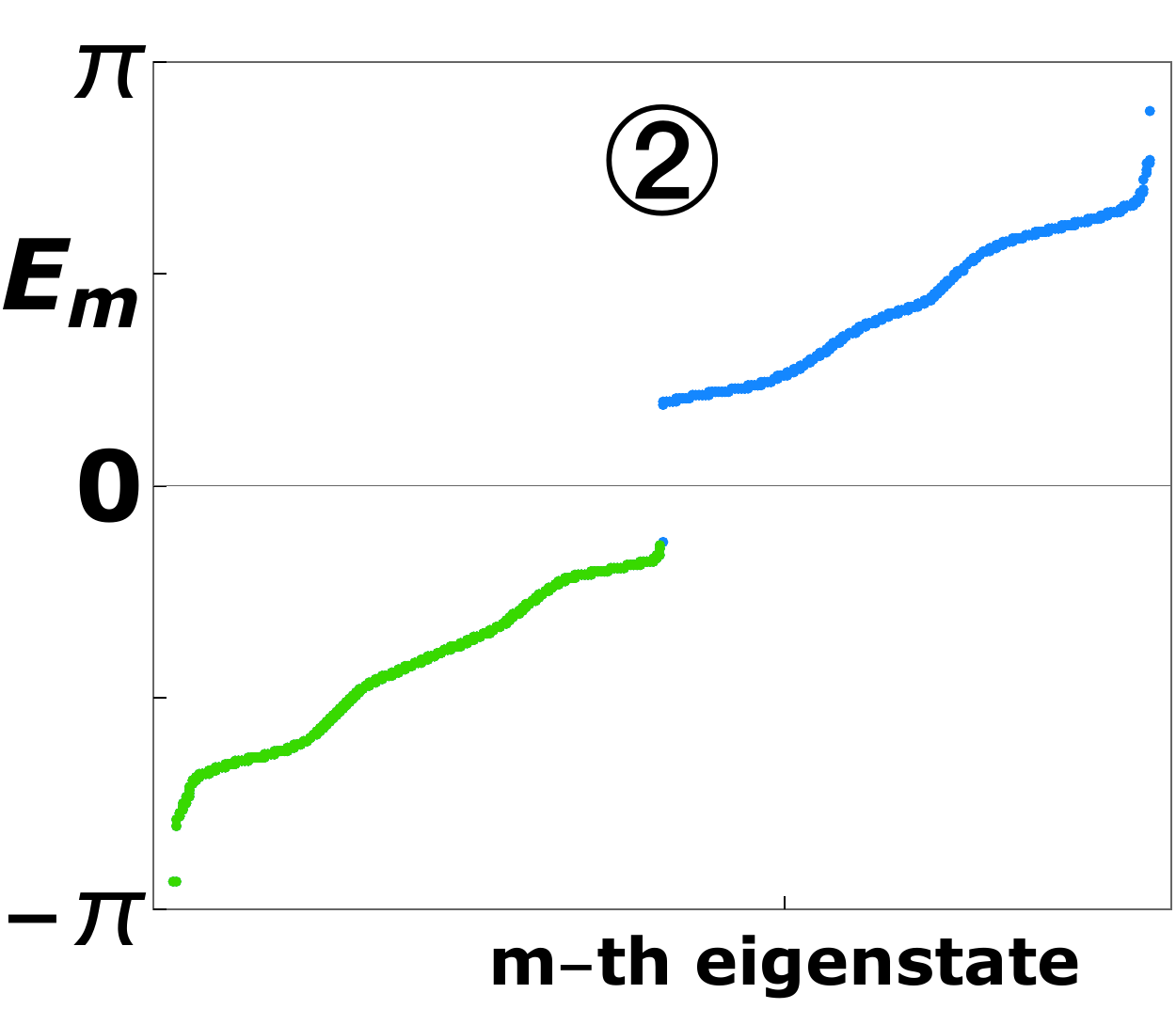}
			\includegraphics[width=3cm]{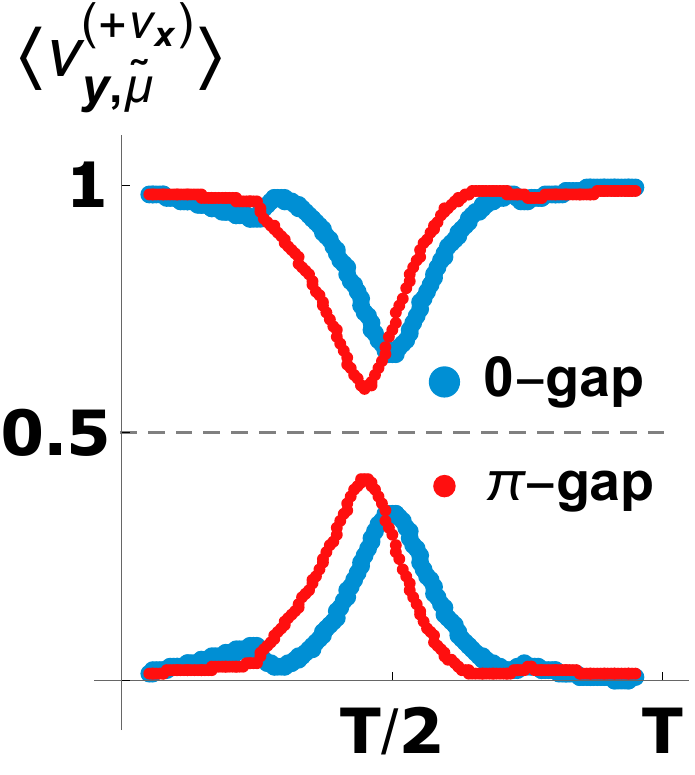}\\
			(b) Perturbing extended model in Fig.~\ref{supp:fig_addmodel}
		} 
		\caption{\label{fig:breakingmirror} When mirror symmetries are broken by a static chemical potential bias $ 0.3\times (2\hbar/T) \tau_3\sigma_0 $, all corner states are gapped out, for both (a) main text model and (b) the extended model in this section. Dynamical polarization also faithfully captures that the 0-gap corner states are gapped out more severely than those in $ \pi $-gap. }
	\end{figure}

	As a double-check for the importance of mirror symmetries, we further include a static chemical potential bias term breaking both mirror symmetries, $ 0.3\times (2\hbar/T) \tau_3\sigma_0 $. From Fig.~\ref{fig:breakingmirror}, we see that the corner states are immediately gapped out, as reflected by both the open-boundary spectrum and dynamical polarization.

	\section{Algebraic details for the solvable model in main text}	
	
	\subsection{Floquet operator for the model in main texts}
	First, we consider the bare evolution operator. Recall that the dimensionless $ (\gamma,\lambda) $ corresponds to $ \frac{T}{2\hbar} (\gamma, \lambda) $, and we would denote $ 2t/T $ simply as $ t $ in the following, so $ t\in[0,2] $. Such a convention is usually adopted in the study of binary drives such that time duration for $ h_1, h_{2\boldsymbol{k}} $ can be regarded as $ 1 $ respectively. When $ t \in[0,1/2] $:
	\begin{align}
	U(\boldsymbol{k},t) = \cos(\sqrt{2}\gamma t) - i\sin(\sqrt{2}\gamma t) \frac{h_1}{\sqrt2};
	\end{align}
	when $ t\in[1/2,3/2] $,
	\begin{align}
	U(\boldsymbol{k},t) &= \left(\cos(\sqrt2\lambda (t-1/2)) - i\sin(\sqrt{2}\lambda (t-1/2)) \frac{h_{2\boldsymbol{k}} }{\sqrt2} \right)  \left(\cos(\gamma/\sqrt2) - i\sin(\gamma/\sqrt2) \frac{h_1}{\sqrt2}\right);
	\end{align}
	and finally, when $ t\in[3/2,2] $,
	\begin{align}
	U(\boldsymbol{k},t) &= \left(\cos(\sqrt2\gamma (t-3/2)) - i\sin(\sqrt2\gamma (t-3/2) ) \frac{h_1}{\sqrt2}\right)
	\left(\cos(\sqrt2\lambda) - i \sin(\sqrt2\lambda) \frac{h_{2\boldsymbol{k}}}{\sqrt2} \right) \left(\cos(\gamma/\sqrt2) - i\sin(\gamma/\sqrt2) \frac{h_1}{\sqrt2}\right).
	\end{align}

	The Floquet operator is the evolution operator at the end of a full period:
	\begin{align}
	& U(\boldsymbol{k},T) = \left(\cos(\frac{\sqrt2}{2}\gamma) - i\sin(\frac{\sqrt2}{2}\gamma ) \frac{h_1}{\sqrt2}\right)
	\left(\cos(\sqrt2\lambda)-i \sin(\sqrt2\lambda) \frac{h_{2\boldsymbol{k}}}{\sqrt2}\right) \left(\cos(\frac{\sqrt2}{2}\gamma) - i\sin(\frac{\sqrt2}{2}\gamma ) \frac{h_1}{\sqrt2}\right).
	\end{align}
	Using the relations
	\begin{align}
	\nonumber
	&\frac{h_1h_{2\boldsymbol{k}} + h_{2\boldsymbol{k}}h_1}{2} = (\cos k_x + \cos k_y)\tau_0\sigma_0,\\
	&\frac{h_1h_{2\boldsymbol{k}}h_1}{2} = \cos k_y \tau_1\sigma_0 + \sin k_x \tau_2\sigma_3 -\cos k_x \tau_2\sigma_2 + \sin k_y \tau_2\sigma_1,
	\end{align}
	we have the analytical form for the Floquet operator as
	\begin{align}\label{supp:equf}
	U(\boldsymbol{k},T) = f_{\boldsymbol{k}} \mathbb{I} + i (g_{1\boldsymbol{k}}\Gamma_1 + g_{2\boldsymbol{k}}\Gamma_2 + g_{3\boldsymbol{k}}\Gamma_3 + g_{4\boldsymbol{k}}\Gamma_4),
	\end{align}
	where
	\begin{align}\nonumber
	f_{\boldsymbol{k}} &= \cos\sqrt2 \gamma \cos\sqrt2\lambda - \sin\sqrt2\gamma\sin\sqrt2\lambda \frac{\cos k_x + \cos k_y}{2} \\
	\nonumber
	g_{1\boldsymbol{k}} &= \frac{1}{\sqrt2}\sin\sqrt2\lambda  
	\sin k_y\\ \nonumber
	g_{2\boldsymbol{k}} &=
	\frac{1}{\sqrt2} \left(
	\sin\sqrt2\gamma \cos\sqrt2\lambda + \cos\sqrt2\gamma \sin\sqrt2\lambda \frac{\cos k_x + \cos k_y}{2} - \sin\sqrt2\lambda \frac{\cos k_x - \cos k_y}{2}
	\right)
	\\ \nonumber
	g_{3\boldsymbol{k}} &= \frac{1}{\sqrt2}\sin\sqrt2\lambda  
	\sin k_x
	\\ \label{fgggg}
	g_{4\boldsymbol{k}} &= - \frac{1}{\sqrt2} \left(
	\sin\sqrt2\gamma \cos\sqrt2\lambda + \cos\sqrt2\gamma \sin\sqrt2\lambda \frac{\cos k_x + \cos k_y}{2} + \sin\sqrt2\lambda \frac{\cos k_x - \cos k_y}{2}
	\right)
	\end{align}
	and $ \mathbb{I} = \tau_0 \sigma_0, \Gamma_{1,2,3} = \tau_2\sigma_{1,2,3}, \Gamma_4 = \tau_1\sigma_0 $ are Dirac matrices being anticommuting with each other $ \{\Gamma_i, \Gamma_j\} = 2\delta_{ij} $, and $ f_{\boldsymbol{k}}, g_{j\boldsymbol{k}} $'s are real numbers satisfying $ f^2_{\boldsymbol{k}} + \sum_{j=1}^4 g_{j\boldsymbol{k}}^2 = 1 $.  
	That means $ U_F $, and therefore its eigenstates, can be parameterized by three $ S^3 $ angles $ (\chi_{\boldsymbol{k}},\theta_{\boldsymbol{k}},\varphi_{\boldsymbol{k}}) $ in addition to the quasienergy $ E_{\boldsymbol{k}} $: $ f_{\boldsymbol{k}}=\cos E_{\boldsymbol{k}}, (g_{1\boldsymbol{k}}, g_{2\boldsymbol{k}}, g_{3\boldsymbol{k}}, g_{4\boldsymbol{k}}) = \sin E_{\boldsymbol{k}} (\sin\chi_{\boldsymbol{k}} \sin\theta_{\boldsymbol{k}}\cos\varphi_{\boldsymbol{k}}, \sin\chi_{\boldsymbol{k}}\sin\theta_{\boldsymbol{k}}\sin\varphi_{\boldsymbol{k}}, $ $\sin\chi_{\boldsymbol{k}}\cos\theta_{\boldsymbol{k}}, \cos\chi_{\boldsymbol{k}}) $. 
	The eigenvalues of $ U(\boldsymbol{k},T) $ can be obtained via $ (U(\boldsymbol{k},T)-f_{\boldsymbol{k}}\mathbb{I})^2 = i^2\sum_{j=1}^4 g_{j\boldsymbol{k}}^2 = i^2(1-f_{\boldsymbol{k}}^2) $, which gives in our case $ U(\boldsymbol{k},T) |E_{\boldsymbol{k}}\rangle = e^{iE_{\boldsymbol{k}}} |E_{\boldsymbol{k}}\rangle $,
	\begin{align}\nonumber
	\exp(iE_{\boldsymbol{k}\pm}) &= \exp(\pm iE_{\boldsymbol{k}}) = f_{\boldsymbol{k}} \pm i \sqrt{1-f^2_{\boldsymbol{k}}},\\ \label{defEk}
	f_{\boldsymbol{k}} = \cos E_{\boldsymbol{k}} &= \cos(\sqrt2\gamma) \cos(\sqrt2\lambda)  -\frac{\cos k_x + \cos k_y}{2} \sin(\sqrt2\gamma) \sin(\sqrt2 \lambda),
	\end{align}
	with each band two-fold degenerate.
	The gap closes at $ k_x, k_y = 0,\pi $ when
	\begin{align}
	f_{\boldsymbol{k}} = \pm1, \qquad \Rightarrow \qquad \sqrt2 |\lambda| = \sqrt2 |\gamma | + m\pi.
	\end{align}
	This gives the topological phase boundaries in the main text.
	
	\subsection{Periodized evolution operator}
	To obtain the periodized evolution operators, one needs to compute the eigenstates of the Floquet operator. Note that all the $ \Gamma $ matrices are Hermitian ones, and the eigenstates of $ U(\boldsymbol{k},T) = \cos E_{\boldsymbol{k}} + i\sin E_{\boldsymbol{k}} M_{\boldsymbol{k}} $ are the same as those of the Hermitian matrix $ M_{\boldsymbol{k}} $: $ U(\boldsymbol{k},T)|E_{\boldsymbol{k}\pm}\rangle = \exp(\pm iE_{\boldsymbol{k}}) |E_{\boldsymbol{k}}\rangle,  M_{\boldsymbol{k}}|E_{\boldsymbol{k}\pm}\rangle = \pm |E_{\boldsymbol{k}}\rangle $, where
	\begin{align}\label{eq:floquet}
	M_{\boldsymbol{k}} = \frac{1}{\sin E_{\boldsymbol{k}}}\sum_{j=1}^4 g_{j\boldsymbol{k}}\Gamma_j = 
	\begin{pmatrix}
	0 & V_{\boldsymbol{k}}^\dagger \\ V_{\boldsymbol{k}} & 0
	\end{pmatrix}, &&
	V_{\boldsymbol{k}} = \frac{1}{\sin E_{\boldsymbol{k}}}(g_{4\boldsymbol{k}}\sigma_0 + i(g_{1\boldsymbol{k}}\sigma_1 + g_{2\boldsymbol{k}}\sigma_2 + g_{3\boldsymbol{k}}\sigma_3)),
	\end{align}
	with unitary matrix $ V_{\boldsymbol{k}}V_{\boldsymbol{k}}^\dagger = \sigma_0 $. The eigenstates then can be easily constructed as
	\begin{align}\label{eq:ufeig10}
	& |E_{\boldsymbol{k}+\uparrow} \rangle = 
	\frac{1}{\sqrt2}\begin{pmatrix}
	\begin{pmatrix}
	1 \\ 0
	\end{pmatrix}
	\\
	V_{\boldsymbol{k}} 
	\begin{pmatrix}
	1 \\ 0
	\end{pmatrix}
	\end{pmatrix}
	, &&
	|E_{\boldsymbol{k}+\downarrow} \rangle = \frac{1}{\sqrt2}
	\begin{pmatrix}
	\begin{pmatrix}
	0 \\ 1
	\end{pmatrix}
	\\
	V_{\boldsymbol{k}} 
	\begin{pmatrix}
	0 \\ 1
	\end{pmatrix}
	\end{pmatrix}
	& |E_{\boldsymbol{k}-\uparrow} \rangle = \frac{1}{\sqrt2}
	\begin{pmatrix}
	\begin{pmatrix}
	-1 \\ 0
	\end{pmatrix}
	\\
	V_{\boldsymbol{k}} 
	\begin{pmatrix}
	1 \\ 0
	\end{pmatrix}
	\end{pmatrix}, &&
	|E_{\boldsymbol{k}-\downarrow} \rangle = \frac{1}{\sqrt2}
	\begin{pmatrix}
	\begin{pmatrix}
	0 \\ -1
	\end{pmatrix}
	\\
	V_{\boldsymbol{k}} 
	\begin{pmatrix}
	0 \\ 1
	\end{pmatrix}
	\end{pmatrix},
	\end{align}
	with corresponding eigenvalues $ E_{\boldsymbol{k}} $ in Eq.~(\ref{defEk}) defined within $[0,\pi] $. The $ \uparrow, \downarrow $ denotes two degenerate bands with the same eigenvalue. Then,
	\begin{align}
	[U(\boldsymbol{k},T)]_{\varepsilon=0}^{-t/T} &= 
	e^{-iE_{\boldsymbol{k}}t/T} (|E_{\boldsymbol{k}+\uparrow}\rangle
	\langle E_{\boldsymbol{k}+\uparrow} | + |E_{\boldsymbol{k}+\downarrow}\rangle
	\langle E_{\boldsymbol{k}+\downarrow} |)
	+
	e^{-i(2\pi-E_{\boldsymbol{k}})t/T} (|E_{\boldsymbol{k}-\uparrow}\rangle
	\langle E_{\boldsymbol{k}-\uparrow} | + |E_{\boldsymbol{k}-\downarrow}\rangle
	\langle E_{\boldsymbol{k}-\downarrow} | ),\\
	[U(\boldsymbol{k},T)]_{\varepsilon=\pi}^{-t/T} &= 
	e^{-iE_{\boldsymbol{k}}t/T} (|E_{\boldsymbol{k}+\uparrow}\rangle
	\langle E_{\boldsymbol{k}+\uparrow} | + |E_{\boldsymbol{k}+\downarrow}\rangle
	\langle E_{\boldsymbol{k}+\downarrow} |)
	+
	e^{iE_{\boldsymbol{k}}t/T} (|E_{\boldsymbol{k}-\uparrow}\rangle
	\langle E_{\boldsymbol{k}-\uparrow} | + |E_{\boldsymbol{k}-\downarrow}\rangle
	\langle E_{\boldsymbol{k}-\downarrow} | )
	\end{align} 
	where
	\begin{align}
	|E_{\boldsymbol{k}+\uparrow}\rangle
	\langle E_{\boldsymbol{k}+\uparrow} | + |E_{\boldsymbol{k}+\downarrow}\rangle
	\langle E_{\boldsymbol{k}+\downarrow} | = \frac{1}{2} (\tau_0\sigma_0 + M_{\boldsymbol{k}}), &&
	|E_{\boldsymbol{k}-\uparrow}\rangle
	\langle E_{\boldsymbol{k}-\uparrow} | + |E_{\boldsymbol{k}-\downarrow}\rangle
	\langle E_{\boldsymbol{k}-\downarrow} | = 
	\frac{1}{2} (\tau_0\sigma_0 - M_{\boldsymbol{k}}) 
	\end{align}
	Thus, they can be simplified as
	\begin{align}\label{bigq}
	[U(\boldsymbol{k},T)]_{\varepsilon=0}^{-t/T} = 
	e^{-i \pi t/T} \left(
	\cos\frac{(\pi-E_{\boldsymbol{k}}) t}{T} \tau_0\sigma_0 + i\sin\frac{(\pi-E_{\boldsymbol{k}})t }{T} M
	\right),
	&& [U(\boldsymbol{k},T)]_{\varepsilon=\pi}^{-t/T} = \cos\frac{E_{\boldsymbol{k}} t}{T} \tau_0\sigma_0 - i\sin\frac{E_{\boldsymbol{k}}t}{T} M,
	\end{align}	
	from which we easily see that at $ t=T $, they become the inverse of Floquet operator $ U (\boldsymbol{k},T) $.
	In summary, we can apply Eqs.~(\ref{fgggg}), (\ref{defEk}), (\ref{eq:floquet}) and (\ref{bigq}) (with previous convention $ 2t/T\rightarrow t\in[0,2] $) to compute the periodized evolution operator $ U_{\varepsilon}(\boldsymbol{k},t) = U(\boldsymbol{k},t) [U(\boldsymbol{k},T)]_\varepsilon^{-t/T} $ at any moment.

	\section{Numerical algorithms}
	\subsection{Static polarization for Floquet operator $ U(\boldsymbol{k},T) $ in strip geometry}
	To be self-contained, we briefly review the procedure to compute static polarization in previous literature. Corresponding to Fig.~2 in the main text, we consider a strip sample finite along $ x $: $ x_i=1,2,\dots,L_x $, while periodic along $ y $: $ k_y=-\pi, -\pi+\Delta_y,\dots, -\pi+(L_y-1)\Delta_y $, where $ \Delta_y=2\pi/L_y $. The construction of static polarization is based on the eigenstates of a static Hamiltonian $ \hat{H}=\sum_{mx_i,nx_j;k_y} \hat{c}^\dagger_{k_y,mx_i} |0\rangle  [H(k_y)]_{mx_i,nx_j} \langle 0| \hat{c}_{k_y,nx_j} $, or a time-independent Floquet operator $\hat{U}(T) = \sum_{mx_i,nx_j;k_y}  \hat{c}^\dagger_{k_y,mx_i} |0\rangle  [U_F(k_y)]_{mx_i,nx_j} \langle 0 |  \hat{c}_{k_y,nx_j} $,
	\begin{align}
	\sum_{nx_j} [H(k_y)]_{(mx_i,nx_j)} [u_{\alpha,k_y}]_{nx_j}  = E_{\alpha,k_y} [u_{\alpha,k_y}]_{mx_i}, && \text{or } && \sum_{nx_j}[U_F(k_y)]_{(mx_i,nx_j)} [u_{\alpha,k_y} ]_{nx_j} = e^{ iE_{\alpha,k_y}} [u_{\alpha,k_y} ]_{mx_i}. 
	\end{align}
	Here $ m,n $ are sublattice indices, while $ \alpha=1,\dots,N_{\text{occ}} $ is the band index. In the case of half-filling with 4 sublattices, $ N_{\text{occ}} = 2L_x $. The position operator projected to occupied bands, $ \hat{y}_{\text{occ}} $ discussed in the main text, 
	is related to the Wilson loop
	\begin{align}
	W_{y,k_y} = F_{k_y+(L_x-1)\Delta_y} \dots F_{k_y+\Delta_y} F_{k_y}, &&
	[F_{k_y}]_{\alpha \beta } = \sum_{mx_i} [u_{\alpha,k_y+\Delta_y }]_{mx_i}^* [u_{\beta,k_y}]_{mx_i}.
	\end{align}
	It can be diagonalized as $ [W_{y,k_y}]_{\alpha \beta } = \sum_{j=1}^{N_{\text{occ}}}  [\nu_j(k_y)]_\alpha e^{2\pi i \nu_j} [\nu_j^*(k_y)]_\beta $. 
	Then, $ \hat{y}_{\text{occ}} |w_j(y_i) \rangle = e^{i\Delta_y(y_i+\nu_j)} |w_j(y_i) \rangle $, 
	where the edge Wannier functions around unit cells $ y_i=1,2,\dots,L_y $ are
	\begin{align}
	|w_j(y_i)\rangle 
	= \frac{1}{\sqrt{L_y}}\sum_{k_y\alpha mx_i} \hat{c}^\dagger_{k_y,mx_i} |0\rangle [\nu_j(k_y)]_\alpha [u_{\alpha,k_y}]_{mx_i} e^{-ik_y y_i}.
	\end{align}
	Finally, the $ x_i $ resolved tangential polarization reads
	\begin{align}
	p_y(x_i) 
	= \frac{1}{2\pi L_y} \sum_{jk_ymy_i} | \langle 0| \hat{c}_{k_y,mx_i} | w_j(y_i)\rangle|^2  \nu_j 
	= \frac{1}{2\pi L_y} \sum_{jk_ym} 
	\left| \sum_{\alpha} [\nu_j(k_y)]_\alpha [u_{\alpha,k_y}]_{mx_i}
	\right|^2 \nu_j.
	\end{align}

	\subsection{Simulation of dynamics for non-interacting particles}
	
	Here we consider the many-body dynamics of free fermions or bosons. The Floquet operator can be writen as
	\begin{align}
	U_F = e^{iH_F}, &&  H_{F} = \Psi^\dagger {\cal H} \Psi, && 
	\Psi = (c_1,\dots,c_N)^T,
	\end{align}
	with $ c_j $ the fermion operator at site $ j $ and $ {\cal H}_{F} $ an $ N\times N $ matrix that can be decomposed into its eigenbasis
	\begin{align}
	{\cal H} = |E_\alpha\rangle E_\alpha \langle E_\alpha |.
	\end{align}
	Note $ |E_\alpha\rangle $ and $ e^{iE_{\alpha}} $ are eigenstates and eigenvalues of the Floquet operator in the first quantized form.
	Here the effective Floquet Hamiltonian $ H_F $ gives the same dynamics as $ U_F $ {\em only} at stroboscopic time $ t=NT, N\in\mathbb{Z} $. Since we do not look at evolution during one period, the branch cut does not matter unlike in the main text.
	
	For an observable $ A $ that can also be expressed in the bilinear form (such as the density $ n_j = c_j^\dagger c_j $)
	\begin{align}
	A = \Psi^\dagger {\cal A} \Psi, && A(NT) = (U_F^\dagger)^N A (U_F)^N = e^{iH_FN} A e^{-iH_FN} ,
	\end{align}
	note $ [AB,C] = A\{B,C\} - \{A,C\}B $ for fermions and $ [AB,C] = A[B,C] + [A,C]B $ for bosons, in both cases we have
	\begin{align}
	(U_F^\dagger)^N \Psi_\mu (U_F)^N = \sum_{n=0}^\infty \frac{(iN)^n}{n!} [(\Psi_\alpha^\dagger {\cal H}_{\alpha\beta} \Psi_\beta)^{(n)}, \Psi_\mu]= (e^{-i{\cal H}N})_{\mu\beta} \Psi_\beta,
	\end{align}
	and therefore 
	\begin{align}
	A(NT) = \Psi^\dagger \left( e^{i{\cal H}N} {\cal A} e^{-i{\cal H}N} \right) \Psi.
	\end{align}
	Now, if the Floquet Hamiltonian matrix can be diagonalized by the unitary matrix $ G $,
	\begin{align}
	G^\dagger {\cal H} G = \text{diag}\{E_1, E_2, \dots, E_D\},
	\end{align}
	where $ D $ is the dimension of  $ {\cal H} $,	we have
	\begin{align}
	A(NT) = \sum_{\alpha,\beta=1}^D \Gamma^\dagger_\alpha G_\alpha^\dagger {\cal A}_{\alpha\beta} G_\beta \Gamma_\beta e^{i(E_\alpha - E_\beta)N}, && \Gamma = G^\dagger \Psi.
	\end{align}
	
	For our purposes, consider the initial state being a Fock one, $ |\psi_{\text{ini}} \rangle = \prod_{i\in\text{corner 2$ \times $ 2}} c_{i}^\dagger |0\rangle $. Then
	\begin{align}
	\langle\psi_{\text{ini}} | A(NT) |\psi_{\text{ini}} \rangle = \sum_{i=1}^N n_i^{(0)} G_{i\alpha} e^{iE_\alpha N} \left( G^\dagger {\cal A} G \right)_{\alpha\beta } e^{-iE_\beta N} G^\dagger_{\beta i},
	\end{align}
	where $ \alpha,\beta $ are summed over all eigenstates, and $ n_i^{(0)} $ (being a real number) is the initial particle number at site $ i $. The above formula applies to both free bosons and fermions when the initial state is a Fock one, with the restriction that for spinless fermions, $ n_i^{(0)}\le1 $ due to Pauli's principle of exclusion.

	\section{Additional details for experimental proposals}

	\subsection{Lattice setup and driving scheme}

	\begin{figure}
		[h]
		\parbox{5cm}{
			\includegraphics[width=5cm]{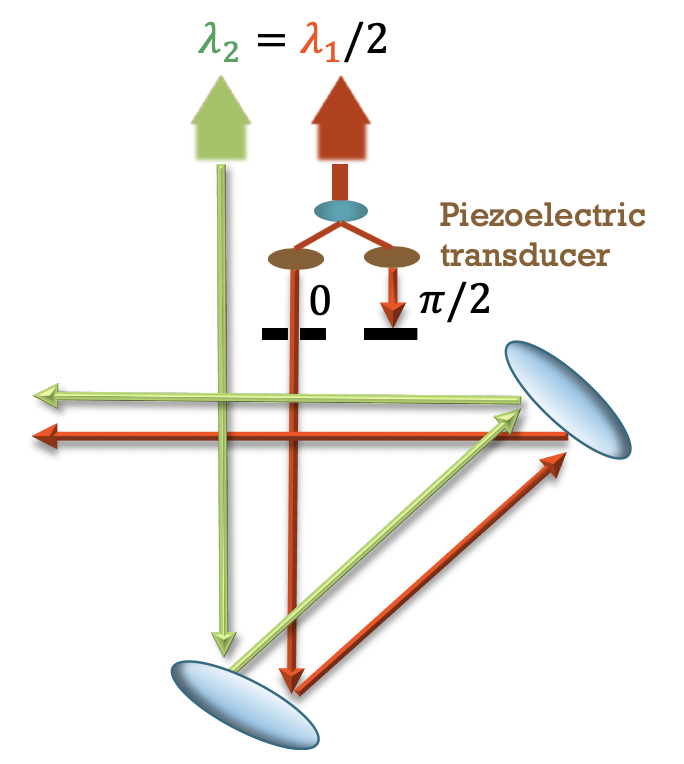}\\ (a) Laser and driving}
		\parbox{3cm}{\quad \\ \qquad \\
			\includegraphics[width=3cm]{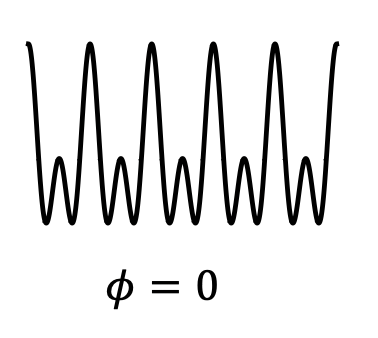}\\
			\includegraphics[width=3cm]{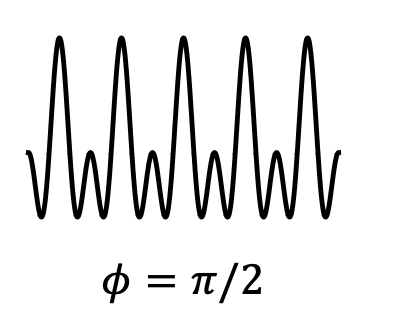} \\  (b) Total potential
		}
		\parbox{4cm}{
			\begin{align*}
			V(x,y) &= V_1 \cos^2 (\frac{2\pi}{\lambda_1} x) + V_2 \cos^2 (\frac{2\pi}{\lambda_2} x + \phi(t) )\\
			& \qquad + (x\rightarrow y)\\
			\phi(t) &= \left\{
			\begin{array}{ll}
			0, & t\in [nT, nT+T/2)\\
			\pi/2, & t\in[nT+T/2+ (n+1)T)
			\end{array}
			\right.
			\end{align*}
		}
		\caption{\label{suppfig:lattsetup} Schematic plot for experimental design.}
	\end{figure}
	
	Since the engineering and driving of bipartite square lattice with gauge fields are highly mature in cold atom experiments (see, i.e.~\cite{Wu2016,Dai2016,Liu2009,Aidelsburger2011,Aidelsburger2014}), we only provide a brief discussion on the lattice setup here. In later subsections, we focus most of our discussions on the detection of anomalous Floquet insulators, which is relatively less explored previously. 
	
	A schematic plot for the laser setup is shown in Fig.~\ref{suppfig:lattsetup}, which is inspired by the USTC experiment~\cite{Wu2016} realizing the much more challenging non-Abelian gauge field in 2D. The $ \pi $-flux can be produced by either shaking or Raman scheme. The bichromatic laser beams, often chosen as red ($ \lambda_1=1064 $nm) and green ($ \lambda_2=532 $nm) ones, has a relative phase difference $ \phi(t) $.  Such a phase can be precisely controlled, i.e. by a piezoelectric transducer(PZT). In practice, it may be most convenient to use a beamsplitter to split the red laser beam, such that each passes through different PZT fixed to static $ \phi=0, \pi/2 $ respectively. Then, the driving consist of opening/closure for the laser beam paths for the two red beams.  Driving it according to Fig.~\ref{suppfig:lattsetup} produces our model.

	\begin{figure}
		[h]
		\parbox{6cm}{
			\includegraphics[width=4cm]{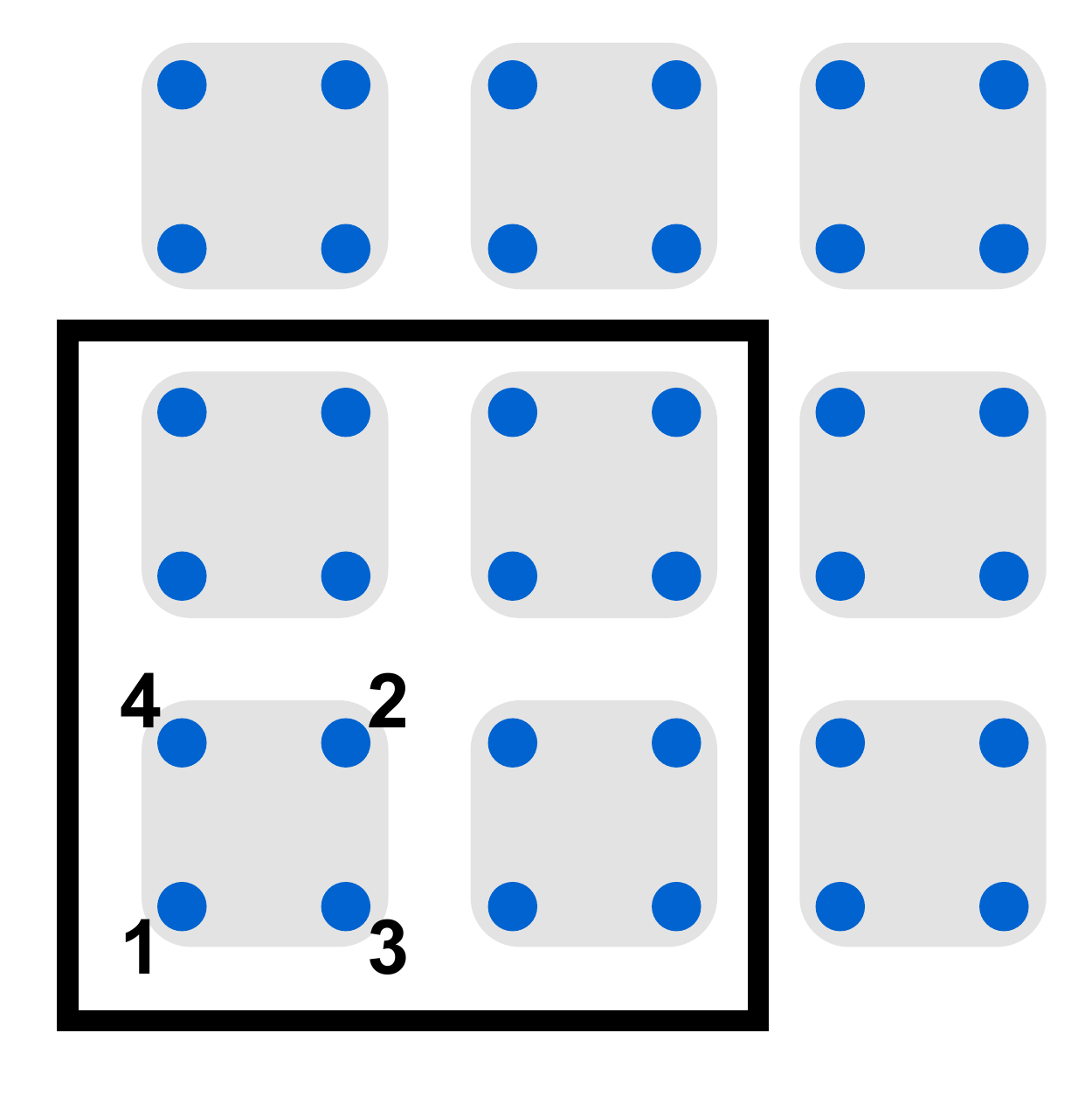}\\ (a)}
		\parbox{6cm}{
			\includegraphics[width=4.5cm]{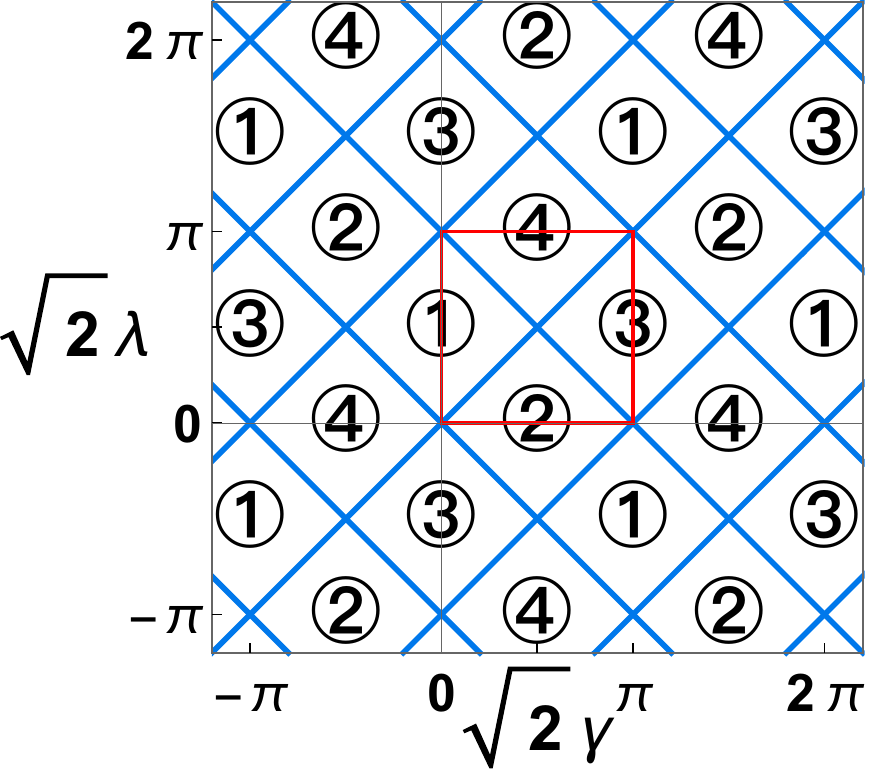} \\ (c)}
		\\
		\parbox{4cm}{
			\includegraphics[width=4cm]{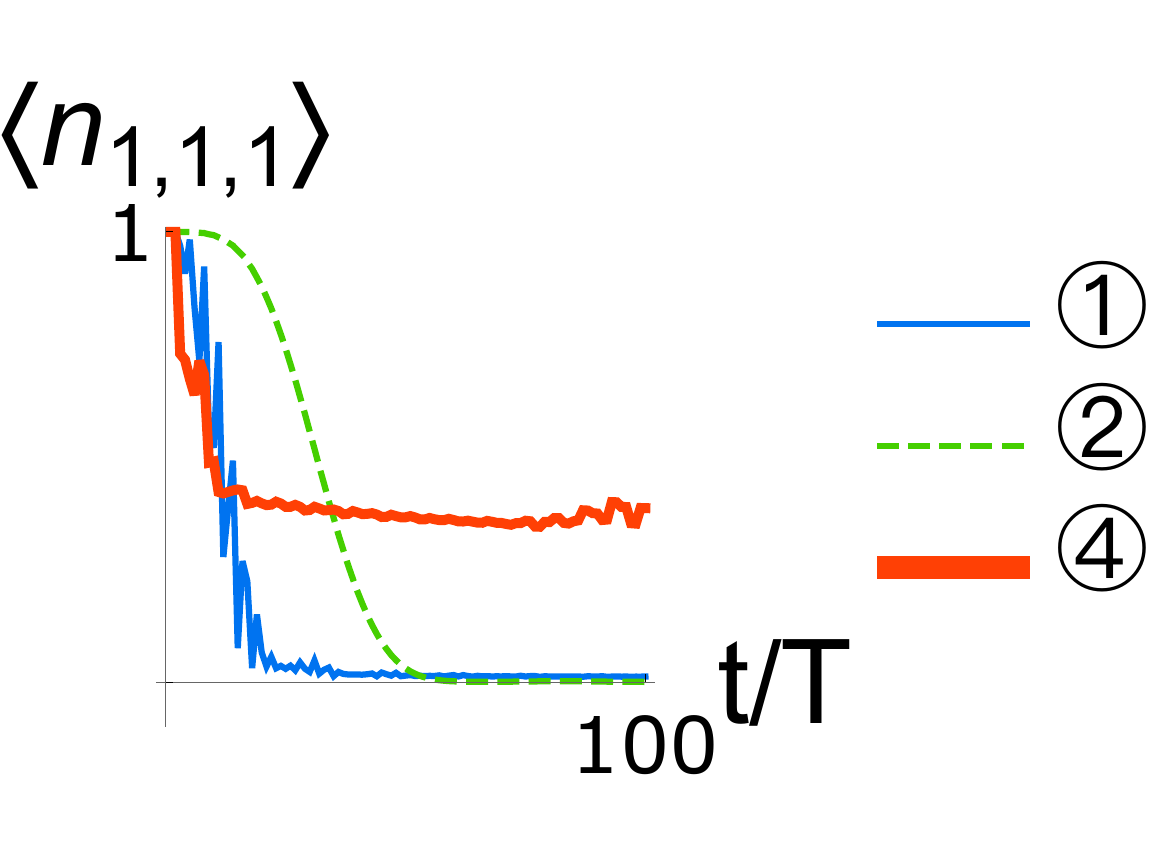}\\ (b1)}
		\parbox{4cm}{
			\includegraphics[width=4cm]{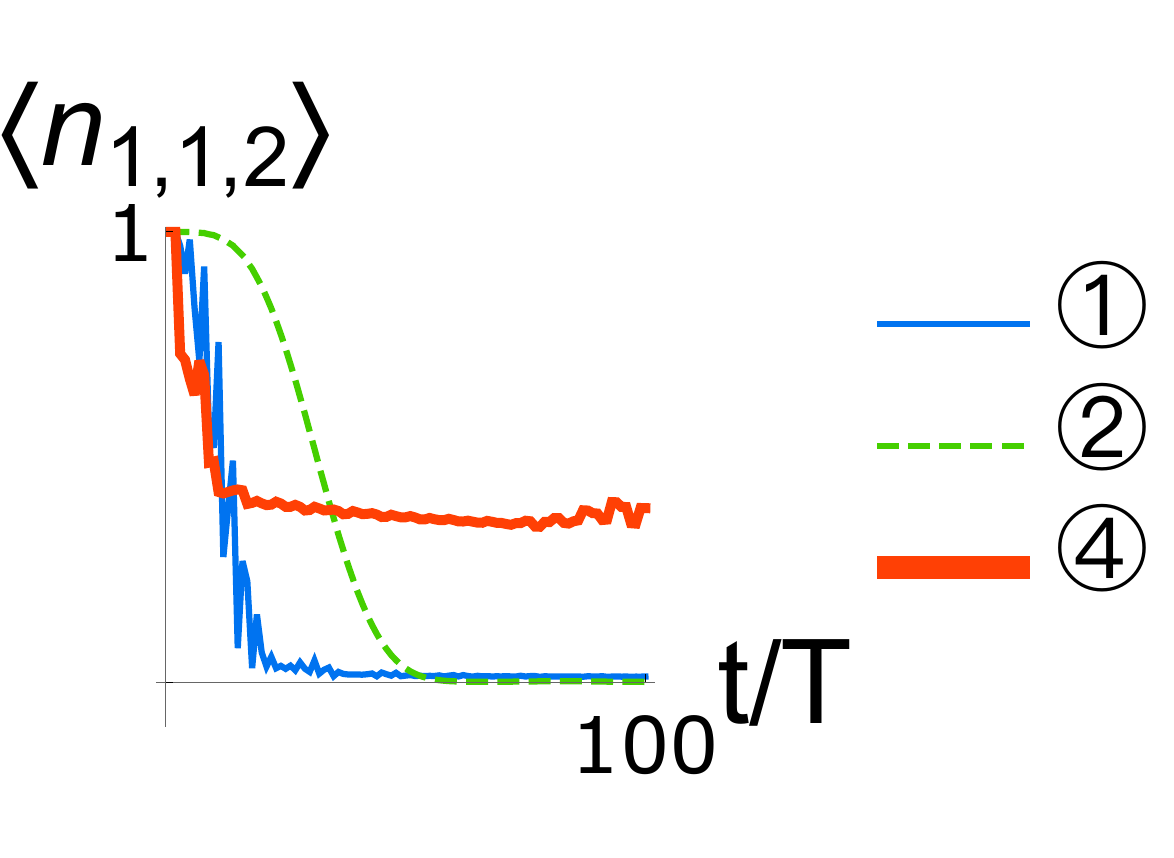}\\ (b2)}
		\parbox{4cm}{
			\includegraphics[width=4cm]{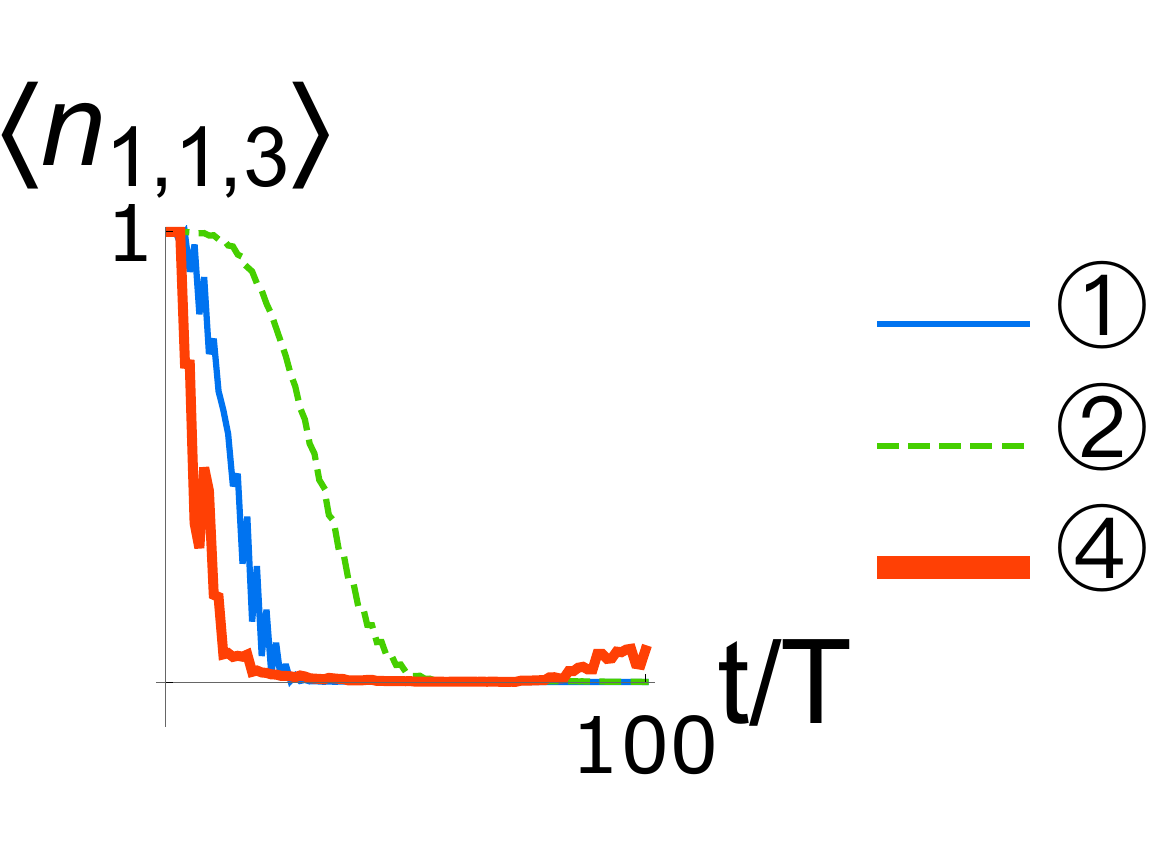}\\ (b3)}
		\parbox{4cm}{
			\includegraphics[width=4cm]{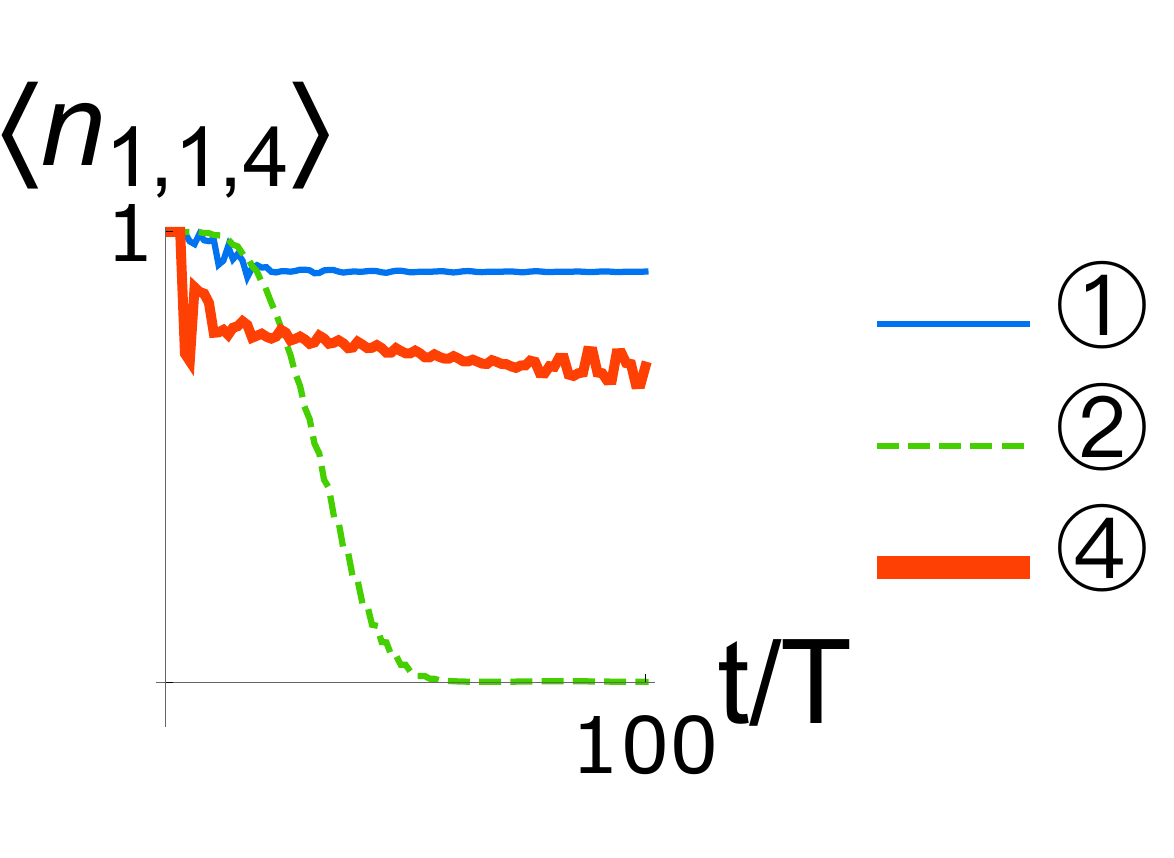}\\ (b4)}
		\caption{\label{fig:expt}The dynamics of density at each sublattice in the corner unit cell. The black frame in (a) denotes the unit cells with initial populations, and the sublattice indices for the corner unit cell is also shown. (b1)--(b4) shows the dynamics of density at the four sublattice sites of the corner unit cell $ n_{x,y,s} $, with $ x=1, y=1, s=1,2,3,4 $. The lattice size is $ 20\times20 $ as in the main text. (c) Parameters in Floquet systems are periodic and the phase diagram in Fig.~2(a) of the main text (red square) starts to repeat once parameters go beyond those presented there. Recall $ \lambda,\gamma $ are already made dimensionless by comparing to driving frequency in the main text.}
	\end{figure}
	\subsection{Sublattice density dynamics at the corner unit cell}
	In the main text, we show the dynamics of averaged density in the corner unit cell. Here, we further demonstrate the dynamics of density within the corner cell for each sublattices respectively. The corner states in phase \pfour\, has more weight in sublattices 1 and 2, compared with those in phase \pone\, (different phases are defined in Fig.~2 of the main text). This result is understandable, because to accommodate two corner modes in phase \pfour, it requires more than one site (i.e. sublattice 4) for spinless particles.

	\subsection{Influence of smooth edge}
	\begin{figure}
		[h]
		\begin{tabular}{c||c|c|c}
			$ \xi $ & $ V(x,y)/V_0 $ & Phase \pfour & Phase \ptwo \\\hline\hline
			0.050 (Sharp boundary) & 
			\parbox{2.2cm}{
				\includegraphics[width=2.2cm]{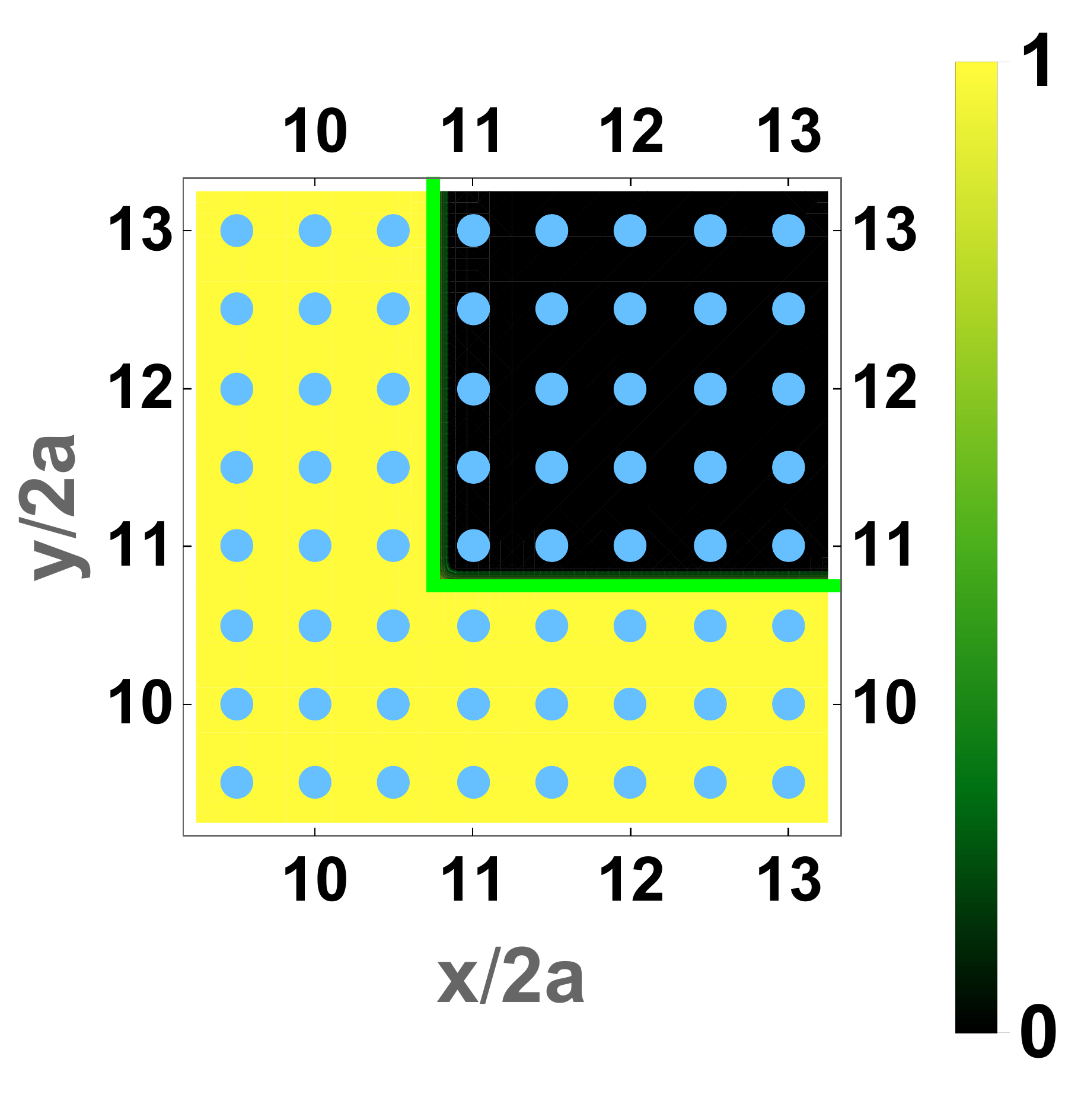}\\
				(a1) } 
			& \parbox{4.4cm}{
				\includegraphics[width=2.2cm,angle=90]{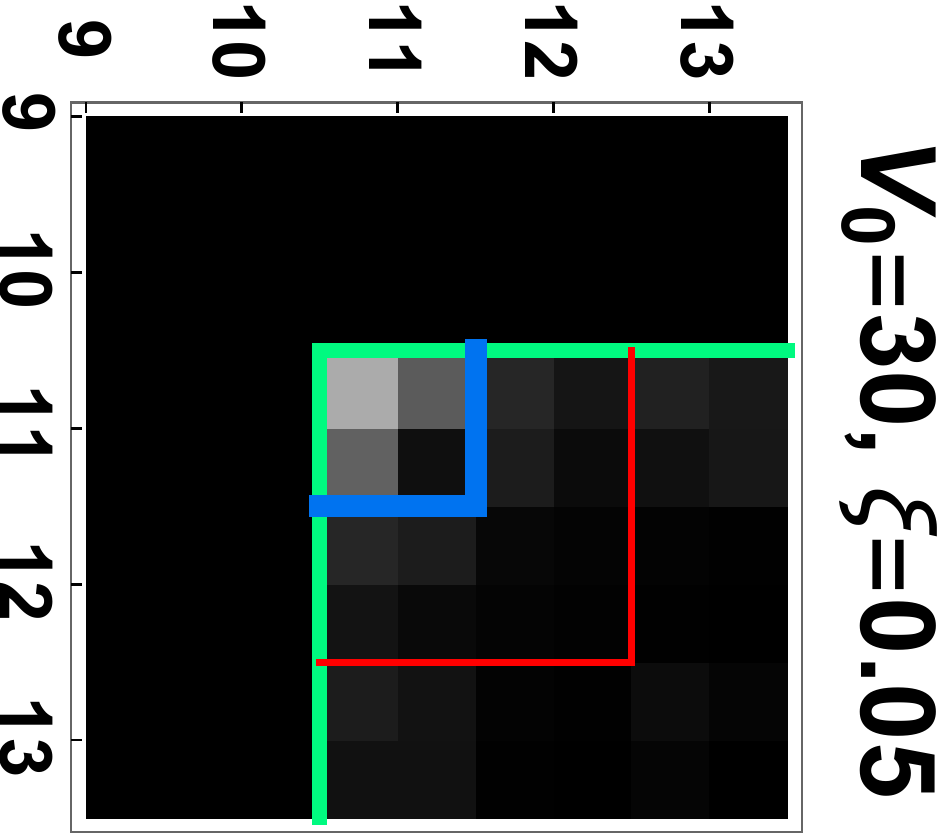} 
				\includegraphics[width=2.2cm]{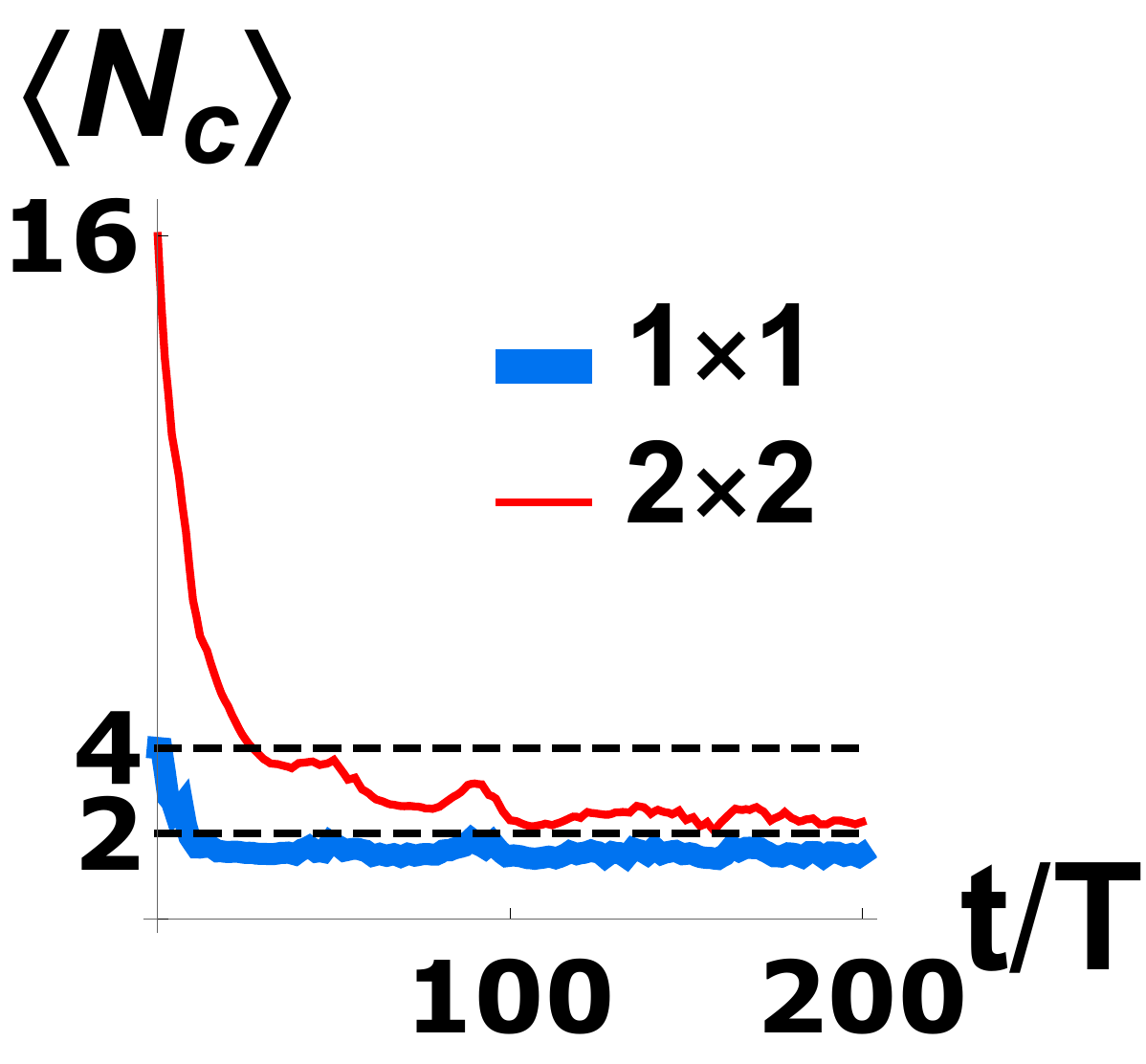}\\
				(a2) }
			&  
			\parbox{4.4cm}{
				\includegraphics[width=2.2cm,angle=90]{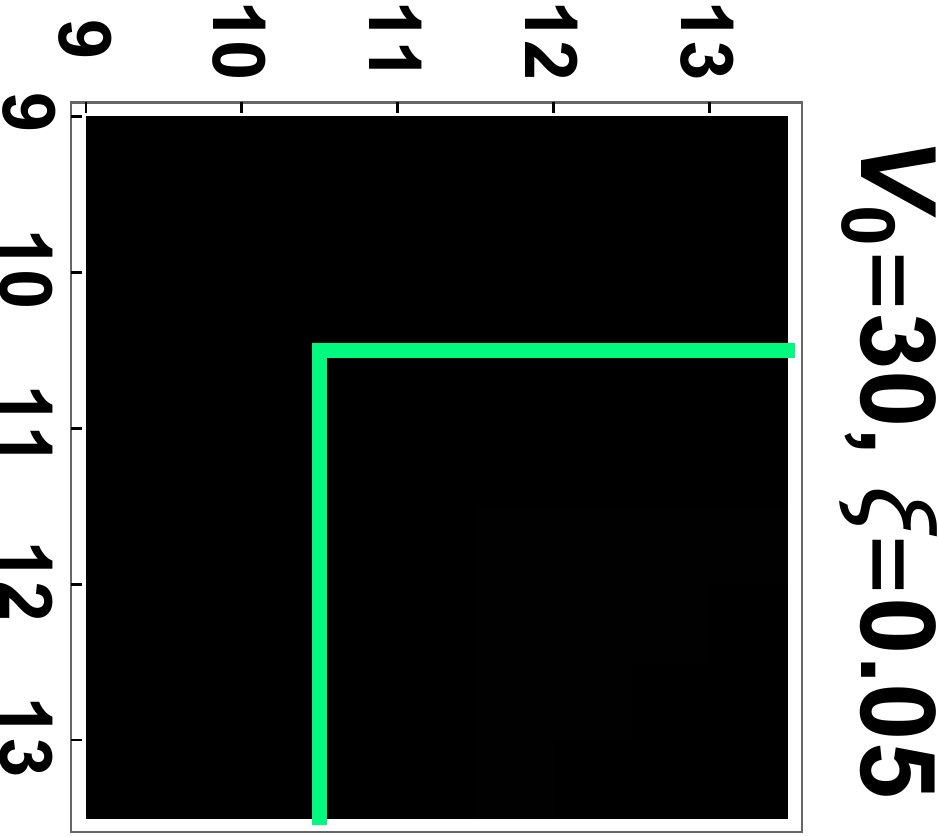} 
				\includegraphics[width=2.2cm]{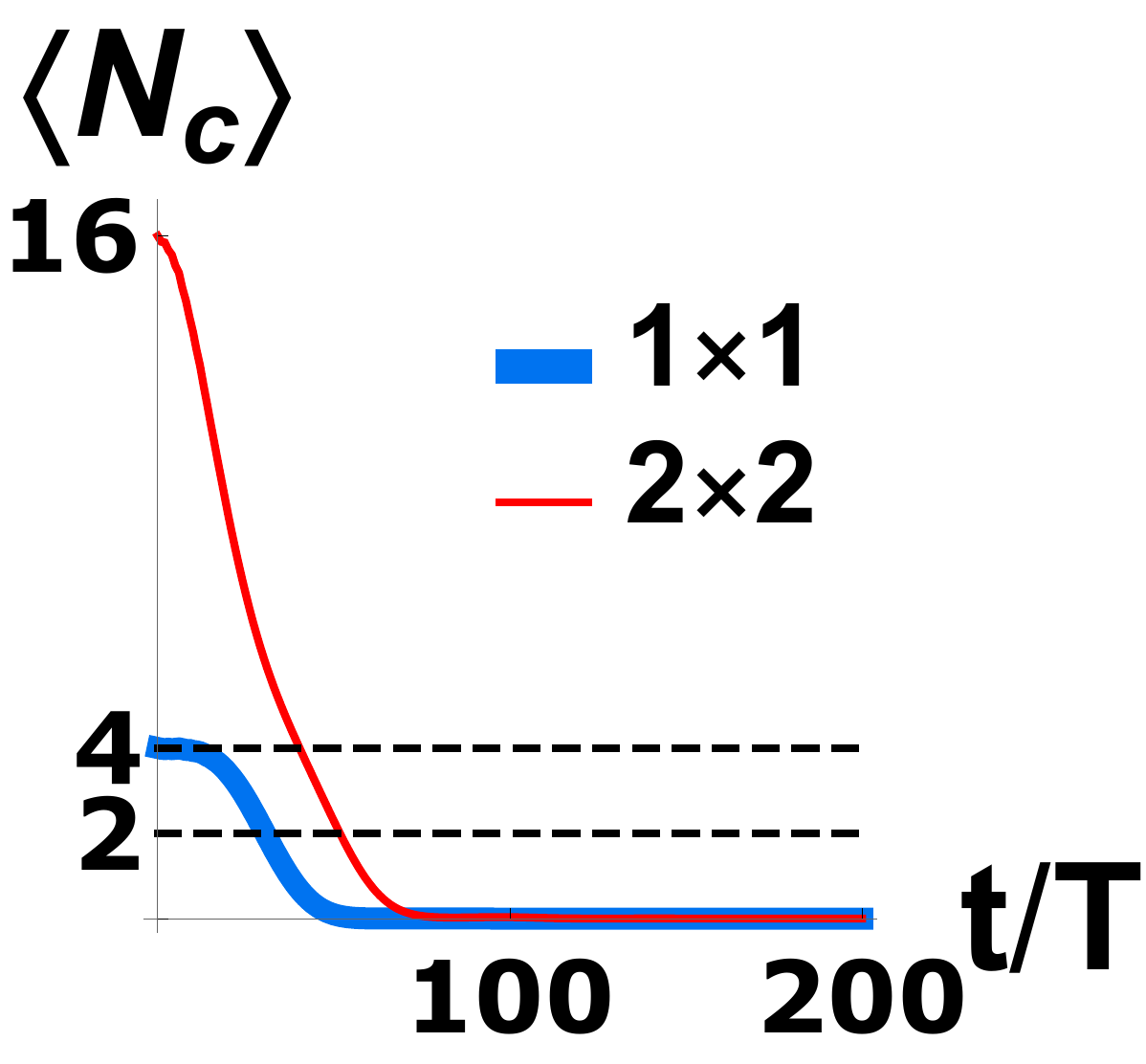}\\
				(a3) }
			\\\hline
			0.125 (Boundary $ \sim 1a $) & 
			\parbox{2.2cm}{
				\includegraphics[width=2.2cm]{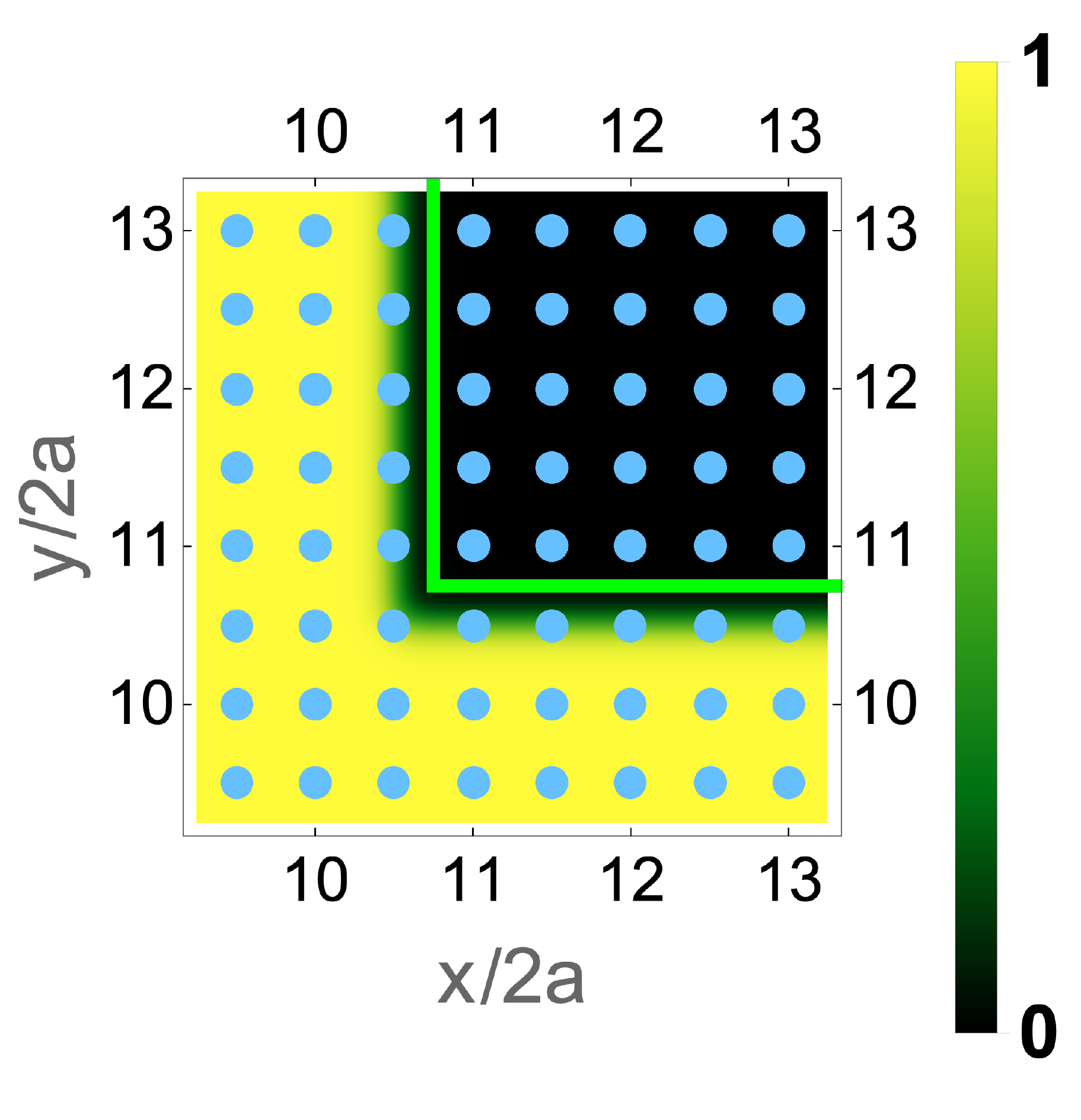}\\
				(b1) } 
			& 
			\parbox{4.4cm}{
				\includegraphics[width=2.2cm,angle=90]{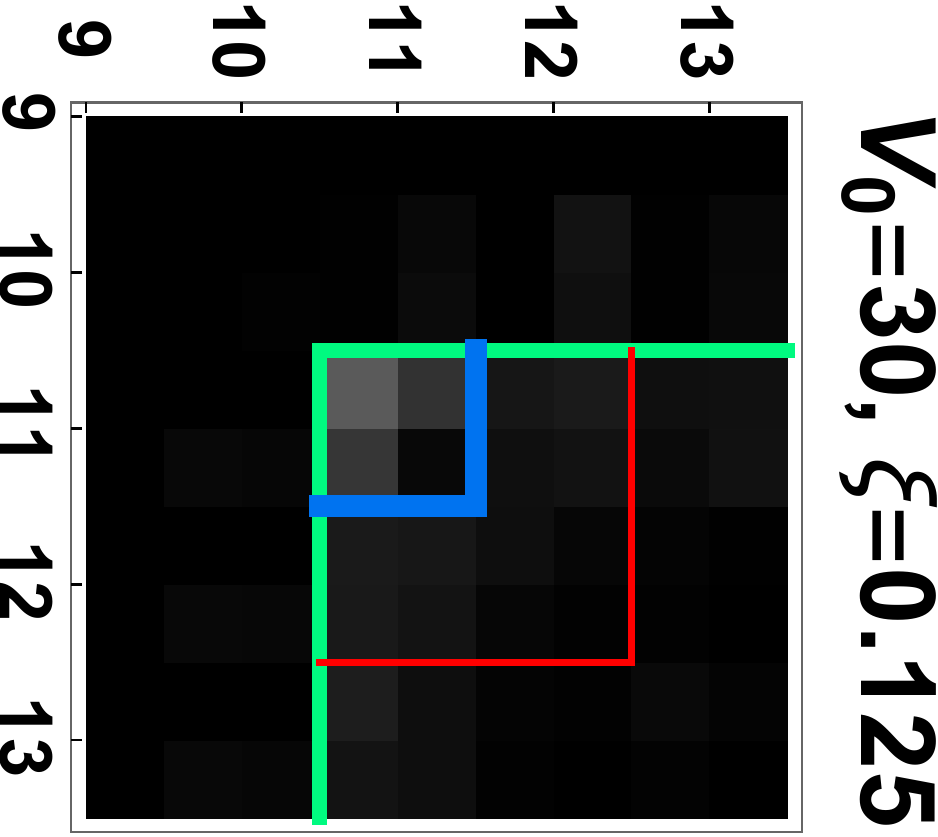} 
				\includegraphics[width=2.2cm]{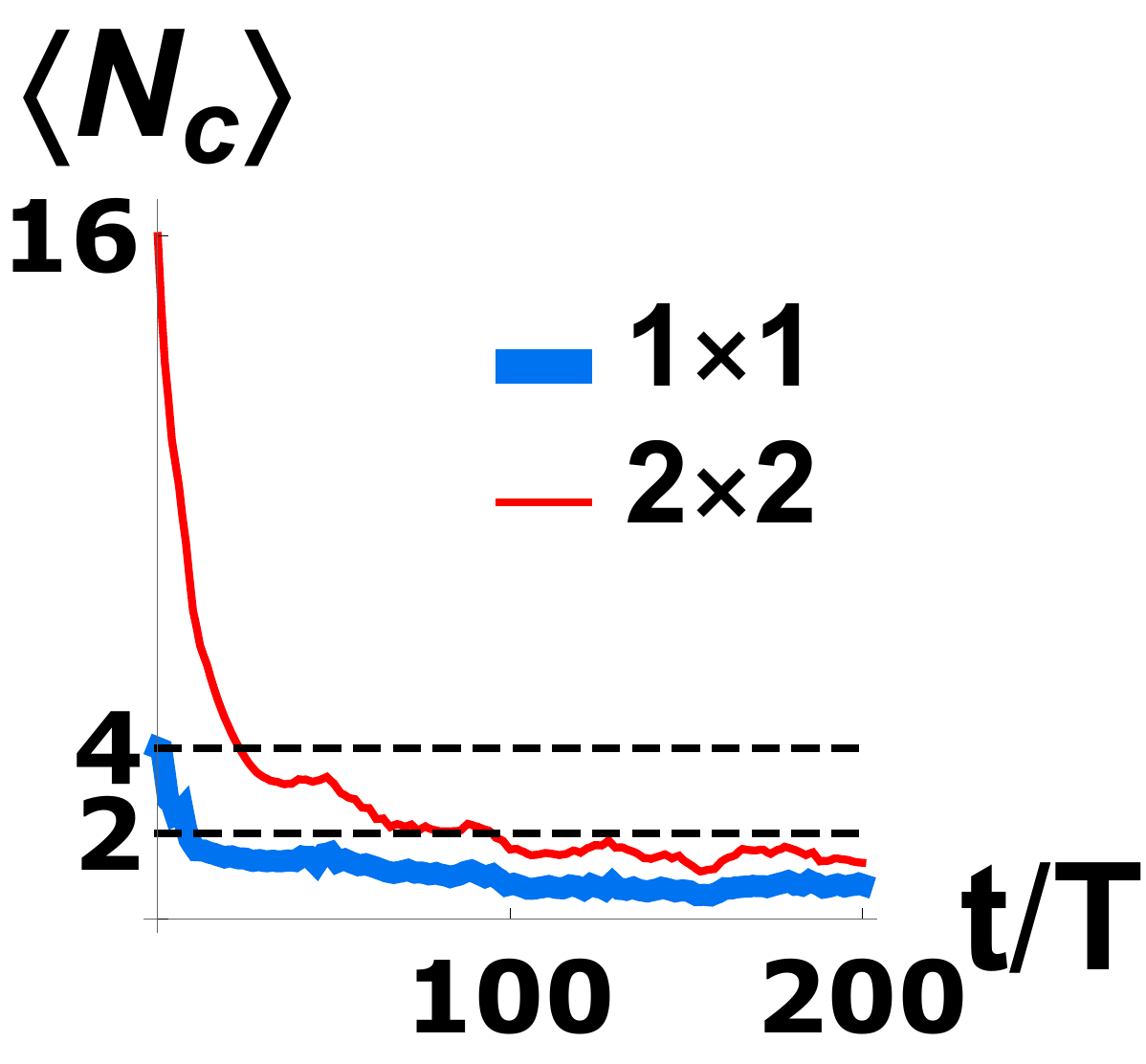}\\
				(b2) }
			& 
			\parbox{4.4cm}{
				\includegraphics[width=2.2cm,angle=90]{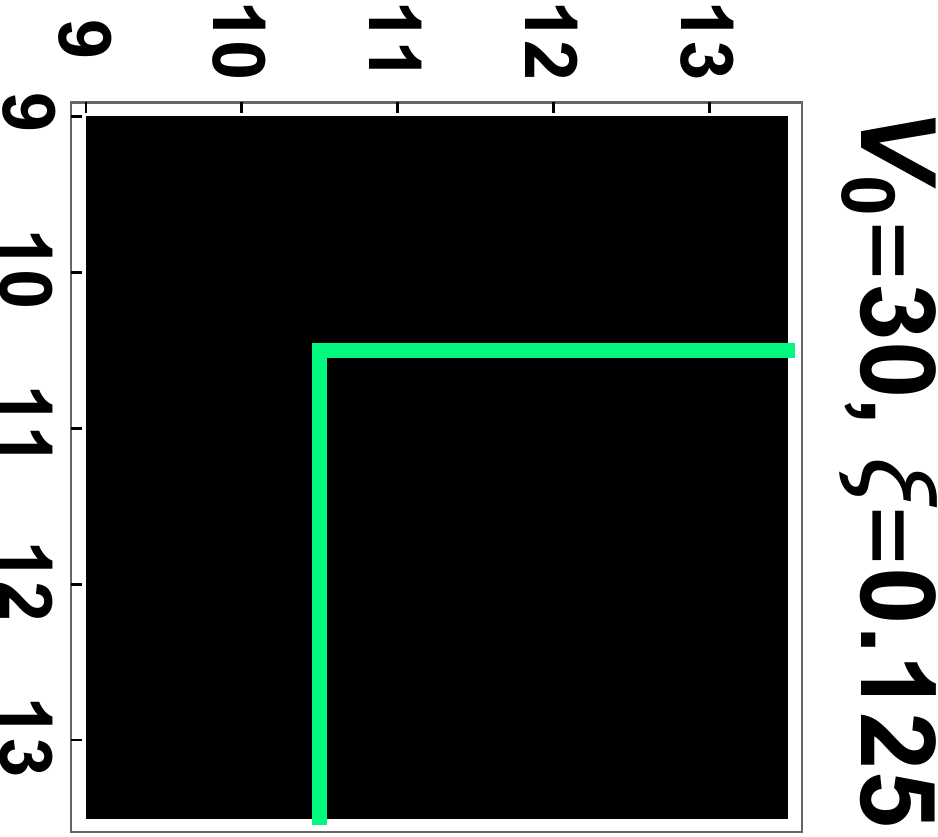} 
				\includegraphics[width=2.2cm]{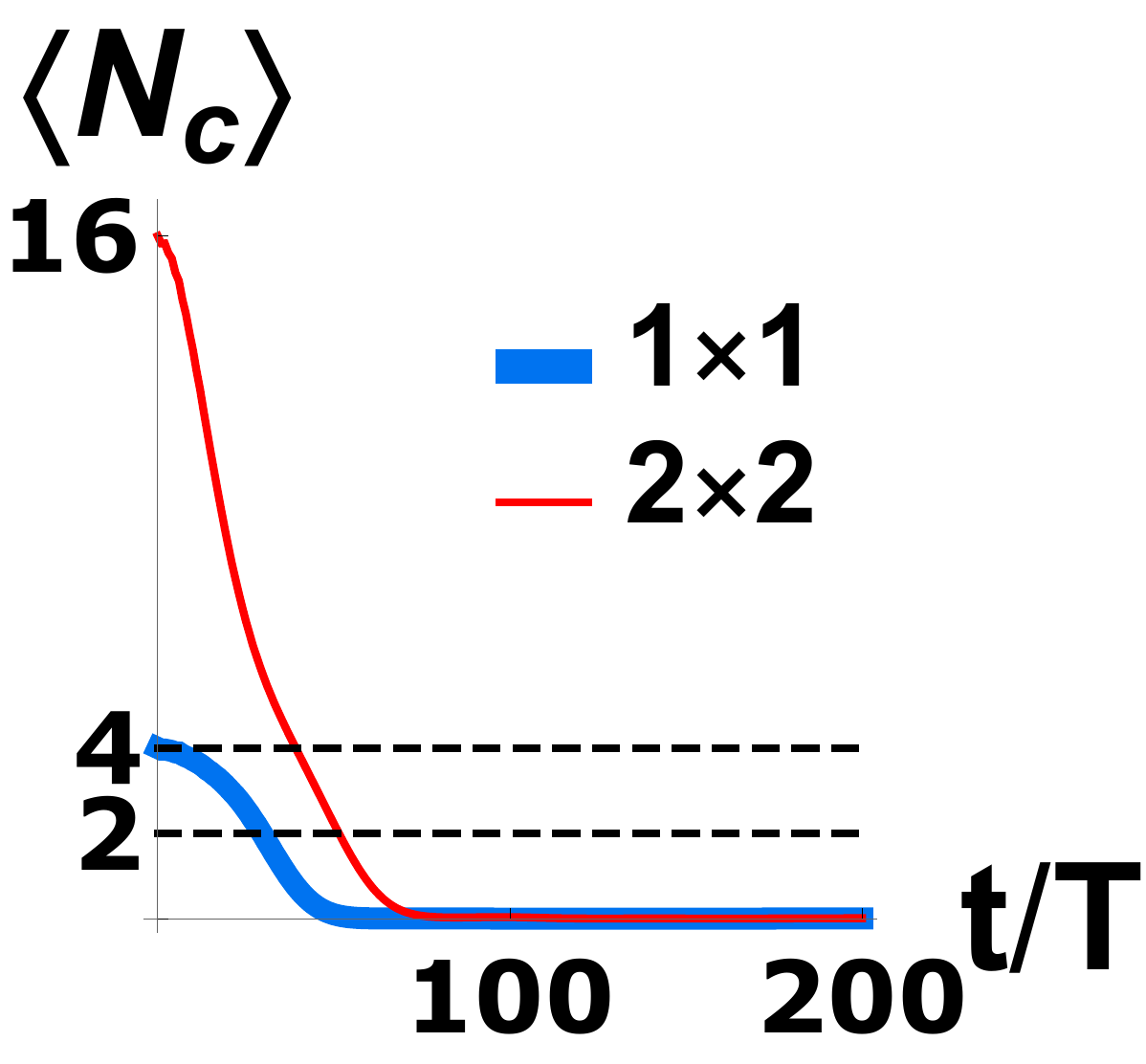}\\
				(b3) }
			\\ \hline
			0.250 (Boundary $ \sim 2a $) & \parbox{2.2cm}{
				\includegraphics[width=2.2cm]{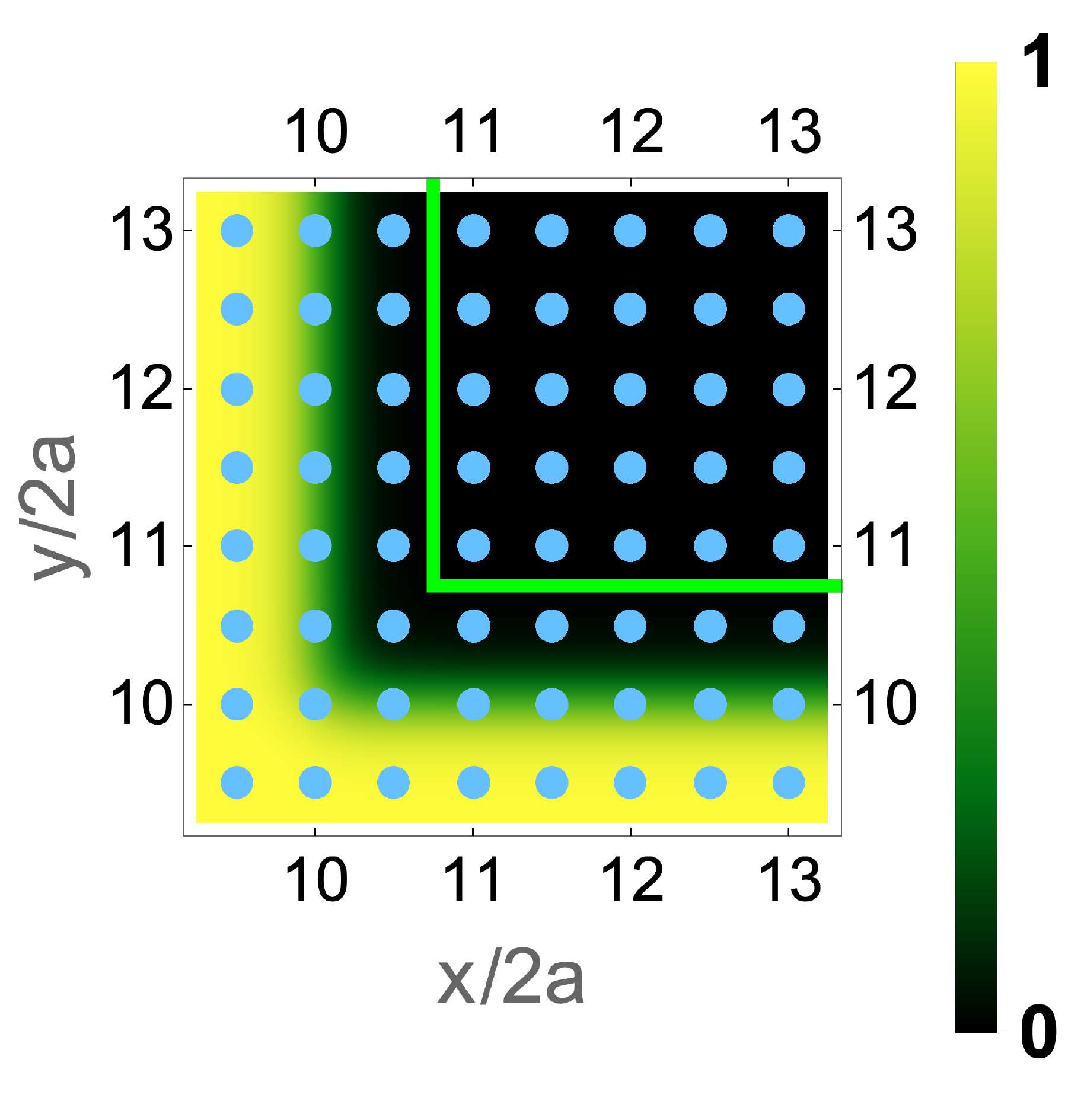}\\
				(c1) } 
			&
			\parbox{4.4cm}{
				\includegraphics[width=2.2cm,angle=90]{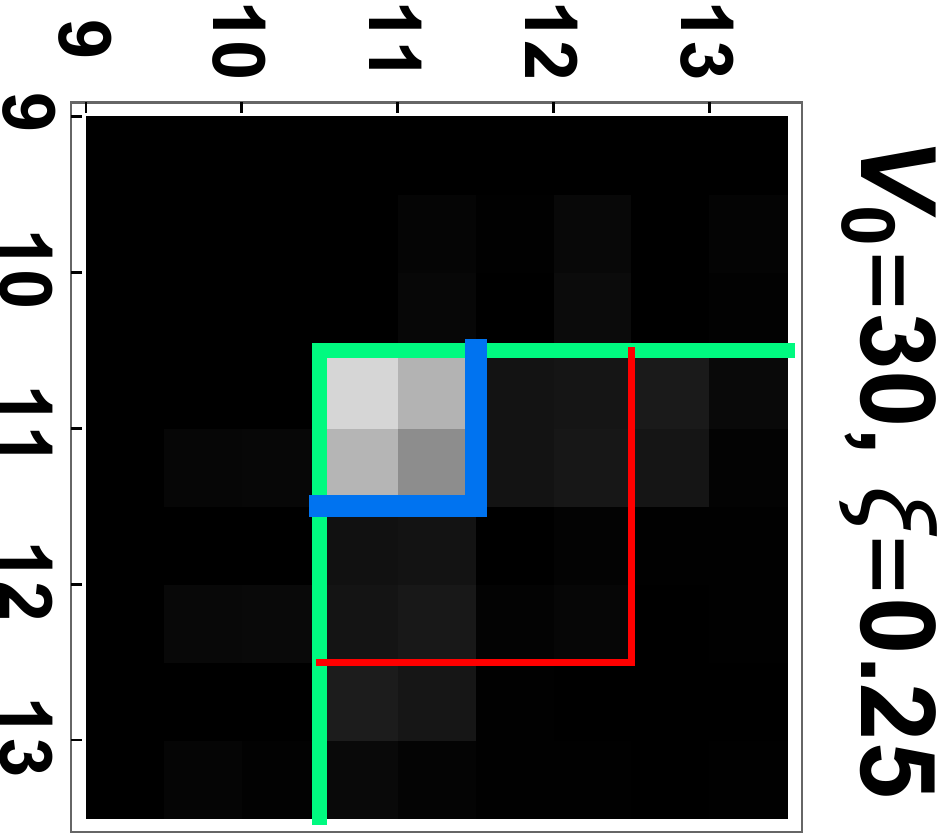} 
				\includegraphics[width=2.2cm]{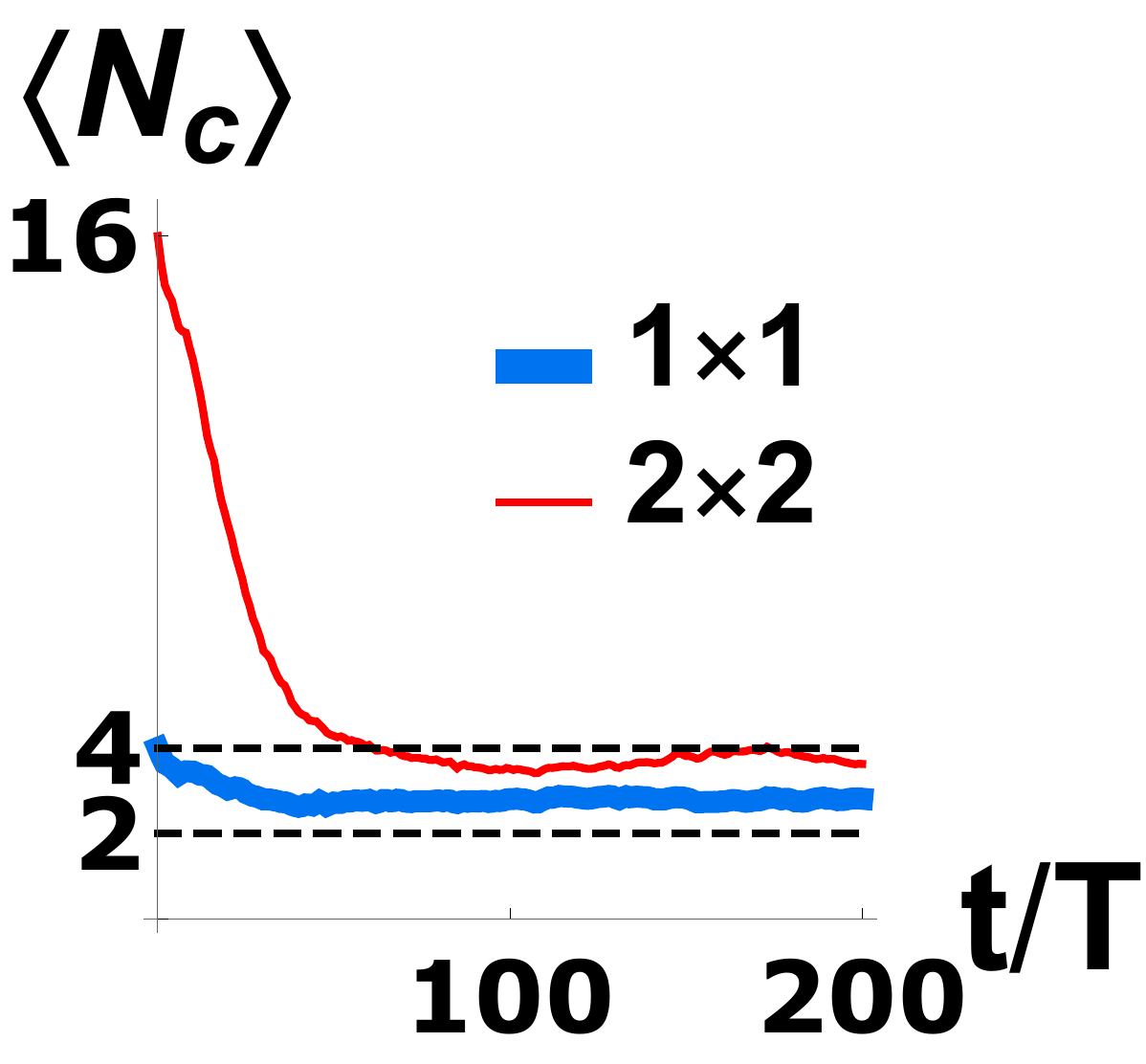}\\
				(c2) }
			&
			\parbox{4.4cm}{
				\includegraphics[width=2.2cm,angle=90]{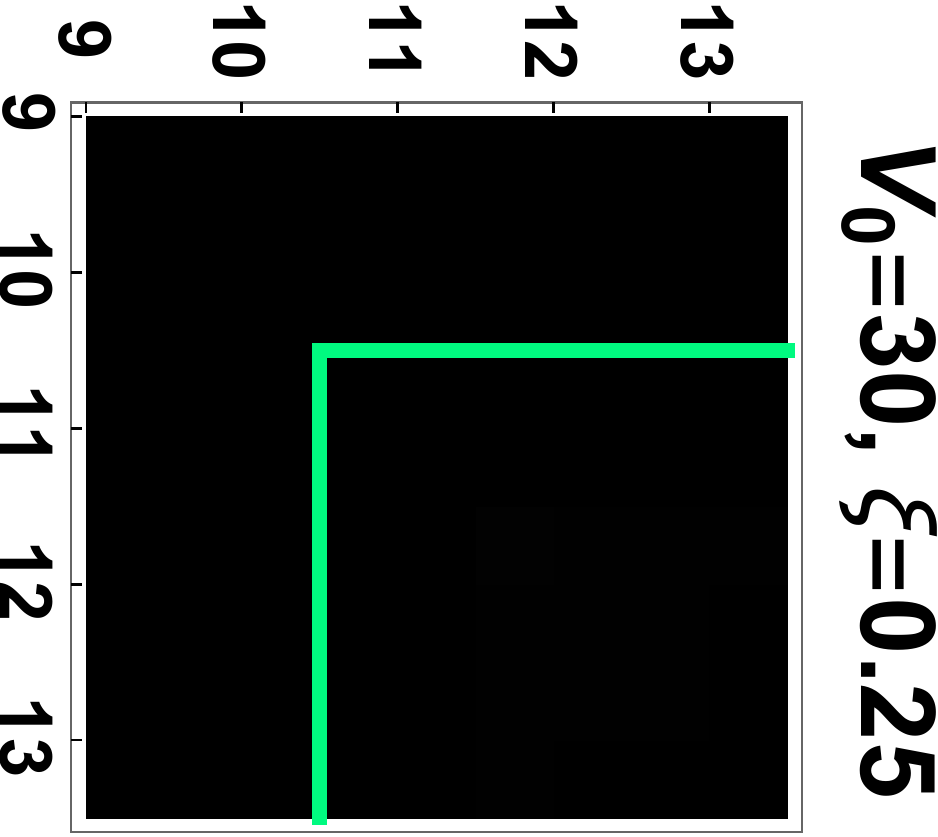} 
				\includegraphics[width=2.2cm]{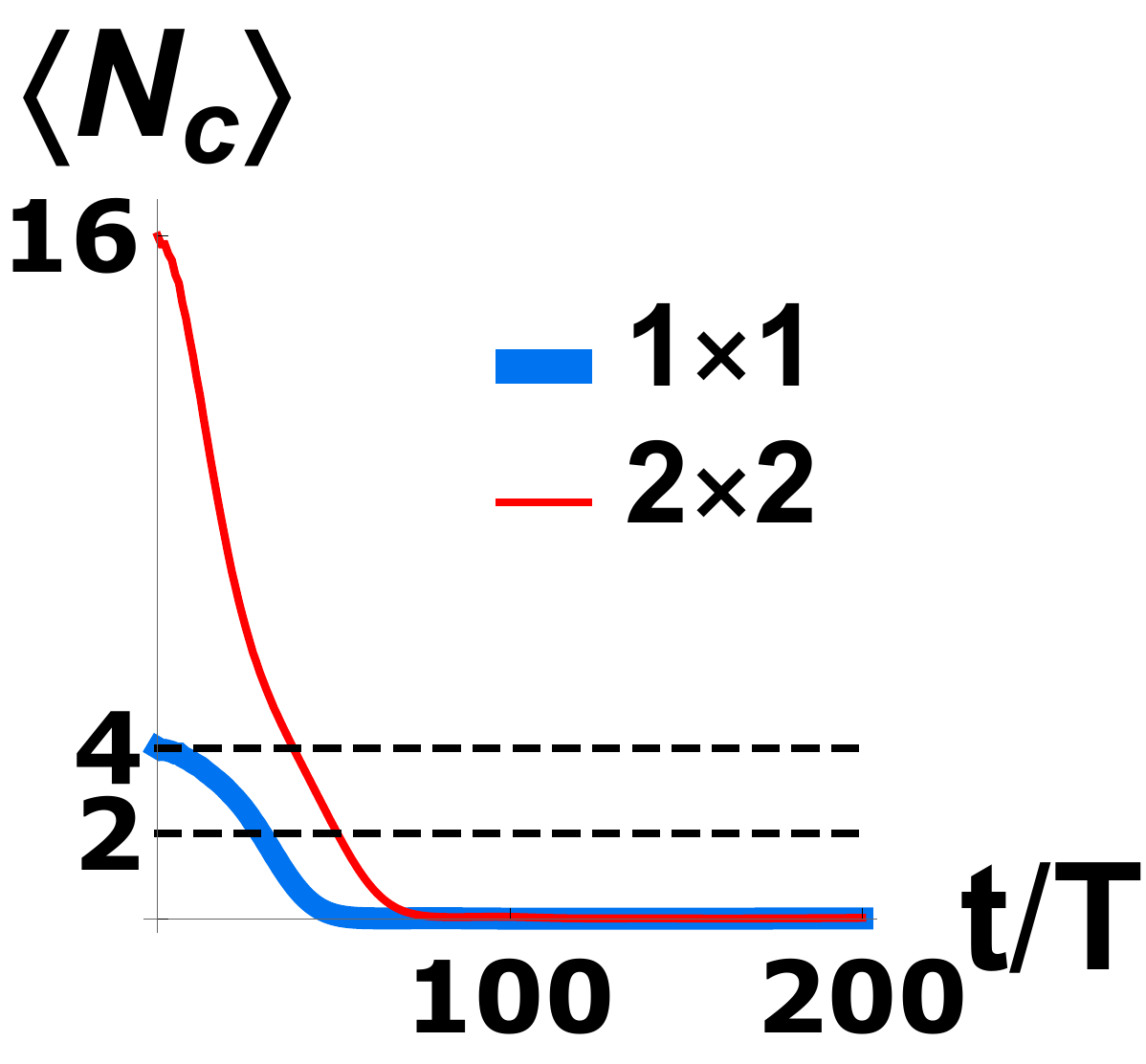}\\
				(c3) }
			\\\hline
		\end{tabular}
		\caption{\label{fig:exptfinite} The influence of finite barrier height and sharpness of boundaries. All unspecified parameters are the same as in Fig.~4 of the main text. For comparison, in phase \pfour\, the parameters for hopping $ \lambda = 0.95\pi/\sqrt2 \approx 2.11 $, so $ V_0 = 30 \approx 14.2\lambda $, which is fairly within the capability of current technology. Here, $ \langle N_c\rangle $ is the total particle number at $ t=100T $ within the $ 1\times 1 $ (blue) or $ 2\times 2 $ (red) unit cells near the corner. The initial state is the same as in the main text, with the whole corner $ 2\times 2 $ region fully populated.}
	\end{figure}
	In the main text, we consider ideal situation of sharp edges for the corner particle dynamics. For real experiments, even though the digital mirror device could engineer the chemical potential site-by-site, the synthetic gauge field (which leads to the $ \pi $-flux in our model) is usually engineered by shaking or moving lattice (Raman scheme) methods. Such perturbation would periodically drive the centers of lattice sites; due to the high frequencies, it corresponds to an effective static ``smearing" of sharp changes for chemical potentials in nearby lattice sites, causing a more smooth boundary. To further simulate the experimental situation, we examine the effect of smooth edge potentials taking the form
	\begin{align}
	V(x,y) = V_0 \left(  \frac{1+\tanh [(x-x_0)/\xi]}{2} \frac{1+\tanh [(y-y_0)/\xi]}{2} \right),
	\end{align}
	which exhibits a smooth boundary for particles in the region $ x>x_0 $ and $ y>y_0 $, see Fig.~\ref{fig:exptfinite} (b1)--(b3). Here $ x=2a(n_x + m_{x,s}), y=2a(n_y+m_{y,s}) $, where $ a $ is the lattice constant, $ (n_x, n_y)\in N $ denote the unit-cells, and $ m_{x,s}, m_{y,s} =0 $ or $ 0.5 $ are the displacements for sublattices $ s=1,2,3,4 $. 
	
	The smoothness of boundaries is controlled by the parameter $ \xi $, and the step-function potential is recovered when $ \xi\rightarrow 0 $. In real experimental situation, the shaking amplitude is usually smaller than 1 lattice constant, which corresponds to $ \xi\sim0.125 $. We put the $ x_0, y_0 $ at the system center, 
	\begin{align}
	x_0/2a=L_x/2+1-4\xi/2a, &&  y_0=L_y/2+1 - 4\xi/2a ,
	\end{align}
	where $ L_{x,y} $ are numbers of unit cells. The reason for shifting $ x_0, y_0 $ by $ 4\xi $ is to avoid chemical potential bias near the boundary within the bulk $ x\ge x_0, y\ge y_0 $; such an in-bulk bias would break the chiral symmetry and annihilate corner states. 
	
	Here, we consider the influence of barrier height $ V_0 $ and the smoothness $ \xi $ on corner particle dynamics. The result of in main text would correspond to $ V_0\rightarrow \infty $ and $ \xi\rightarrow 0 $ limits, which is fairly approximated by the $ V_0=30, \xi=0.05 $ results in Fig.~\ref{fig:exptfinite}(a1)--(a3). Further decreasing the sharpness of the potential from $ \xi=0.05\rightarrow 0.125 $, which spans 1 lattice sites for the boundary region, would lower the visibility of corner states but the situations in phase \pfour\, and \ptwo\, are still qualitatively different. An interesting phenomenon occurs when the sharpness is lowered to certain extent, i.e. to $ \xi=0.25 $ in Tab.~\ref{fig:exptfinite}(c1)--(c3), where the boundary region spans 2 lattice sites. We see that the topological phase shows an {\em increase} of the corner state visibility. This is because the corner region would harbor the corner states both from inside the bulk region ($ x>x_0, y>y_0 $) and outside, which results in doubling of corner states. (Note that the sign of potential $ V_0 $ does not matter, as one can factor out its sign which corresponds to an overall sign of the Hamiltonian. Due to the particle-hole symmetry, such an overall sign does not change the particle dynamics). In all situations, we see that the topologically trivial phase \ptwo\, is quite insensitive to the sharpness of the boundary.

	\subsection{Tomography for Floquet-Bloch band  through time-of-flight for four-band model}

	Compared with the original band tomography method~\cite{Hauke2014,Flaeschner2016}, we encounter two major differences. First, our system involves a slow Floquet driving such that the evolution cannot be represented by an effective static Hamiltonian as in the original fast-driving scheme. Thus, without specific preparations, particles will not equilibrate into the ``lowest" bands. To perform tomography, it is necessary to find a way to concentrate particles into one set of degenerate bands. Second, there are four bands in the model and the two sets of bands are doubly degenerate respectively, unlike in the original case involving only 2 bands without any degeneracy. We tackle these two differences in the following.
	
	\subsubsection{Populating ``lower'' Floquet-Bloch bands under slow driving}
	Since the purpose here is to distinguish the normal HOTI from the anomalous FHOTI, it is adequate to choose certain representative parameter regions in phase \pone\, and \pfour. 
	
	For phase \pone, it is straightforward to notice that when $ \gamma=0 $, the Floquet model reduces to a static one described by $ h_{2\boldsymbol{k}} $ alone, with artificial ``identity evolution" during $ t_1 $ and $ t_3 $. Also, we note that $ h_{2\boldsymbol{k}} $ alone produces two completely flat bands. Thus, one can first equilibrate the system under static $ h_{2\boldsymbol{k}} $ at a temperature larger than the band width (which is zero), but smaller than the band gap $ \lambda $, such that all momentum states at the well-defined lowest two bands are equally populated. Then, one deviates from the static regime by slowly introducing small $ \lambda h_1 $, which does not close the quasi-energy gap and therefore particles would still concentrate in the lower Floquet bands. For optimal effects, we choose $ \sqrt2\lambda = \pi/2 $ such that the quasi-energy band gap is maximal.
	
	For phase \pfour, there is no nearby static fixed points with large quasi-energy gaps. But notice the periodicity of Floquet operators in term of controlling parameters $ \sqrt2\lambda $ and $ \sqrt2\lambda + n\pi $ in Eq.~(\ref{supp:equf}). For instance, $ \sqrt2\lambda = \pi $ in phase \pfour\, has Floquet operators identical to $ \sqrt2\lambda=0 $ in \ptwo. The latter case is a static one evolving under only $ h_1 $. Thus, one can take advantage of the one-to-one correspondence between parameters. First, equilibrate the system into lower bands of $ h_1 $. Then, suddenly start the driving with $ \sqrt2\lambda = \pi $ (and $ \sqrt2\gamma=\pi/2 $ such that the quasi-energy band gap is maximal). Since in this cases the Floquet eigenstates are exactly the same as static ones at $ \lambda=0 $, and due to the large quasi-energy gap, the populations in two Floquet bands would largely remain unaffected. Then, similar to the situation in phase \pone, we can slowly deviate slightly from the fixed point $ \sqrt2(\gamma,\lambda) = \pi(0.5,1) $. The population bias in two sets of bands due to various experimental defects can be further determined by band mapping technique after quenching back to the static limit $ \lambda=0 $.

	\subsubsection{Four-band tomography with lowest two bands being degenerate}
	Here we generalize the framework in Ref.~\cite{Tarnowski2017} to 4-bands. First, we introduce the three angles in $ S^3 $, $ (\theta_{\boldsymbol{k}}, \varphi_{\boldsymbol{k}}, \lambda_{\boldsymbol{k}}) $, characterizing the eigenstates of $ U_F $ as mentioned in the main text. Note the Floquet operator can be written in the form of Eq.~(\ref{eq:floquet}). To lighten notations we suppress most of $ \,_{\boldsymbol{k}} $ indices. Write $ V $ and its eigenstates as
	\begin{align}\nonumber
	& V = \cos\chi+ i \sin\chi
	\begin{pmatrix}
	\cos\theta & \sin\theta e^{-i\varphi} \\
	\sin\theta e^{i\varphi} & -\cos\theta
	\end{pmatrix},
	\\ \nonumber
	&\cos\chi_{} = \frac{g_{4}}{\sin E_{}}, 
	\quad
	\sin\chi = \frac{g}{\sin E}, 
	\quad
	\cos\theta = \frac{g_3}{g},
	\quad
	\sin\theta\cos\varphi = \frac{g_1}{g}, 
	\quad
	\sin\theta \sin\varphi = \frac{g_2}{g},
	\quad
	g \equiv \sqrt{g_1^2+g_2^2+g_3^2}
	\\ \label{eq:angleV}
	& V|\uparrow\rangle = e^{+i\chi_{\boldsymbol{k}}} |\uparrow\rangle, \quad 
	|\uparrow\rangle = 
	\begin{pmatrix}
	\cos\frac{\theta}{2} \\ e^{i\varphi} \sin\frac{\theta}{2}
	\end{pmatrix},
	\qquad\qquad
	V|\downarrow\rangle = e^{-i\chi_{\boldsymbol{k}}} |\downarrow\rangle, \quad
	|\downarrow\rangle = 
	\begin{pmatrix}
	-e^{-i\varphi}\sin\frac{\theta}{2} \\ \cos\frac{\theta}{2}
	\end{pmatrix}.
	\end{align}
	Then, similar to the construction in Eq.~(\ref{eq:ufeig10}), the Floquet eigenstates can be written as $ U_F|E_\pm \rangle = e^{\pm i E} |E_\pm\rangle, |E_\pm\rangle =  \frac{1}{\sqrt2} \begin{pmatrix}
	\pm |\alpha\rangle \\ V |\alpha\rangle
	\end{pmatrix}  $, where $ |\alpha\rangle $ can be an arbitrary SU(2) spinor. But here we choose the two eigenstates of $ V $ rather than $ (1,0) $ and $ (0,1) $ for $ |\alpha\rangle $, so the Floquet eigenstates are
	\begin{align}
	|E_{\boldsymbol{k}\pm\uparrow} \rangle = \frac{1}{\sqrt2} 
	\begin{pmatrix}
	\pm \cos\frac{\theta}{2} \\ \pm e^{i\varphi}\sin\frac{\theta}{2} \\
	e^{i\chi}\cos\frac{\theta}{2} \\ e^{i(\chi+\varphi)}\sin\frac{\theta}{2}
	\end{pmatrix}, &&
	|E_{\boldsymbol{k}\pm\downarrow}\rangle = \frac{1}{\sqrt2} 
	\begin{pmatrix}
	\mp \sin\frac{\theta}{2} \\
	\pm e^{i\varphi}\cos\frac{\theta}{2}\\
	- e^{-i\chi} \sin\frac{\theta}{2} \\
	e^{-i(\chi-\varphi)} \cos\frac{\theta}{2}
	\end{pmatrix}, &&
	\left\{ 
	\begin{array}{l}
	U_F|E_{\boldsymbol{k}+\uparrow,\downarrow}\rangle = e^{+iE_{\boldsymbol{k}}} |E_{\boldsymbol{k}+\uparrow,\downarrow}\rangle, \\ U_F|E_{\boldsymbol{k}-\uparrow,\downarrow}\rangle = e^{-iE_{\boldsymbol{k}}} |E_{\boldsymbol{k}-\uparrow,\downarrow}\rangle\
	\end{array}
	\right.
	\end{align}
	The angles $ (\theta_{\boldsymbol{k}}, \varphi_{\boldsymbol{k}},\lambda_{\boldsymbol{k}}) $ above are the ones used in the main text. With these solutions, we can write the transformation between fermion operators at sublattice $ i $: $ c_{\boldsymbol{k}i} $, and those at certain band $ c_{\boldsymbol{k}\pm,\mu}, \mu=\uparrow, \downarrow $, as
	\begin{align}
	\begin{pmatrix}
	c_{\boldsymbol{k}1} \\
	c_{\boldsymbol{k}2} \\
	c_{\boldsymbol{k}3} \\
	c_{\boldsymbol{k}4}
	\end{pmatrix} = 
	\frac{1}{\sqrt2} 
	\begin{pmatrix}
	\cos\frac{\theta}{2} & -\sin\frac{\theta}{2} & -\cos\frac{\theta}{2} & \sin\frac{\theta}{2} \\
	\sin\frac{\theta}{2} e^{i\varphi} &
	\cos\frac{\theta}{2} e^{i\varphi} & 
	-\sin\frac{\theta}{2} e^{i\varphi} & -\cos\frac{\theta}{2} e^{i\varphi} \\
	\cos\frac{\theta}{2} e^{i\chi} & 
	- \sin\frac{\theta}{2}e^{-i\chi} &
	\cos\frac{\theta}{2} e^{i\chi} &
	-\sin\frac{\theta}{2} e^{-i\chi} \\
	\sin\frac{\theta}{2} e^{i(\chi+\varphi)} & \cos\frac{\theta}{2} e^{-i(\chi-\varphi)} &
	\sin\frac{\theta}{2} e^{i(\chi+\varphi)} & \cos\frac{\theta}{2} e^{-i(\chi-\varphi)} &
	\end{pmatrix}
	\begin{pmatrix}
	c_{\boldsymbol{k}+\uparrow} \\
	c_{\boldsymbol{k}+\downarrow}\\
	c_{\boldsymbol{k}-\uparrow} \\
	c_{\boldsymbol{k}-\downarrow}
	\end{pmatrix}
	\end{align}
	During time-of-flight, four plain waves formed in the four sublattices respectively will interfere with each other. The interference pattern can be described as (up to normalization constants for the total particle number)
	\begin{align}
	n_{\boldsymbol{k}} = \langle \psi_{\text{ini}}| d_{\boldsymbol{k}}^\dagger d_{\boldsymbol{k}} |\psi_{\text{ini}}\rangle, && 
	d_{\boldsymbol{k}} = \frac{c_{\boldsymbol{k}1}+c_{\boldsymbol{k}2}+c_{\boldsymbol{k}3}+c_{\boldsymbol{k}4}}{2}, && \{d_{\boldsymbol{k}}, d_{\boldsymbol{k}}^\dagger \} = 1, && \int \frac{d\boldsymbol{k}}{(2\pi)^2} n_{\boldsymbol{k}} = \frac{1}{2} \text{ (half-filling)}. 
	\end{align}
	Here, we consider the initial state where the two ``$ - $" bands are completely filled, while the two "$ + $" bands are empty, i.e. $ |\psi_{\text{ini}}\rangle = (\prod_{\boldsymbol{k}} c_{\boldsymbol{k}-\uparrow}^\dagger) (\prod_{\boldsymbol{k}} c_{\boldsymbol{k}-\downarrow}^\dagger) |0\rangle $. Then, $ \langle c_{\boldsymbol{k}-\uparrow}^\dagger c_{\boldsymbol{k}-\uparrow}\rangle = \langle c_{\boldsymbol{k}-\downarrow}^\dagger c_{\boldsymbol{k}-\downarrow}\rangle = 1 $, and other terms all vanish. As such, $ n_{\boldsymbol{k}} $ is the sum of two independent interference patterns of $ |E_{\boldsymbol{k}-\uparrow}\rangle $ and $ |E_{\boldsymbol{k}-\downarrow}\rangle $,
	\begin{align}\nonumber
	n^{(4)}_{\boldsymbol{k}} &= \frac{1}{8} \left|
	-\cos\frac{\theta}{2} - \sin\frac{\theta}{2}e^{i\varphi} + \cos\frac{\theta}{2} e^{\chi} + \sin\frac{\theta}{2} e^{i(\chi+\varphi)}
	\right|^2 + 
	\frac{1}{8} \left|
	\sin\frac{\theta}{2} - \cos\frac{\theta}{2} e^{i\varphi} - \sin\frac{\theta}{2}e^{-i\chi} + \cos\frac{\theta}{2} e^{-i(\chi-\varphi)}
	\right|^2\\
	&= \frac{1}{2}(1-\cos\chi_{\boldsymbol{k}}) = \frac{1}{2} \left(1-\frac{g_{4\boldsymbol{k}}}{\sin E_{\boldsymbol{k}}} \right).
	\end{align}
	
	A direct time-of-flight, as we see above, does not involve enough information to back up all three angles $ (\theta_{\boldsymbol{k}}, \varphi_{\boldsymbol{k}}, \lambda_{\boldsymbol{k}}) $. So, similar to~\cite{Hauke2014,Flaeschner2016}, we introduce a quench to deep lattice regime where the states evolve under static onsite chemical potentials $ \mu_{1,2,3,4} $ only, before doing time-of-flight. As mentioned in the main text, we need three chemical potential profiles. For instance, take $ H_{\text{quench}1} = \frac{\Delta_1}{2} \tau_3\sigma_0 $ and evolve for time $ \tau_1 $, then
	\begin{align}
	|E_{\boldsymbol{k}-\uparrow} \rangle \rightarrow
	\frac{e^{-i\Delta_1\tau_1/2}}{\sqrt2} 
	\begin{pmatrix}
	- \cos\frac{\theta}{2} \\ - e^{i\varphi}\sin\frac{\theta}{2} \\
	e^{i\Delta_1 t}\times e^{i\chi}\cos\frac{\theta}{2} \\ e^{i\Delta_1 t}\times e^{i(\chi+\varphi)}\sin\frac{\theta}{2}
	\end{pmatrix}
	, &&
	|E_{\boldsymbol{k}\pm\downarrow}\rangle \rightarrow \frac{e^{-i\Delta_1\tau_1/2}}{\sqrt2} 
	\begin{pmatrix}
	\sin\frac{\theta}{2} \\
	- e^{i\varphi}\cos\frac{\theta}{2}\\
	-e^{i\Delta_1 t}\times e^{-i\chi} \sin\frac{\theta}{2} \\
	e^{i\Delta_1 t}\times e^{-i(\chi-\varphi)} \cos\frac{\theta}{2}
	\end{pmatrix},
	\end{align}
	A time-of-flight after such a procedure produces
	\begin{align}\nonumber
	n_{\boldsymbol{k}}^{(1)} &= \frac{1}{2} (1-\cos\chi_{\boldsymbol{k}} \cos\Delta_1 \tau_1 +  \sin\chi_{\boldsymbol{k}} \sin\theta_{\boldsymbol{k}} \cos\varphi_{\boldsymbol{k}} \sin \Delta_1 \tau_1 )\\ 
	&\xrightarrow{\Delta_1\tau_1 = \pi/2} \frac{1}{2}(1+\sin\chi_{\boldsymbol{k}} \sin\theta_{\boldsymbol{k}}  \cos\varphi_{\boldsymbol{k}}) = \frac{1}{2} \left( 1-\frac{g_1}{\sin E_{\boldsymbol{k}}} \right)
	\end{align}
	Similarly, take $ H_{\text{quench 2}} = \frac{\Delta_2}{4}(\tau_3\sigma_0 + \tau_0\sigma_3 + \tau_3\sigma_3) = \frac{\Delta_2}{4}\text{diag}(3,-1,-1,-1) $ and evolve for time $ \tau_2 $, we have 
	\begin{align}\nonumber
	n_{\boldsymbol{k}}^{(2)} &= \frac{1}{4}(2 - \cos\lambda_{\boldsymbol{k}} (1+\cos\Delta_2\tau_2) - \sin\lambda_{\boldsymbol{k}} ( \sin\theta_{\boldsymbol{k}}\sin\varphi_{\boldsymbol{k}}(1-\cos\Delta_2\tau_2) + (\cos\theta_{\boldsymbol{k}}+ \sin\theta_{\boldsymbol{k}} \cos\varphi_{\boldsymbol{k}})\sin\Delta_2\tau_2 )) \\
	& \xrightarrow{\Delta_2\tau_2=\pi} \frac{1}{2} (1-\sin\lambda_{\boldsymbol{k}} \sin\theta_{\boldsymbol{k}} \sin\varphi_{\boldsymbol{k}} ) = \frac{1}{2} \left( 1-\frac{g_{2\boldsymbol{k}}}{\sin E_{\boldsymbol{k}}} \right)
	\end{align}
	Finally, quenching to $ H_{\text{quench 3}} = \frac{\Delta_3}{2} \tau_3\sigma_3  $ for time $ \tau_3 $, the corresponding time-of-flight signatures will be
	\begin{align}\nonumber
	n_{\boldsymbol{k}}^{(3)} &= \frac{1}{2}(1-\cos\chi_{\boldsymbol{k}} \cos\Delta_3\tau_3 + \sin\chi_{\boldsymbol{k}} \cos\theta_{\boldsymbol{k}} \sin\Delta_3\tau_3 )\\
	&\xrightarrow{\Delta_3\tau_3=\pi/2} \frac{1}{2} (1+\sin\chi_{\boldsymbol{k}}\cos\theta_{\boldsymbol{k}}) = \frac{1}{2} \left(1-\frac{g_{3\boldsymbol{k}}}{\sin E_{\boldsymbol{k}}} \right)
	\end{align}
	Up to this point, we see that $ n_{\boldsymbol{k}}^{(1,2,3,4)} $ contain adequate information to back up the matrix $ V $ in Eq.~(\ref{eq:umv}), and therefore the eigenstates for $ U_F $. The three angles $ (\chi_{\boldsymbol{k}}, \theta_{\boldsymbol{k}}, \varphi_{\boldsymbol{k}}) $ can be written as
	\begin{align}
	\chi_{\boldsymbol{k}} = \arccos \left(1-2n_{\boldsymbol{k}}^{(4)} \right), &&
	\theta_{\boldsymbol{k}} = \arccos\frac{2n_{\boldsymbol{k}}^{(3)}-1}{\sqrt{1-(2n_{\boldsymbol{k}}^{(4)}-1)^2 }}, &&
	\varphi_{\boldsymbol{k}} = \arg\left((2n_{\boldsymbol{k}}^{(1)}-1) + i(1-2n_{\boldsymbol{k}}^{(2)}) \right)
	\end{align}

\end{document}